\documentclass[longauth]{aa}

\usepackage[varg]{txfonts}
\usepackage{graphicx}
\usepackage{url}
\usepackage{natbib}
\usepackage{color}
\usepackage{multirow}

\bibpunct{(}{)}{;}{a}{}{,} 

\begin{document}


\title{Stellar populations of galaxies in the ALHAMBRA survey\thanks{Based on observations collected at the Centro Astron\'omico Hispano Alem\'an (CAHA) at Calar Alto, operated jointly by the Max-Planck Institut für Astronomie and the Instituto de Astrof\'isica de Andaluc\'ia (CSIC)} up to $z \sim 1$}
\subtitle{III. The stellar content of the quiescent galaxy population during the last $8$~Gyr}

%
\author{L.~A.~D\'iaz-Garc\'ia\inst{\ref{a1},\ref{a0}}
\and A.~J.~Cenarro\inst{\ref{a01}}
\and C.~L\'opez-Sanjuan\inst{\ref{a01}}
\and I.~Ferreras\inst{\ref{a2}}
\and A.~Fern\'andez-Soto\inst{\ref{a5},\ref{a6}}
\and R.~M.~Gonz\'alez~Delgado\inst{\ref{a3}}
\and I.~M\'arquez\inst{\ref{a3}}
\and J.~Masegosa\inst{\ref{a3}}
\and I.~San Roman\inst{\ref{a1}}
\and K.~Viironen\inst{\ref{a01}}
\and S.~Bonoli\inst{\ref{a01}}
\and M.~Cervi\~no\inst{\ref{a3},\ref{a4}}
\and M.~Moles\inst{\ref{a1}}
\and D.~Crist\'obal-Hornillos\inst{\ref{a01}}
\and E.~Alfaro\inst{\ref{a3}}
\and T.~Aparicio-Villegas\inst{\ref{a7}}
\and N.~Ben\'itez\inst{\ref{a3}}
\and T.~Broadhurst\inst{\ref{a8},\ref{a9}}
\and J.~Cabrera-Ca\~no\inst{\ref{a10}}
\and F.~J.~Castander\inst{\ref{a11}}
\and J.~Cepa\inst{\ref{a4},\ref{a12}}
\and C.~Husillos\inst{\ref{a3}}
\and L.~Infante\inst{\ref{a13},\ref{a14}}
\and J.~A.~L.~Aguerri\inst{\ref{a4},\ref{a12}}
\and V.~J.~Mart\'inez\inst{\ref{a16},\ref{a17},\ref{a6}}
\and A.~Molino\inst{\ref{a15}}
\and A.~del~Olmo\inst{\ref{a3}}
\and J.~Perea\inst{\ref{a3}}
\and F.~Prada\inst{\ref{a3}}
\and J.~M.~Quintana\inst{\ref{a3}}
}

%

\institute{Centro de Estudios de F\'isica del Cosmos de Arag\'on (CEFCA), Plaza San Juan 1, Floor 2, E--44001 Teruel, Spain\label{a1}\\  \email{ladiaz@asiaa.sinica.edu.tw}
\and Academia Sinica Institute of Astronomy \& Astrophysics (ASIAA), 11F of Astronomy-Mathematics Building, AS/NTU, No.~1, Section 4, Roosevelt Road, Taipei 10617, Taiwan\label{a0}
\and Centro de Estudios de F\'isica del Cosmos de Arag\'on (CEFCA) - Unidad Asociada al CSIC, Plaza San Juan 1, Floor 2, E--44001 Teruel, Spain\label{a01}
\and Mullard Space Science Laboratory, University College London, Holmbury St Mary, Dorking, Surrey RH5 6NT, United Kingdom\label{a2}
\and Instituto de F\'isica de Cantabria (CSIC-UC), E-39005 Santander, Spain\label{a5}
\and Unidad Asociada Observatorio Astron\'omico (IFCA-UV), E-46980, Paterna, Spain\label{a6}
\and IAA-CSIC, Glorieta de la Astronom\'ia s/n, 18008 Granada, Spain\label{a3}
\and Instituto de Astrof\'isica de Canarias, V\'ia L\'actea s/n, 38200 La Laguna, Tenerife, Spain\label{a4}
\and Observat\'orio Nacional-MCT, Rua Jos\'e Cristino, 77. CEP 20921-400, Rio de Janeiro-RJ, Brazil\label{a7}
\and Department of Theoretical Physics, University of the Basque Country UPV/EHU, 48080 Bilbao, Spain\label{a8}
\and IKERBASQUE, Basque Foundation for Science, Bilbao, Spain\label{a9}
\and Departamento de F\'isica At\'omica, Molecular y Nuclear, Facultad de F\'isica, Universidad de Sevilla, 41012 Sevilla, Spain\label{a10}
\and Institut de Ci\`encies de l'Espai (IEEC-CSIC), Facultat de Ci\`encies, Campus UAB, 08193 Bellaterra, Spain\label{a11}
\and Departamento de Astrof\'isica, Facultad de F\'isica, Universidad de La Laguna, 38206 La Laguna, Spain\label{a12}
\and Instituto de Astrof\'{\i}sica, Universidad Cat\'olica de Chile, Av. Vicuna Mackenna 4860, 782-0436 Macul, Santiago, Chile\label{a13}
\and Centro de Astro-Ingenier\'{\i}a, Universidad Cat\'olica de Chile, Av. Vicuna Mackenna 4860, 782-0436 Macul, Santiago, Chile\label{a14}
\and Observatori Astron\`omic, Universitat de Val\`encia, C/ Catedr\`atic Jos\'e Beltr\'an 2, E-46980, Paterna, Spain\label{a16}
\and Departament d'Astronomia i Astrof\'isica, Universitat de Val\`encia, E-46100, Burjassot, Spain\label{a17}
\and Instituto de Astronom{\'{\i}}a, Geof{\'{\i}}sica e Ci\'encias Atmosf\'ericas, Universidade de S{\~{a}}o Paulo, S{\~{a}}o Paulo, Brazil\label{a15}
}

\date{Received ? / Accepted ?}

%
\abstract{}  
         {We aim at constraining the stellar population properties of quiescent galaxies. These properties reveal how these galaxies evolved and assembled since $z\sim1$ up to the present time.}  
         {Combining the ALHAMBRA multi-filter photo-spectra with the SED-fitting code MUFFIT, we build a complete catalogue of quiescent galaxies via the dust-corrected stellar mass vs colour diagram. This catalogue includes stellar population properties, such as age, metallicity, extinction, stellar mass and photometric redshift, retrieved from the analysis of composited populations based on two independent sets of SSP models. We develop and apply a novel methodology to provide, for the first time, the analytic probability distribution functions (PDFs) of mass-weighted age, metallicity, and extinction of quiescent galaxies as a function of redshift and stellar mass. We adopt different star formation histories to discard potential systematics in the analysis.}     
         {The number density of quiescent galaxies is found to increase since $z\sim1$, with a more substantial variation at lower mass. Quiescent galaxies feature extinction $A_V<0.6$, with median values in the range $A_V = 0.15$--$0.3$. At increasing stellar mass, quiescent galaxies are older and more metal rich since $z\sim1$. A detailed analysis of the PDFs reveals that the evolution of quiescent galaxies is not compatible with passive evolution and a slight decrease is hinted at median metallicity $0.1$--$0.2$~dex. The intrinsic dispersion of the age and metallicity PDFs show a dependence with stellar mass and/or redshift. These results are consistent with both sets of SSP models and the alternative SFH assumptions explored. Consequently, the quiescent population must undergo an evolutive pathway including mergers and/or remnants of star formation to reconcile the observed trends, where the ``progenitor'' bias should also be taken into account.}     
         {}  
         
%

%
\titlerunning{The stellar content of quiescent galaxies since $z\sim 1$}
%
\authorrunning{L.~A.~D\'iaz-Garc\'ia et al.}

\maketitle


%


\section{Introduction}\label{sec:introduction}

Over the past two decades, many authors found that galaxies lie on two well differentiated groups or colour distributions \citep[e.~g.][]{Bell2004,Baldry2004,Williams2009,Ilbert2010,Peng2010,Arnouts2013,Moresco2013,Fritz2014}. This bimodality can be interpreted in terms of differences of either the stellar content of the galaxies, or variations in their evolutive pathways. In this sense, red galaxies typically exhibit evolved stellar populations with low levels of star formation, and are termed quiescent, passive or even ``dead'' galaxies. The formation and evolution of the so-called quiescent galaxies remains a challenge to date, as these galaxies started to form stars at very early epochs, shutting down their star formation at later times by mechanisms that are still open to debate \citep{Faber2007,Peng2010,Ilbert2013,Peng2015,Barro2016}. One of the most extended and promising approaches to determine the star formation history (SFH) of quiescent galaxies is the study of the evolution of their stellar population content with cosmic time.

Many authors focused on the SFH of galaxies in low-redshift samples. This is typically termed the ``archaeological'' approach or fossil record method. This method has been extensively used to assess the stellar content of galaxies through either integrated properties or spatially-resolved observations \citep[e.~g.][]{Cid2005,Gallazzi2005,Thomas2005,Rogers2010,delaRosa2011,Trevisan2012,FerreMateu2013,Conroy2014,Trujillo2014,Belli2015,Mcdermid2015,
GonzalezDelgado2015,Citro2016,Zheng2016,Goddard2017,GonzalezDelgado2017}. The analysis is based on the (full) SED fitting or on targeted spectral indices that are sensitive to parameters such as age, metallicity, $\alpha$ enhancement, IMF, etc. These methods usually adopt stellar population models with different SFHs, including bursts of various durations. Alternatively to the fossil record method, the comparison between the stellar populations of similar samples at high and low redshifts (termed the ``look-back'' approach) provides complementary constraints to galaxy evolution \citep[e.~g.][]{Schiavon2006,vanDokkum2008,SanchezBlazquez2009,Choi2014,Gallazzi2014,Jorgensen2014,Fagioli2016,Gargiulo2016,Kriek2016,Siudek2017}. Whilst the ``look-back'' studies constitute a direct comparison between distributions of galaxies at different redshift, any interpretation of the results is limited by the ``progenitor'' bias\citep[a term introduced by][]{vanDokkum2001}. Consequently, samples of galaxies at high redshift are biased subsets of the nearby counterparts, because the former only includes the oldest members of the current distributions. In fact, recent studies point out that there is an increasing number of quiescent galaxies since $z\sim3$, supporting the idea that samples of quiescent galaxies at high or intermediate redshift are largely biased when compared to their low redshift counterpart \citep[e.~g.][]{Drory2009,Pozzetti2010,Ilbert2010,Brammer2011,Cassata2011,Davidzon2013,Ilbert2013,Moustakas2013,Moresco2013,Tomczak2014}. Other recent results advocate a reduction in the number of massive star-forming galaxies \citep[][]{Bell2007,Davidzon2013,Ilbert2013,Moustakas2013}, a hypothesis that also explains the observational size growth of massive quiescent galaxies \citep[e.~g.][]{vanDokkum2008b,Shankar2009,Belli2015,Fagioli2016,Gargiulo2016,Mcdermid2015,Williams2016} and the scatter in the red sequence \citep[RS, ][]{Harker2006,Ruhland2009}.

Some of these quiescent galaxies are very old \citep[see e.~g.][]{Jorgensen2013,Whitaker2013} and would have undergone a very efficient process of star formation, followed by fast quenching, because the sequence of quiescent galaxies is already in place at $z\sim3$ \citep{vanDokkum2003,Kriek2006,Kriek2008,vanDokkum2010,Whitaker2011,Ilbert2013,Whitaker2013}. Some authors have dedicated large efforts to study their evolution over a long period of time, amongst others, through the study of their star formation rates \citep[][]{Papovich2006,Martin2007,Zheng2007,PerezGonzalez2008,Damen2009,Barro2014b}, studying the evolution of their number density with cosmic time \citep{Cimatti2006,Ferreras2009a,Ferreras2009b,Ilbert2010,Pozzetti2010,Brammer2011,Ilbert2013,Moustakas2013}, or attempting to reconstruct their SFH by fossil record methods \citep[][]{Heavens2004,Thomas2005,Jimenez2007,Barro2014a,Mcdermid2015,GonzalezDelgado2017}. Overall, there is good agreement in that the evolution of these galaxies strongly depends on the stellar mass \citep[largely studied at low and intermediate redshift, e.~g.][]{Ferreras2000,Kauffmann2003, Gallazzi2005, Thomas2005, SanchezBlazquez2006, Jimenez2007, Kaviraj2007, Panter2008, Vergani2008, Ferreras2009a, Ferreras2009c, vanDokkum2010b,delaRosa2011,GonzalezDelgado2014b, Peng2015, Whitaker2017} and more slightly on the environment or morphology \citep[e.~g.][]{Thomas2005,Ferreras2006,Rogers2010,LaBarbera2014,GonzalezDelgado2015,Mcdermid2015,GonzalezDelgado2017}. In this sense, the more massive galaxies were formed at earlier epochs owing to a more efficient and quicker process of star formation, i.e. ``downsizing'' \citep{Cowie1996}. In addition, there is good agreement on the presence of a tight positive correlation between the gas-phase metallicity and the stellar mass \citep[e.~g.][]{Tremonti2004,Savaglio2005,Erb2006,Lee2006}, that can be also extended to total stellar metallicity \citep{Gallazzi2005,Gallazzi2014,GonzalezDelgado2014,Peng2015}, a trend called the stellar mass-metallicity relation (MZR, with distinction between the gas-phase metallicity and the total stellar metallicity). This relation has been confirmed in studies at intermediate redshift \citep[e.~g.][]{SanchezBlazquez2009,Gallazzi2014,Jorgensen2017}. Nevertheless, some authors point out that there are more relevant parameters than stellar mass as drivers of the stellar populations of galaxies \citep[][]{DiazGarcia2019}, such as the stellar surface density \citep[][]{Kauffmann2003,Franx2008} or central velocity dispersion \citep{Trager2000,Gallazzi2006,Graves2010,Cappellari2013}.

Although strong, in situ, star formation episodes are widely accepted to be a relevant channel contributing to galaxy formation, observations suggest that other mechanisms, such as mergers, can also contribute significantly \citep[see e.~g.][]{Toomre1977,Schweizer1992,Barnes1996,Trager2000,Benson2003,Croton2006,Khochfar2006,Somerville2008,Hopkins2008,Ferreras2009a,Hopkins2009,
vanderWel2009,Skelton2012,DiazGarcia2013,LopezSanjuan2013,Ferreras2014}. Once star formation is quenched or strongly reduced, these mechanisms may be important drivers of galaxy evolution. A detailed analysis of the stellar content of quiescent galaxies can unveil these mechanisms, as well as how galaxies evolve once they slow down or quench their star formation activity. In fact, the evolution of median values of stellar population properties is clearly relevant, but also the intrinsic dispersions of these values, which can be tightly related to mechanisms modifying the stellar content of galaxies. 

In this context, the state-of-the-art multi-filter surveys, e.~g COMBO-17 \citep{Wolf2003}, MUSYC \citep{Gawiser2006}, COSMOS \citep{Scoville2007}, ALHAMBRA \citep{Moles2008}, CLASH \citep{Postman2012}, SHARDS \citep{PerezGonzalez2013}, J-PAS \citep{Benitez2014} and J-PLUS \citep{Cenarro2018}, can provide an alternative way to explore the stellar content of galaxies through SED-fitting techniques \citep{Mathis2006,Koleva2008,Walcher2011,DiazGarcia2015,RuizLara2015} beyond the present day Universe. These photometric surveys, typically deeper than spectroscopy, can easily observe large samples of galaxies at intermediate redshift ($z\sim 1$--$2$). This allows us to set milestones on the assembly of the stellar content of quiescent galaxies, offering a more continuous view of galaxy formation than the fossil record approach, as they offer a sequence of ``snapshots'' across cosmic time. Moreover, multi-filter photometric surveys that combine narrow and medium bands -- whose data are defined as photo-spectra -- offer remarkable advantages: (i) There is no sampling bias other than the photometric depth of the detection band; (ii) Independent photometric calibration of each band; (iii) The photometric depth is usually much deeper than spectroscopic surveys with similar telescopes; (iv) Photometry is not affected by aperture bias, as dynamical apertures are used to collect all the flux from the sources; (v) Large scale multi-filter surveys provide a photo-spectrum at each pixel on the sky, enabling us to spatially (2D) explore resolved sources \citep[][]{Ferreras2005,SanRoman2017}; (vi) Large samples of galaxies across a wide redshift range allow unbiased statistical studies, where the various systematics can be mitigated owing to the large number of sources.

This work is part of a series of papers focused on the formation and evolution of quiescent galaxies since $z\sim1$, making use of multiple observables (e.~g.~comoving number densities, stellar population properties, and sizes). In this paper, we study the stellar content of quiescent galaxies from the ALHAMBRA multi-filter survey to constrain their properties, and we also set limits on the range of values found. For the first time, we build the probability distribution functions (PDF) of stellar age, metallicity, and dust extinction in quiescent galaxies since $z\sim1$, including their number densities in our analysis to provide a general picture of how these galaxies evolve once star formation is quenched. This is a unique opportunity to explore alternative mechanisms, e.~g.~as a result of their closest environment or by mergers, modifying the stellar content of galaxies that may remain unnoticed under an efficient in-situ star formation of the host galaxy.

This paper is structured as follows. In Sect.~\ref{sec:sample}, we briefly explain the selection of the quiescent galaxy sample from the ALHAMBRA survey, and we also provide basic details of the ALHAMBRA data, SED-fitting techniques and the main ingredients to determine the stellar population properties involved in this work. The comoving number densities of quiescent galaxies from ALHAMBRA are presented in Sect.~\ref{sec:number}. The main results of this work, namely, the constraints found on the stellar content of quiescent galaxies since $z\sim1$ and their evolution with redshift are detailed in Sect.~\ref{sec:results}. We discuss and compare our results in Sects.~\ref{sec:discussion_evolution} and \ref{sec:previous}, respectively. The conclusions are briefly summarized in Sect.~\ref{sec:conclusions}.

Throughout this work a $\Lambda$CDM cosmology is adopted, with $H_0 = 71$~km~s$^{-1}$, $\Omega_\mathrm{M}=0.27$, and $\Omega_\mathrm{\Lambda}=0.73$. Stellar masses are quoted in solar mass units $[\mathrm{M_\odot}]$ and magnitudes in the AB-system \citep{Oke1983}. In this work, we assume \citet{Chabrier2003} and Kroupa Universal \citep[][]{Kroupa2001} initial stellar mass functions (IMF, more details in Sect.~\ref{sec:method}).


\section{The sample of quiescent galaxies}\label{sec:sample}

Our parent catalogue is the sample of quiescent galaxies of \citet[][hereafter DG17]{DiazGarcia2017a}. This catalogue is complete in stellar mass and in magnitude, down to $I=23$ and it contains $\sim8\,500$ quiescent galaxies from the multi-filter ALHAMBRA survey\footnote{\url{http://www.alhambrasurvey.com}} over a redshift range $0.1 \le z \le 1.1$. To select quiescent galaxies, DG17 performed a dust-corrected stellar mass-colour diagram (MCDE) on a general sample of $\sim90\,000$ galaxies. This diagram has been shown to be a reliable diagnosgtic to substantially reduce the contamination of dusty star-forming galaxies (details in DG17). The DG17 catalogue includes stellar population properties, retrieved by use of different SSP models via SED-fitting. The properties include mass- and luminosity-weighted ages and metallicities, stellar masses, dust extinction, photo-$z$, rest-frame luminosities, colours corrected for extinction, and the parameter uncertainties are also provided. Below, we briefly detail the main features of the ALHAMBRA data set (Sect.~\ref{sec:data}) and the methodology used to retrieve the stellar population properties of quiescent galaxies (Sect.~\ref{sec:method}).


\subsection{The ALHAMBRA photometric data}\label{sec:data}

The ALHAMBRA survey provides fluxes in $23$ photometric bands\footnote{\url{http://svo2.cab.inta-csic.es/theory/fps3/}} \citep[][PSF corrected]{Coe2006}, $20$ in the optical range $\lambda\lambda\ 3500$--$9700$~\AA\ \citep[top hat medium band filters, $FWHM\sim300$~\AA, overlapping close to zero between contiguous bands; see][]{Aparicio2010} and $3$ in the NIR spectral window $\lambda\lambda\ 1.0$--$2.3$~{$\mu$m} \citep[$J$, $H$, and $K_\mathrm{s}$ bands; further details in][]{Cristobal2009}, for each source in $7$ non-contiguous fields along the northern hemisphere. The current effective area of the survey is $\sim2.8$~deg$^2$, acquired at the $3.5$~m telescope of the Calar Alto Observatory\footnote{\url{http://www.caha.es}} (CAHA). The observations in the optical range were performed with the wide-field camera LAICA\footnote{\url{http://www.caha.es/CAHA/Instruments/LAICA}} ($4$\ CCDs of $4096 \times 4096$ pixels and pixel scale $0.225\arcsec~\mathrm{pixel^{-1}}$) and with Omega-$2000$\footnote{\url{http://www.caha.es/CAHA/Instruments/O2000}} ($1$CCD with $2048 \times 2048$ pixels and plate scale $0.45\arcsec~ \mathrm{pixel^{-1}}$) in the NIR regime. We adopt the ALHAMBRA Gold catalogue\footnote{\url{http://cosmo.iaa.es/content/alhambra-gold-catalog}} \citep{Molino2014} as the reference photometric data set. This catalogue contains $\sim95\,000$ galaxies imaged in $20+3$ optical and NIR bands, respectively. The Gold catalogue provides accurate photometry (non-fixed aperture), needed to undertake stellar population studies \citep[][]{DiazGarcia2015}, and it is supplemented with precise photo-$z$ predictions ($\sigma_z \sim 0.012$), down to $I=23$. 


\subsection{Stellar population properties of quiescent galaxies}\label{sec:method}

In order to retrieve the stellar population parameters of quiescent galaxies in the DG17 catalogue, we ran the code MUFFIT \citep[][]{DiazGarcia2015}. This code has proven a reliable tool to constrain the stellar content of galaxies from multi-filter photometric data \citep[][]{DiazGarcia2015}. MUFFIT builds composite models of stellar populations (mixtures of two SSPs) from SSP sets. In this work we separately use two independent sets of SSP models to construct two samples of composite models of stellar populations, allowing us to assess potential systematics caused by the differing model prescriptions between these sets. The first set comprises the \citet{Bruzual2003} SSP models (hereafter BC03; Padova $1994$ tracks, with stellar age from $0.06$ to $14$~Gyr, and metallicities $[\mathrm{M/H}]=-1.65$, $-0.64$, $-0.33$, $0.09$, $0.55$), assuming a \citet{Chabrier2003} initial mass function. The second set comprises the EMILES\footnote{\url{http://miles.iac.es}} SSP models \citep[$\lambda\lambda~1\,680$~\AA--$5$~{$\mu\mathrm{m}$};][] {Vazdekis2016}, i.~e.~the UV and NIR extension of MIUSCAT models \citep{Vazdekis2003,Vazdekis2010,Vazdekis2012}. In the EMILES models, the two sets of theoretical isochrones adopted by the authors were taken into account: the scaled-solar isochrones of \citet[][hereafter Padova00]{Girardi2000} and \citet[][BaSTI in the following]{Pietrinferni2004}. For this set, we took $22$ ages in the range of $0.05$--$14$~Gyr and metallicities $[\mathrm{M/H}]=-1.31$, $-0.71$, $-0.40$, $0.00$, $0.22$ for Padova00 and $[\mathrm{M/H}]=-1.26$, $-0.96$, $-0.66$, $-0.35$, $0.06$, $0.26$, $0.40$ for BaSTI, both with the Kroupa Universal IMF \citep{Kroupa2001}. Notice that as shown by DG17, stellar masses of quiescent galaxies retrieved through MUFFIT and EMILES SSP models are systematically higher, $\sim0.1$~dex, than those derived from the BC03 models. For this reason, compatible stellar mass bins between EMILES and BC03 predictions differ by $0.1$~dex throughout this work. In all sets of SSP models, dust attenuation was added as a foreground screen\footnote{Therefore, our dust modelling does not distinguish between extinction and attenuation, and we decided to use the former throughout this paper.} to the SSPs with values in the range $A_V=0.0$--$3.1$, following the Milky Way extinction law of \citet{Fitzpatrick1999}, assuming a fixed value of $R_V=3.1$. In addition, we only use SSP models with cosmologically consistent ages to produce the sample of composite models of stellar populations, that is, they cannot be older than the age of the Universe at any redshift, adopting a $\Lambda$CDM cosmology. However, this constraint on age does not alter our results significantly.

Discrepancies between luminosity- and mass-weighted parameters from composite stellar population models can be particularly relevant \citep[][]{Ferreras2004,Serra2007,Rogers2010}. Throughout this work, the mass-weighted ages and metallicities ($\mathrm{Age_M}$ and $\mathrm{[M/H]_M}$, respectively) are preferred to the luminosity-weighted ones. The mass-weighted parameters are more physically motivated and representative of the total stellar content of the galaxy. Moreover, mass-weighted properties are not linked to a definition of luminosity weight, which may differ amongst different studies. However, luminosity-weighted parameters are also estimated. Lookback times were established following the recipes by \citet[][]{Hogg1999}. Hereafter, we define the formation epoch as the sum of the mass-weighted age and lookback time ($\mathrm{Age_M} + t_\mathrm{LB}$).


\section{Number density of quiescent galaxies}\label{sec:number}

Comoving number densities, $\rho_\mathrm{N}$, can be a powerful tool to set constraints on the evolution of quiescent galaxies, as well as to provide hints about the characteristic time scales of the processes involved. To derive comoving number densities, we made use of the $1/V_\mathrm{max}$ formalism \citep{Schmidt1968} in the subsamples that are complete in stellar mass (see DG17 for details). The errors of $\rho_\mathrm{N}$ are estimated by the error propagation of the $1/V_\mathrm{max}$ method, given by Poisson errors \citep[in accordance with similar previous assumptions, e.~g. ][]{Marshall1985,Ilbert2005}. It is worth mentioning that additional uncertainties owing to cosmic variance are also included in the error budget \citep[see also][]{LopezSanjuan2015}, for which we followed the recipe detailed in \citet[][]{Moster2011}. Our estimations point out that the relative cosmic variance of the ALHAMBRA sample stays at a $5$--$7$~\% fraction.

\begin{figure}
\centering
\resizebox{\hsize}{!}{\includegraphics[trim= 0 1.51cm 0 0,clip=True]{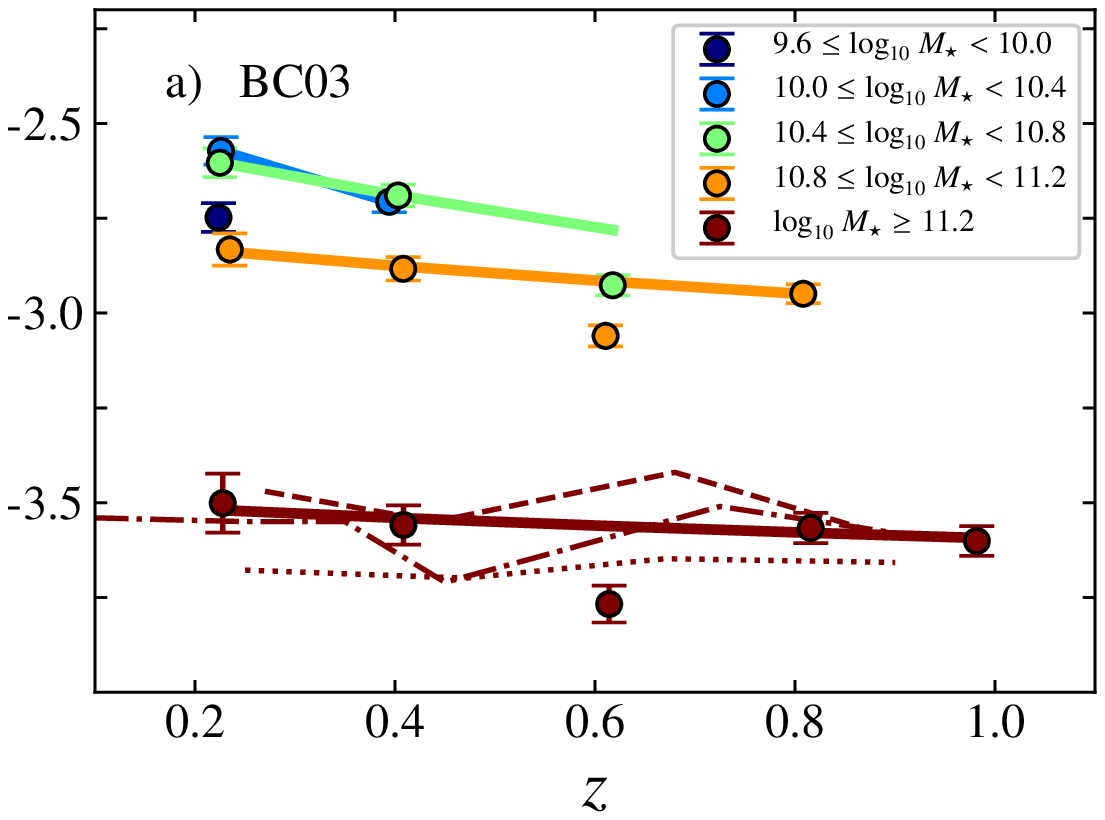}}\\
\resizebox{\hsize}{!}{\includegraphics[trim= 0 1.51cm 0 0.4cm,clip=True]{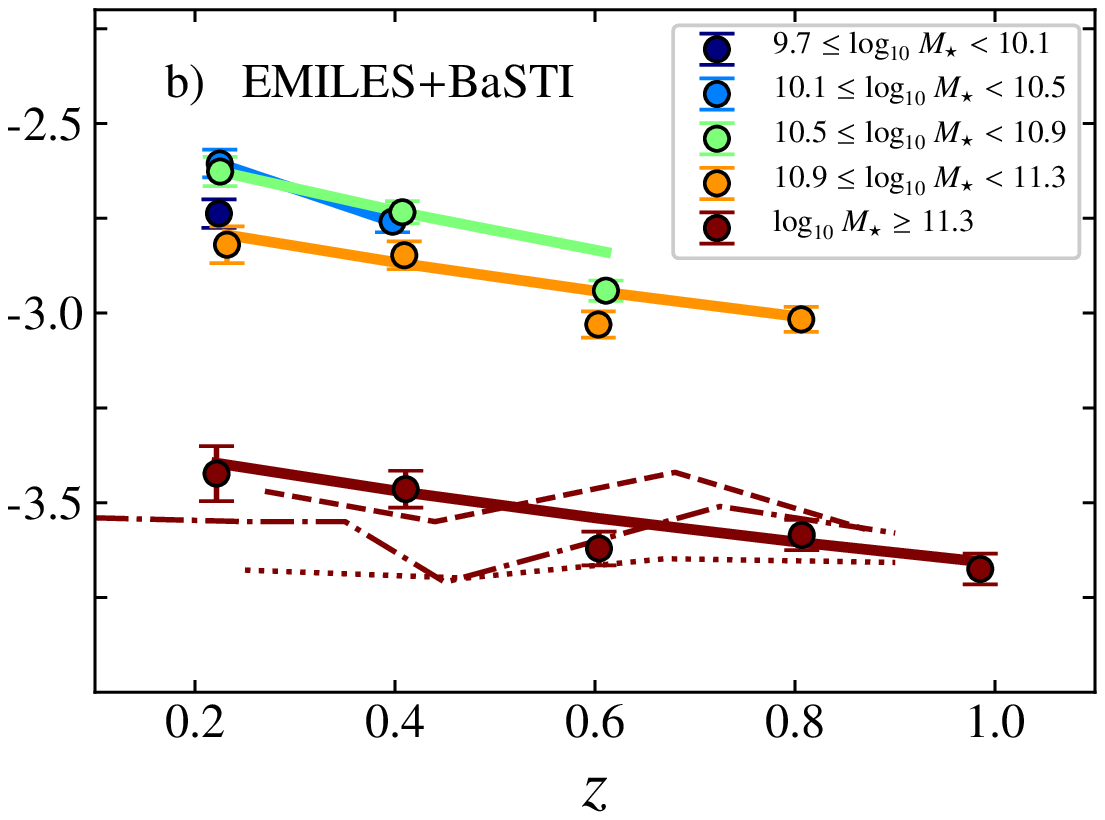}}\\
\resizebox{\hsize}{!}{\includegraphics[trim= 0 0 0 0.4cm,clip=True]{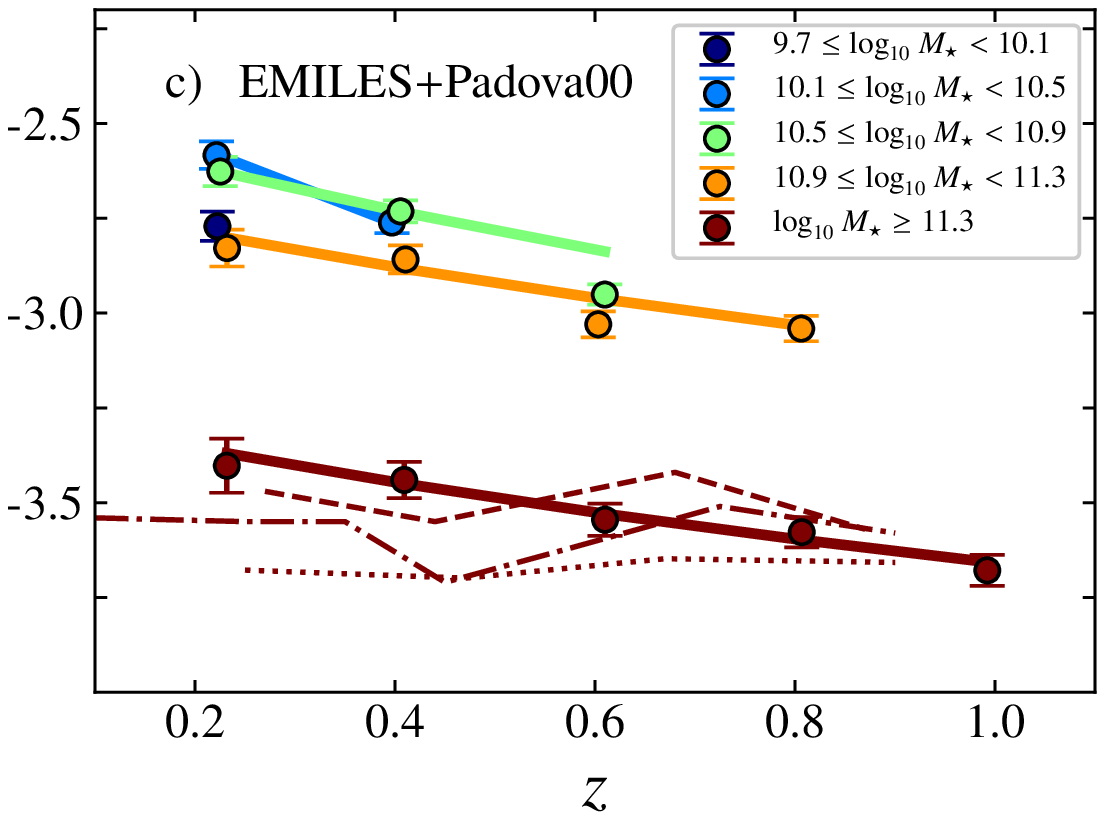}}
\caption{Evolution of the number density ($Y$--axis) of quiescent galaxies in ALHAMBRA with redshift ($X$--axis), for different stellar mass bins (see $inner$-$panels$). Coloured dots illustrate number densities when using a 1/$V_\mathrm{max}$ formalism, whereas solid lines show the best-fit to Eq.~(\ref{eq:number}). \textit{From top to bottom}, number densities obtained from BC03, EMILES+BaSTI, and EMILES+Padova00 SSP models (\textit{panels a}, \textit{b}, and \textit{c}, respectively). Over-plotted, we show the evolution on the number density of massive quiescent galaxies, $\log_{10} M_\star \ge 11$, with redshift from the previous work of \citet[][dashed line]{Pozzetti2010}, \citet[][dotted line]{Moresco2013}, and \citet[][dash-dot line]{Moustakas2013}. In all cases, the number densities at the $0.5<z<0.7$ bin are excluded from the fit as explained in the text.}
\label{fig:density}
\end{figure}

\begin{figure*}
\centering
\resizebox{\hsize}{!}{\includegraphics{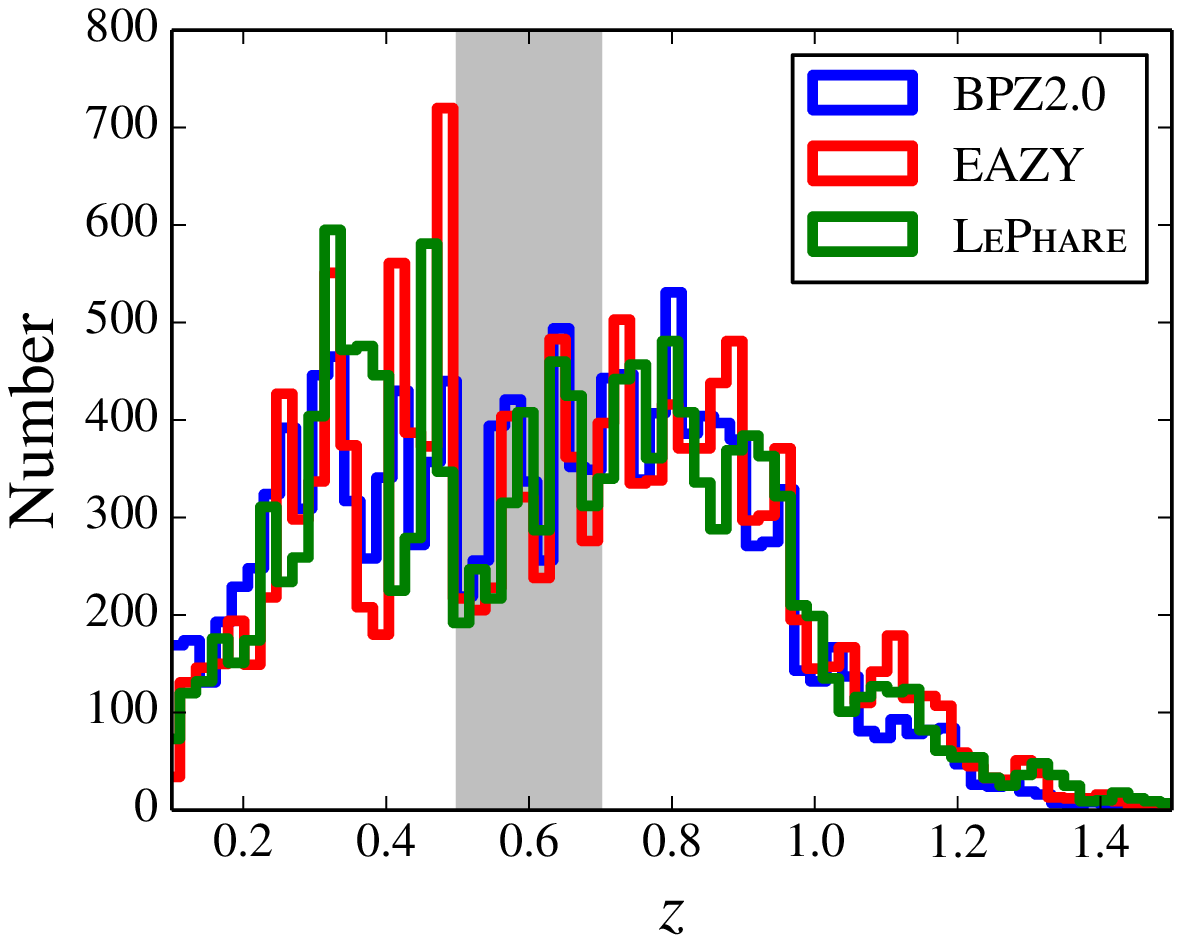}\includegraphics{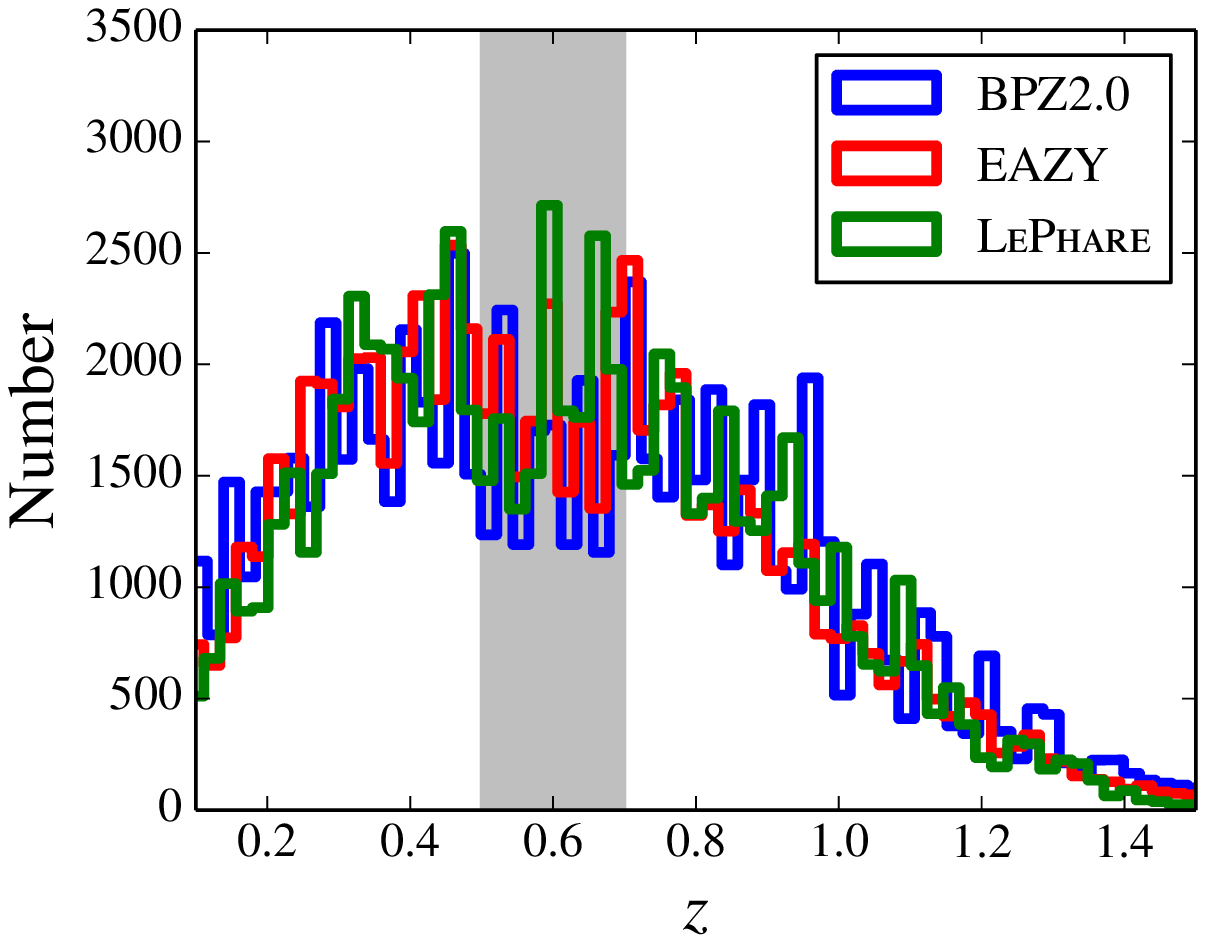}}
\caption{Photometric-redshift distributions of quiescent (\textit{left panel}) and star-forming (\textit{right panel}) galaxies from the ALHAMBRA Gold catalogue down to $0.1 \le z \le 1.5$ using the codes BPZ2.0, EAZY, and LePHARE (references in text). To guide the eye, grey area encloses the redshift bin $0.5 \le z \le 0.7$.}
\label{fig:eazy1}
\end{figure*}

The evolution of the number density of quiescent galaxies is summarised in Table~\ref{tab:density} and illustrated in Fig.~\ref{fig:density}. We found that there is a generalised lack of galaxies at $z\sim0.6$, independent of the stellar mass and more noticeable in the BC03 SSP models (see panel a in Fig.~\ref{fig:density}). In order to ensure that this lack of galaxies is not a systematic introduced by our technique, we checked that the number density of the whole population of galaxies in ALHAMBRA also present a lack of galaxies. To support this idea and as sanity check, we studied the distribution of photo-$z$ provided by the parent Gold catalogue instead of the one provided by MUFFIT, and those provided making use of other independent photo-$z$ codes: EAZY \citep[][]{Brammer2008} and LePHARE \citep[][]{Arnouts2002,Ilbert2006}. For EAZY, we allowed the combination of its default templates; whereas for LePHARE we chose the COSMOS SED templates, that include dust extinction. Similarly to BPZ \citep{Benitez2000}, both codes can apply constraints on the redshift distribution during the $\chi^2$ fitting procedure, which has benn shown to improve the photo-$z$ estimates \citep[e.~g.][]{Benitez2000,Ilbert2006,Brammer2008}. We assumed the default priors of each code: for EAZY the prior is on the $R$ band, and for LePHARE, it is applied on the $I$ band. After running EAZY and LePHARE, the retrieved photo-$z$ distributions are analysed separately for the quiescent and star-forming subsamples, in order to discard the hypothesis that the galaxy deficit in the distribution is driven by a selection bias. The photo-$z$ distribution of our quiescent sample, see left panel in Fig.~\ref{fig:eazy1}, shows a remarkable agreement amongst the three different photo-$z$ codes. In fact, the three codes find a prominent lack of galaxies at $0.5 < z < 0.7$ (see grey region in Fig.~\ref{fig:eazy1}), and the rest of structures are also similar, independently of the code used. Discrepancies between cumulative distribution functions, CDFs, of photo-$z$ distributions do not exceed a value of $0.05$ (a $\lesssim 5$~\% fraction) at $0.1 < z < 1.5$. Indeed at $0.5 < z < 0.7$, the discrepancies between CDFs are even smaller with values $\lesssim 3$~\%. Regarding the star-forming photo-$z$ distribution, as shown on the right panel of Fig.~\ref{fig:eazy1}, there are little discrepancies amongst the output codes. Considering star-forming galaxies at $0.5 < z < 0.7$, the three photo-$z$ codes also produce a galaxy deficit (maximum discrepancy of $\sim5$~\% amongst CDFs). For the LePHARE case, this decrement in galaxy number may be restricted at $0.5 < z < 0.6$. Finally, we retrieved the photo-$z$ distributions for each of the seven ALHAMBRA fields separately, checking that this lack of galaxies at $0.5 < z < 0.7$ does not appear in all the pointings systematically. Consequently, we discarded the idea that this deficit is associated to MUFFIT systematics or a biased selection. This result shows that even though ALHAMBRA comprises seven uncorrelated fields on the sky, some large scale structures are still noticeable in this survey.

\begin{table*}
\caption{Logarithm of the number density, $\log_{10}\rho_\mathrm{N}[h^3\ \mathrm{Mpc}^{-3}]$, for the quiescent galaxies in our sample as a function of stellar mass and redshift.}
\label{tab:density}
\centering
\begin{tabular}{crccccc}
\hline\hline
& & $0.1\le z < 0.3$ & $0.3 \le z < 0.5$ & $0.5 \le z < 0.7$ & $0.7 \le z < 0.9$ & $0.9 \le z \le 1.1$ \\
\hline
\parbox[t]{2mm}{\multirow{7}{*}{\rotatebox[origin=c]{90}{BC03}}} &&&&&& \\
& $9.6 \le \log_{10}M_\star < 10.0$ & $-2.75 \pm 0.04$ & -- & -- & -- & -- \\
& $10.0 \le \log_{10}M_\star < 10.4$ & $-2.57 \pm 0.04$ & $-2.71 \pm 0.03$ & -- & -- & -- \\
& $10.4 \le \log_{10}M_\star < 10.8$ & $-2.60 \pm 0.04$ & $-2.69 \pm 0.03$ & $-2.93 \pm 0.03$ & -- & -- \\
& $10.8 \le \log_{10}M_\star < 11.2$ & $-2.83 \pm 0.04$ & $-2.88 \pm 0.03$ & $-3.06 \pm 0.03$ & $-2.95 \pm 0.02$ & -- \\
& $\log_{10}M_\star \ge 11.2$ & $-3.50 \pm 0.08$ & $-3.56 \pm 0.05$ & $-3.77 \pm 0.05$ & $-3.57 \pm 0.04$ & $-3.60 \pm 0.04$ \\
&&&&&& \\
\hline
\parbox[t]{2mm}{\multirow{7}{*}{\rotatebox[origin=c]{90}{BaSTI}}} &&&&&& \\
& $9.7 \le \log_{10}M_\star < 10.1$ & $-2.74 \pm 0.04$ & -- & -- & -- & -- \\
& $10.1 \le \log_{10}M_\star < 10.5$ & $-2.61 \pm 0.04$ & $-2.76 \pm 0.03$ & -- & -- & -- \\
& $10.5 \le \log_{10}M_\star < 10.9$ & $-2.63 \pm 0.04$ & $-2.73 \pm 0.03$ & $-2.94 \pm 0.03$ & -- & -- \\
& $10.9 \le \log_{10}M_\star < 11.3$ & $-2.82 \pm 0.05$ & $-2.85 \pm 0.04$ & $-3.03 \pm 0.03$ & $-3.02 \pm 0.03$ & -- \\
& $\log_{10}M_\star \ge 11.3$ & $-3.42 \pm 0.07$ & $-3.46 \pm 0.05$ & $-3.62 \pm 0.04$ & $-3.59 \pm 0.04$ & $-3.67 \pm 0.04$ \\
&&&&&& \\
\hline
\parbox[t]{2mm}{\multirow{7}{*}{\rotatebox[origin=c]{90}{Padova00}}} &&&&&& \\
& $9.7 \le \log_{10}M_\star < 10.1$ & $-2.77 \pm 0.04$ & -- & -- & -- & -- \\
& $10.1 \le \log_{10}M_\star < 10.5$ & $-2.58 \pm 0.04$ & $-2.76 \pm 0.03$ & -- & -- & -- \\
& $10.5 \le \log_{10}M_\star < 10.9$ & $-2.63 \pm 0.04$ & $-2.73 \pm 0.03$ & $-2.95 \pm 0.03$ & -- & -- \\
& $10.9 \le \log_{10}M_\star < 11.3$ & $-2.83 \pm 0.05$ & $-2.86 \pm 0.04$ & $-3.03 \pm 0.03$ & $-3.04 \pm 0.03$ & -- \\
& $\log_{10}M_\star \ge 11.3$ & $-3.40 \pm 0.07$ & $-3.44 \pm 0.05$ & $-3.54 \pm 0.04$ & $-3.58 \pm 0.04$ & $-3.68 \pm 0.04$ \\
&&&&&& \\
\hline
\end{tabular}
\tablefoot{From top to bottom, number densities obtained for BC03, EMILES+BaSTI, and EMILES+Padova00 SSP models. These values were measured through the $1/V_\mathrm{max}$ formalism of each bin. All the bins are complete in stellar mass, $\mathcal{C}=0.95$, otherwise appear dashed. As detailed in \citet[][]{DiazGarcia2017a}, there is a systematic shift of $\sim0.1$~dex between the stellar masses of quiescent galaxies using BC03 and EMILES SSP models. All the values were obtained setting $h=1$.}
\end{table*}

\begin{table*}
\caption[Values $\rho_0$ and $\gamma$ that best fit the number density of quiescent galaxies.]{Values $\rho_0$ and $\gamma$ that best fit our number density quantification (see Eq.~(\ref{eq:number})) at different stellar mass bins and SSP models.}
\label{tab:density_par}
\centering
\begin{tabular}{rcccccc}
\hline\hline
& \multicolumn{2}{c}{BC03} & \multicolumn{2}{c}{EMILES+BaSTI} & \multicolumn{2}{c}{EMILES+Padova00} \\
& \multirow{2}{*}{$\log_{10}\rho_0$} & \multirow{2}{*}{$\gamma$} & \multirow{2}{*}{$\log_{10}\rho_0$} & \multirow{2}{*}{$\gamma$} & \multirow{2}{*}{$\log_{10}\rho_0$} & \multirow{2}{*}{$\gamma$} \\
&&&&&& \\
\hline
&&&&&& \\
$10.0 \le \log_{10}M_\star < 10.4$  & $-2.36 \pm 0.10$ & $-2.40 \pm 0.80$ & $-2.37 \pm 0.10$ & $-2.66 \pm 0.80$ & $-2.32 \pm 0.10$ & $-3.05 \pm 0.77$ \\
$10.4 \le \log_{10}M_\star < 10.8$ & $-2.48 \pm 0.10$ & $-1.45 \pm 0.79$  & $-2.47 \pm 0.10$ & $-1.76 \pm 0.80$ & $-2.47 \pm 0.11$ & $-1.76 \pm 0.81$ \\
$10.8 \le \log_{10}M_\star < 11.2$ & $-2.78 \pm 0.05$ & $-0.67 \pm 0.26$  & $-2.68 \pm 0.06$ & $-1.30 \pm 0.33$ & $-2.67 \pm 0.06$ & $-1.40 \pm 0.32$ \\
$\log_{10}M_\star \ge 11.2$ & $-3.49 \pm 0.08$ & $-0.35 \pm 0.33$ & $-3.29 \pm 0.08$ & $-1.24 \pm 0.32$ & $-3.24 \pm 0.07$ & $-1.38 \pm 0.31$ \\
&&&&&& \\
\hline
\end{tabular}
\tablefoot{There is no $\rho_0$ and $\gamma$ fitting values for the lowest stellar mass bin, $9.6 \le \log_{10}M_\star < 10$, because this subsample is only available at the lowest redshift bin, $0.1 \le z < 0.3$. All the values were obtained setting $h=1$.}
\end{table*}

The number density trends are quantified through a redshift-dependent power-law function (solid lines in Fig.~\ref{fig:density}) of the form:
\begin{equation}
\rho_\mathrm{N}(z) = \rho_0\ (1+z)^\gamma\ .
\label{eq:number}
\end{equation}
For the different stellar mass bins, we provide the values $\rho_0$ and $\gamma$ that best fit our number density values in Table~\ref{tab:density_par} (all the number density estimations at $0.5 \le z < 0.7$ were removed from the fit). From Fig.~\ref{fig:density}, we summarize three remarkable results:

\begin{itemize}

\item[$\bullet$]
The number density evolution is well fitted by a power-law function (Eq.~(\ref{eq:number})).
\item[$\bullet$] The number density of quiescent galaxies is found to increase with cosmic time up to the present.
\item[$\bullet$] 
  At the low-mass end, quiescent galaxies have $\gamma$ values that are compatible with a steeper evolution in number density with respect to the massive counterparts, i.~e.~the appearance of low-mass quiescent galaxies is more prominent.
\end{itemize}

The least massive bin in our sample ($9.6 \le \log_{10}M_\star < 10.0$, dark blue dots in Fig.~\ref{fig:density}) greatly reflects the stellar mass range in which the stellar mass function of quiescent galaxies present a local minimum or valley \citep{Drory2009,Tomczak2014}, but owing to completeness reasons we cannot establish the variation of its number density. At higher stellar mass, our fits suggest that the number density of quiescent galaxies $\rho_\mathrm{N}(z)$ grows on average by $\sim52$, $30$, $20$, $12$~\% (in the remaining mass bins, starting from the $10.0 \le \log_{10}M_\star <\ 10.4$ interval) between $z=0.4$ and $z=0.2$ (see Table~\ref{tab:density_par}). It is noticeable that the number densities retrieved with EMILES (see panels b and c in Fig.~\ref{fig:density}) show slightly larger variations than those obtained with the BC03 models (see panel a in Fig.~\ref{fig:density}), illustrating the dependence of $\rho_\mathrm{N}(z)$ on the SSP models adopted to retrieve stellar parameters.

\citet{Ilbert2013} reported that quiescent galaxies of $\log_{10}M_\star > 11.2$~dex, suffer a rapid and efficient increase in number at $1<z<3$. However, these galaxies do not exhibit prominent evolution since $z\sim1$, where the great number density variations of quenched galaxies are more focused on the less massive systems \citep[a result also observed by e.~g.][]{Davidzon2013}, in agreement with our results (see Table~\ref{tab:density_par}). \citet{Ferreras2005,Ferreras2009a} performed an analysis of morphologically-selected early-type galaxies (ETGs) in the GOODS fields, finding a substantial difference between the comoving number density of massive ETGs  and their lower mass counterparts. Between $z=1$ and $z=0.6$, the number density of ETGs was found to increase only by a factor of $0.25$~dex for $\log_{10}M_\star > 11$, and it was even compatible with no evolution at the most massive end \citep[$\log_{10}M_\star > 11.5$, see also][]{Ferreras2009b}. \citet{Pozzetti2010} found that the number density evolution of quiescent galaxies ($\log_{10} M_\star > 11$) is not significant, with a variation $\sim0.1$~dex, since $z\sim0.85$ up to  $z\sim0.25$; while in our work this variation is $\sim0.12$--$0.25$~dex. \citet{Moresco2013} claimed that the number of quiescent galaxies ($\log_{10}M_\star\sim 10.5$) increases by $\sim80$~\% between $z\sim0.65$ and $z\sim 0.2$, compatible with ours ($60$--$75$~\%), and the massive ones ($\log_{10}M_\star>11$) were compatible with no evolution. With a different selection criteria, \citet{Moustakas2013} found out that the number of quiescent galaxies in the mass range $10.0 < \log_{10} M_\star <10.5$ increases by a $60\pm20$~\% fraction between $z\sim 0.6$ and $z\sim 0.2$; at $10.5 < \log_{10} M_\star <11.0$ around $40\pm 10$~\% between $z=0.8$ and $z=0.2$ , and for $11.0 < \log_{10} M_\star <11.5$ the increment is $20\pm10$~\% at $0.2 < z < 1.0$. Using the same redshift range and stellar mass bins as \citet{Moustakas2013}, we find number density variations of $90\pm40$~\%, $60\pm20$~\%, $40\pm20$~\% using BC03 SSPs (and larger variations with EMILES), respectively. In Fig.~\ref{fig:density}, we illustrate the behaviour of the number density at the most massive bin, $\log_{10} M_\star > 11.2$. Note that in our work, we cull dusty star-forming galaxies from the sample of quiescent galaxies via the MCDE, a procedure that differs with respect to the selections of previous studies.

On the other hand, previous studies, such as \citet[][and references therein]{Cerulo2016}, claim that low-mass quiescent cluster galaxies at the faint-end of the stellar mass function do not present a remarkable evolution in number since $z\sim1.5$. However, the authors suggest that this is a consequence of the halo mass, which accelerates the building-up of the passive population of galaxies. In ALHAMBRA, that extends over six uncorrelated fields, we would expect the sample to be dominated in number by field galaxies, explaining why our result differs with respect to others defined in dense environments.


\begin{figure}
\centering
\resizebox{\hsize}{!}{\includegraphics{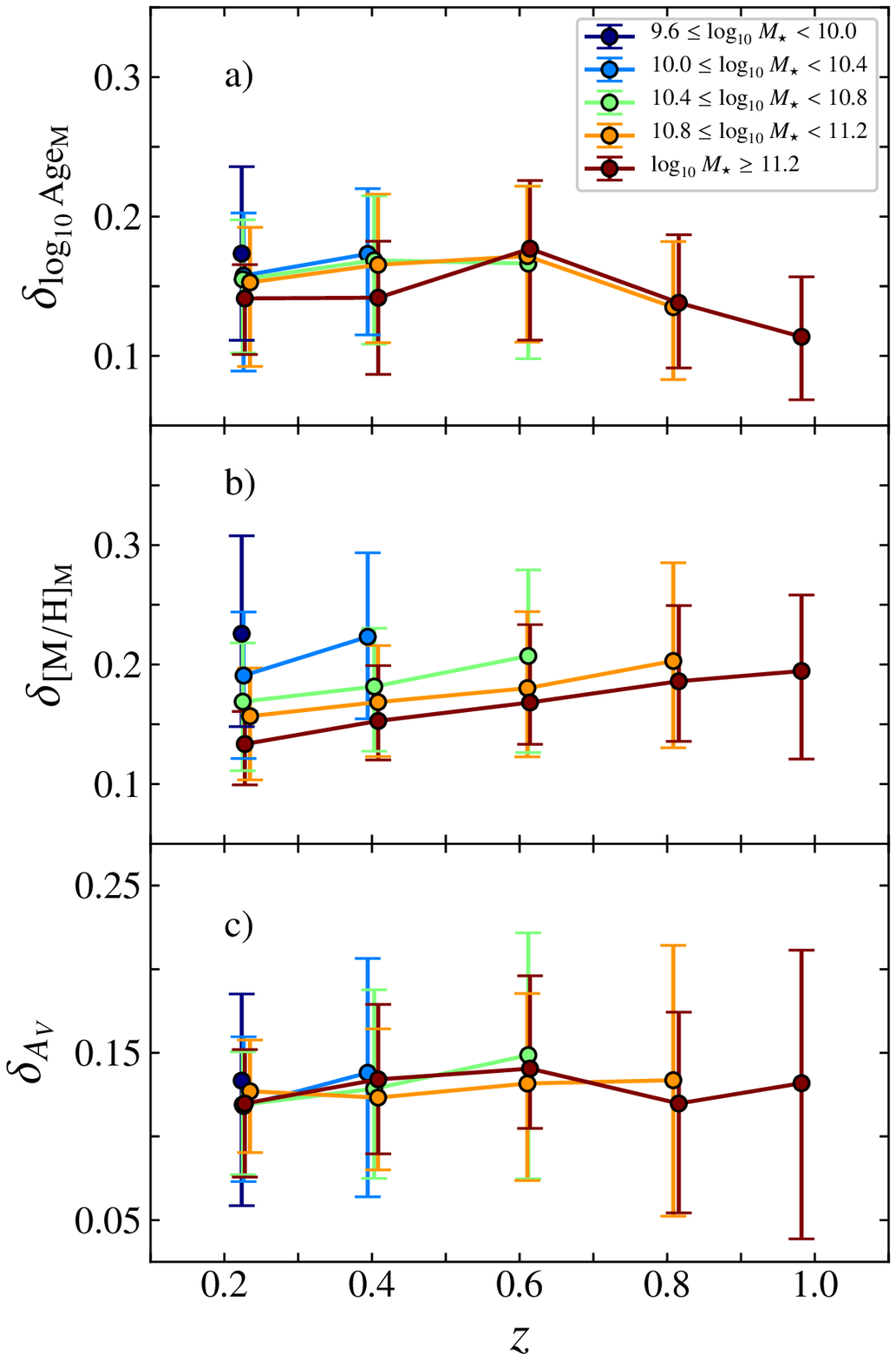}}
\caption{\textit{From top to bottom}, median of the age, metallicity, and extinction uncertainties (\textit{panels a}, \textit{b}, and \textit{c}, respectively) at different redshifts and stellar mass bins (see inner panel). Vertical bars enclose the $68$~\% confidence level of the distribution of uncertainty values.}
\label{fig:uncertainties_sp}
\end{figure}

\section{The stellar populations of quiescent galaxies since $z\sim 1$}\label{sec:results}

This section presents the main results, namely the evolution of the stellar population properties of quiescent galaxies during the last $8$~Gyr of cosmic time. With this aim, we build, for the first time, analytic PDFs of the properties of quiescent galaxies since $z\sim1$ (Sect.~\ref{sec:pdf}). Finally, we describe in detail the observed changes in the stellar content of quiescent galaxies, making use of the redshift dependent PDFs of mass-weighted age (Sect.~\ref{sec:epoch}), metallicity (Sect.~\ref{sec:metallicity}), and extinction (Sect.~\ref{sec:extinction}).


\subsection{Probability distribution functions of stellar population parameters}\label{sec:pdf}

Our sample contains quiescent galaxies covering a wide redshift range, noting that the derived uncertainties of the stellar population parameters have a significant dependence on redshift. Moreover, certain ranges of stellar-population parameters are intrinsically more subject to SED-fitting errors \citep[see fig.~$11$ in][]{DiazGarcia2015}, usually related to well-known degeneracies amongst parameters \citep[such as the age-metallicity degeneracy,][]{Worthey1994,Worthey1999,Peletier2013,DiazGarcia2015}. To illustrate this, the median of the age, metallicity, and extinction uncertainties obtained by MUFFIT and the ALHAMBRA dataset is shown in Fig.~\ref{fig:uncertainties_sp}. In our sample, quiescent galaxies at higher redshift exhibit lower age uncertainties (see panel a in Fig.~\ref{fig:uncertainties_sp}). In fact, this behaviour was also observed using simulations \citep{DiazGarcia2015} and it is a consequence of the age range of SSP models at larger redshifts (they cannot be much larger than the age of the Universe) and because a wider wavelength range of the rest-frame NUV regime is observed. On the other hand, metallicity and extinction uncertainties are larger at increasing redshift (see panels b and c in Fig.~\ref{fig:uncertainties_sp}, respectively). Consequently, the uncertainties can modify the distributions of stellar population parameters in a redshift-dependent way. This behaviour hinders both a direct comparison amongst the distribution of the stellar-population parameters at different redshifts, as well as a precise reconstruction of the intrinsic distribution.

We deal with this issue by performing a maximum likelihood estimator method (MLE) to deconvolve uncertainty effects, and build PDFs of the involved stellar population properties of the quiescent galaxy population (not individual galaxies): mass-weighted ages, metallicities and extinctions. In particular, we adapted the MLE methodology developed by \citet{LopezSanjuan2014}, also used in DG17, to deconvolve observational errors from observed distributions at different stellar mass ranges. In practise, this technique aids in recovering the intrinsic scatter of stellar population distributions from the observed ones, affected by uncertainties. We therefore constrain the most likely set of parameter values that maximizes the probability distributions. As a result, we obtain a set of functional and analytical distributions fitted by log-normal functions, and re-normalize them with their number densities (see Sect.~\ref{sec:number}). For further details of the whole process, we refer interested readers to Appendix~\ref{sec:appendix_pdf}.

Note that, as mentioned in Appendix~\ref{sec:appendix_pdf}, we do not provide the redshift-dependent PDFs for quiescent galaxies at $\log_{10}M_\star < 10.1$, because the reliability of the MLE method is compromised, owing to the low number of sources. Instead, and only for the least massive case, we applied the MLE method assuming no redshift dependence of the PDF parameters (i.~e.~$\mu_2 = \mu_1 = \sigma_2 = \sigma_1 = 0$, details in Appendix~\ref{sec:appendix_pdf}), to set the average values of the median and width of the stellar population parameter distributions at $0.1 \le z < 0.3$. 

In the following, to detail the evolution of the PDFs with redshift, we focus on the evolution of their medians and widths. For this work, we define the width of a PDF as the difference between the $84^\mathrm{th}$ and $16^\mathrm{th}$ percentiles. Note that the width at this point is not a result of uncertainty effects, but the intrinsic dispersion of stellar population properties. The main results of this section, that is, the medians and widths of the PDFs of quiescent galaxies can be found in Figs.~\ref{fig:median_width_age}, \ref{fig:median_width_feh}, and \ref{fig:median_width_av} (showing the mass-weighted age and formation epoch, metallicity, and extinction, respectively). As showed by these figures, the PDF parameters evolve with redshift ($X$-axis in the figures) and depend on the stellar mass range (coloured lines, see insets). Note that the SSP sets of BC03 (first column in figures), EMILES with both BaSTI (second column) and Padova00 (third column) isochrones are included to explore potential systematics due to the use of different population synthesis models.


\subsection{Ages and formation epochs}\label{sec:epoch}

\begin{figure*}
\centering
\resizebox{!}{15.39cm}{\includegraphics[trim = 0 0 0.5cm 0, clip=True]{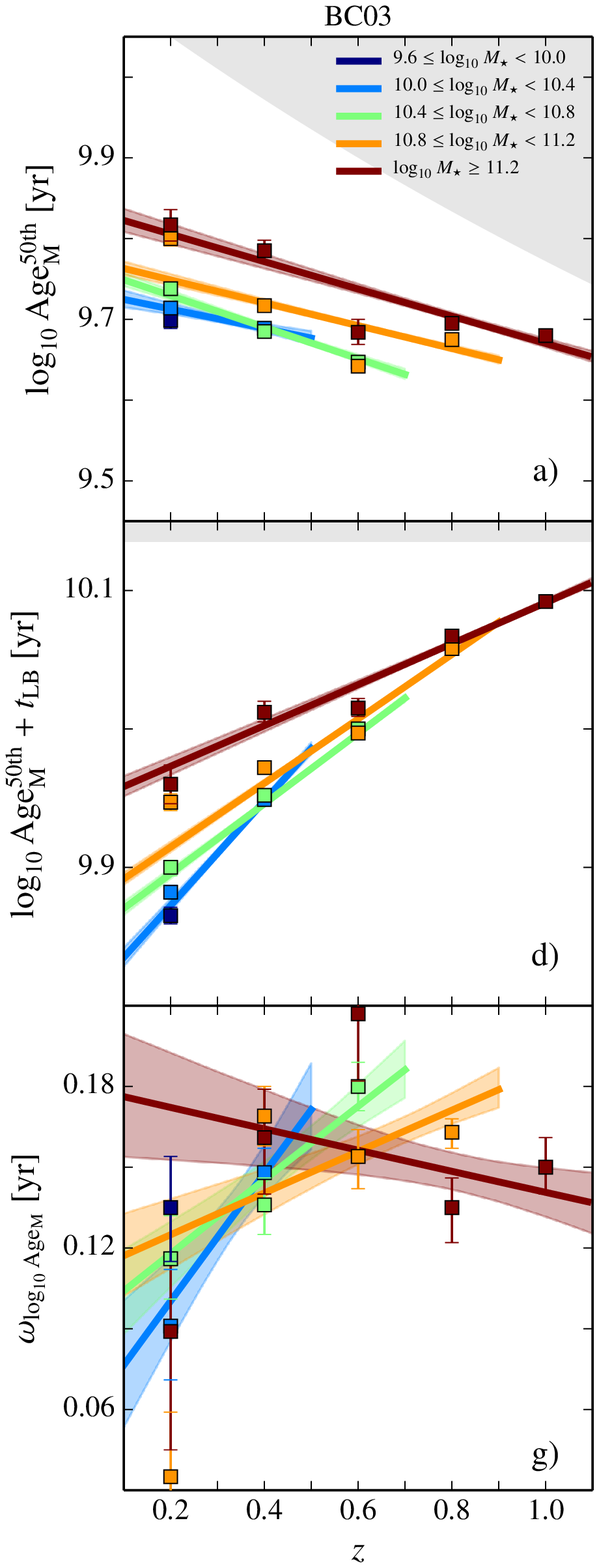}
\includegraphics[trim = 2.33cm 0 0.5cm 0, clip=True]{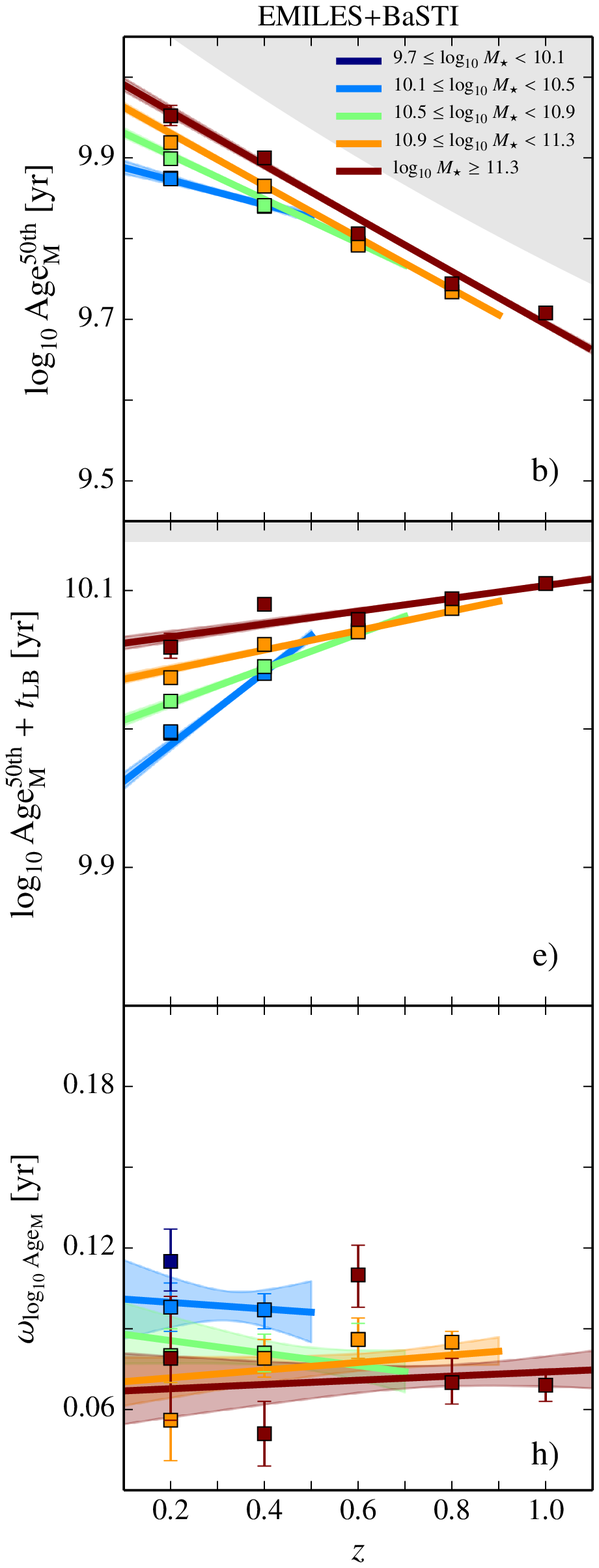}
\includegraphics[trim = 2.33cm 0 0.5cm 0, clip=True]{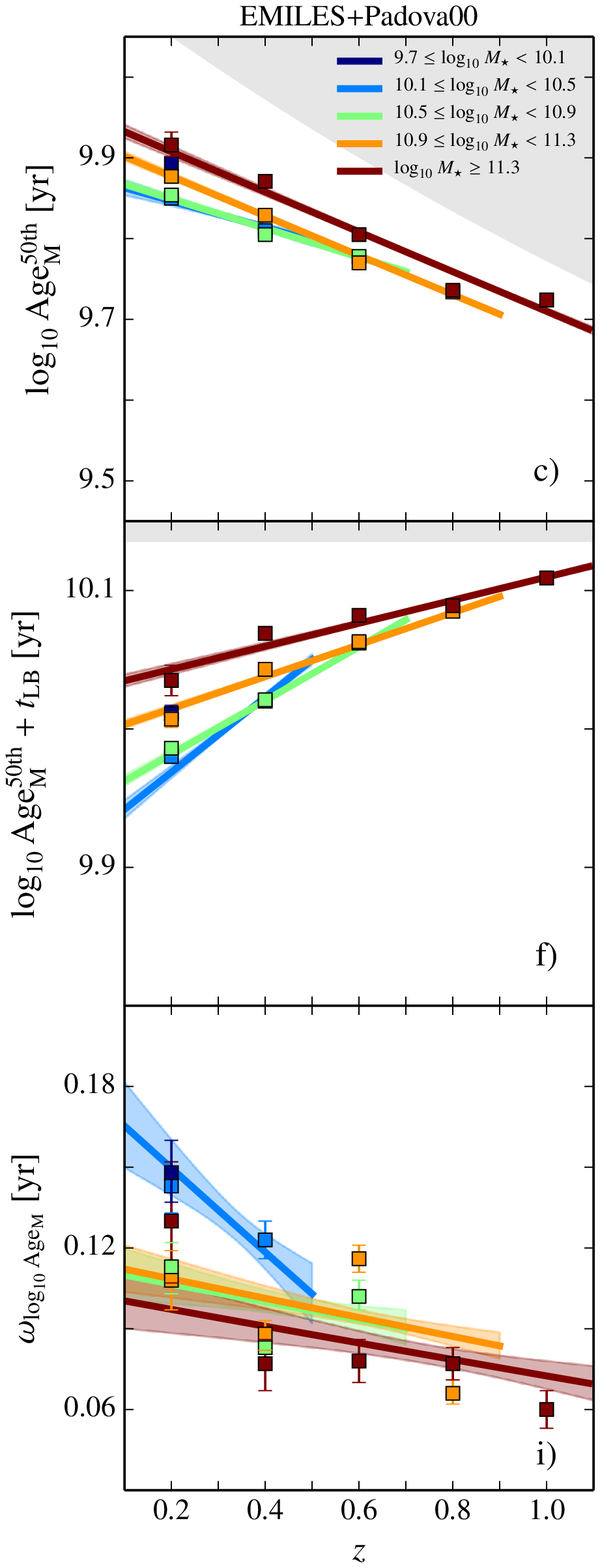}}
\caption{Evolution of the medians of the mass-weighted age (\textit{panels a, b}, and \textit{c}) and formation epoch (\textit{panels d, e}, and \textit{f}) PDFs of the quiescent population along cosmic time for different stellar mass bins. \textit{Panels g, h, and i} show the widths of the mass-weighted age PDFs. \textit{From left to right}, results obtained  using the SSP models of BC03 (\textit{panels a, d}, and \textit{g}), EMILES with BaSTI isochrones (\textit{panels b, e}, and \textit{h}), and EMILES with the Padova00 ones (\textit{panels c, f}, and \textit{i}). The shaded regions enclose the $1~\sigma$ uncertainties of both parameters. Grey region limits the age of the Universe at any redshift. The square-shape markers illustrate the average median and width assuming for the MLE deconvolution $\mu_2 = \mu_1 = \sigma_2 = \sigma_1 = 0$.}
\label{fig:median_width_age}
\end{figure*}

Our results evidence that the mass-weighted age and formation epoch PDFs are correlated with the stellar mass of quiescent galaxies, showing systematic variations that depend on the stellar mass (see panels a--f in Fig.~\ref{fig:median_width_age}). In good agreement with the ``downsizing'' scenario, more massive quiescent galaxies feature a larger stellar content in old stars with respect to the lower mass systems, which are preferentially formed at more recent epochs. The median ages (see panels a, b, and c in Fig.~\ref{fig:median_width_age}) of the quiescent population presents older stellar populations at lower redshifts independently of the stellar mass bin, as expected in passive or quenched systems. Nevertheless, the median of the formation epoch PDFs of quiescent galaxies (panels d, e, and f in Fig.~\ref{fig:median_width_age}) exhibits a continuous and general decrement at lower redshifts for all the stellar masses and SSP models used in this work, especially for BC03 (see panel d in Fig.~\ref{fig:median_width_age}), which is not compatible with pure passive evolution. For the most massive case, $\log_{10}M_\star \ge 11.2$, the formation epoch at $z\sim1.0$ is $\sim12$--$13$~Gyr (equivalent to a formation redshift of $z_\mathrm{f} \sim 4.5$--$6$), whereas at $z=0.2$ the median formation epoch decreases to $\sim9$--$11$~Gyr (i.~e.~corresponding to a formation redshift of $z_\mathrm{f} \sim 1.5$--$3$). At lower stellar masses, the variation of the formation epochs is more pronounced. It is worth mentioning that the mass-weighted ages retrieved from BC03 models are younger than the ones obtained for EMILES with BaSTI ($0.15$--$0.2$~dex or $2$--$3$~Gyr) and Padova00 ($0.05$--$0.10$~dex or $1.5$--$2$~Gyr) isochrones. Notice that the BaSTI isochrones provide ages older than those from Padova00, as the BaSTI isochrones are bluer than the Padova00 ones \citep{Vazdekis2016}.

The width of the mass-weighted age and formation epoch PDFs are roughly the same (see panels g, h, and i in Fig.~\ref{fig:median_width_age}). These are fairly constrained in the range $0.1$--$0.2$~dex ($\sim1$--$2$~Gyr). Using EMILES and BaSTI isochrones, the width of the mass-weighted age PDFs does not present a great dependence with stellar mass. For the EMILES+Padova00 case, we retrieved the same trend, although the PDFs are slightly wider (see panels g, h, and i in Fig.~\ref{fig:median_width_age}). Our results point out that the evolution with redshift of the widths of the age/formation epoch PDFs is mild. In fact, for EMILES and BaSTI there is no evidence of evolution with redshift. Only for BC03 at $\log_{10}M_\star < 10.8$, the widths of the mass-weighted age/formation epoch PDFs decrease at larger cosmic times, with values in the range $0.06$--$0.18$~dex (or $\sim 1$--$2.5$~Gyr). For EMILES, we do not find that the evolution with redshift of the widths of mass-weighted age and formation epoch PDFs strongly depend on mass.

\subsection{Evolution of the metal content}\label{sec:metallicity}

\begin{figure*}
\centering
\resizebox{!}{10.8cm}{\includegraphics[trim = 0 0 0.5cm 0, clip=True]{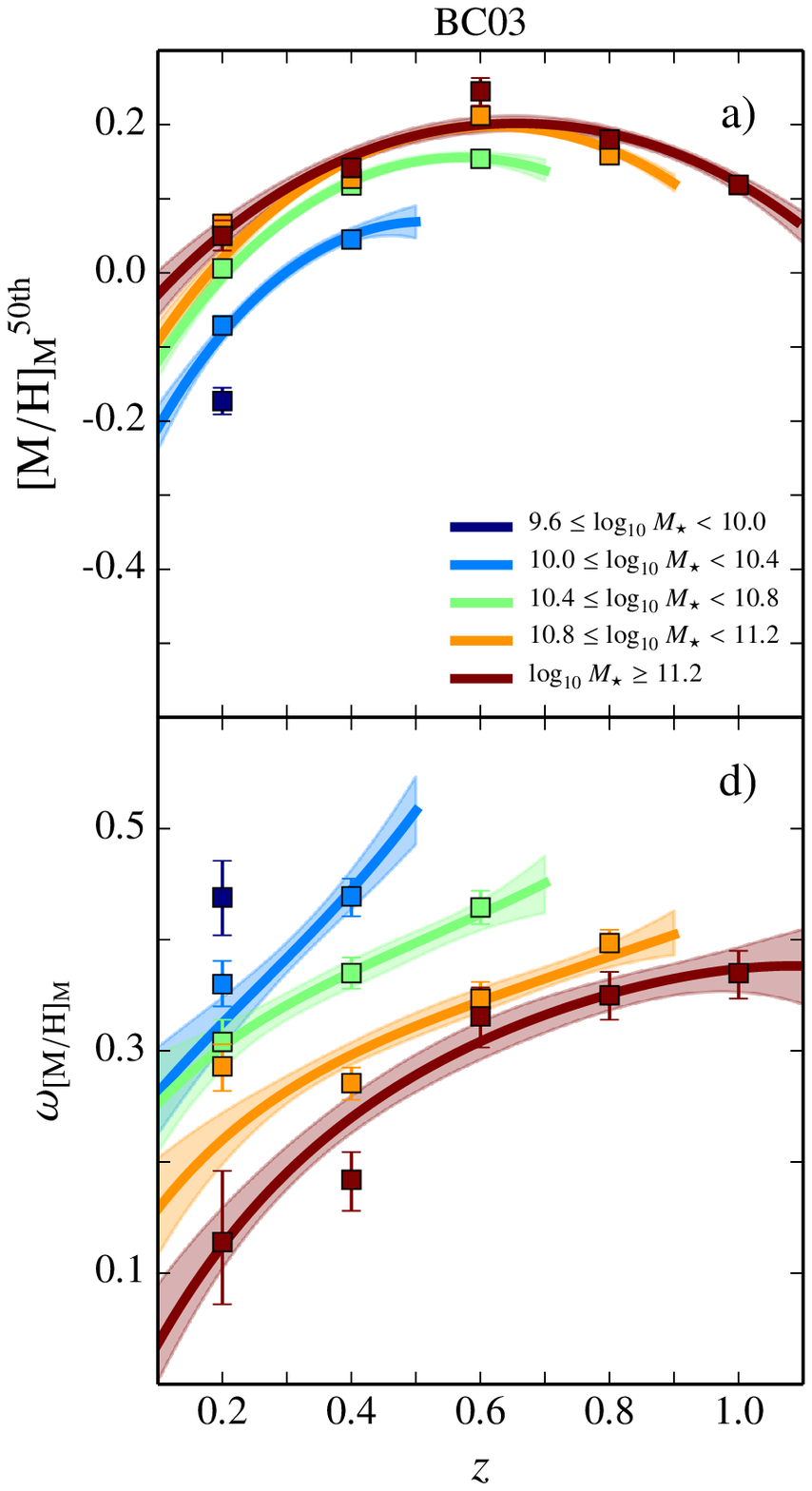}
\includegraphics[trim = 2.33cm 0 0.5cm 0, clip=True]{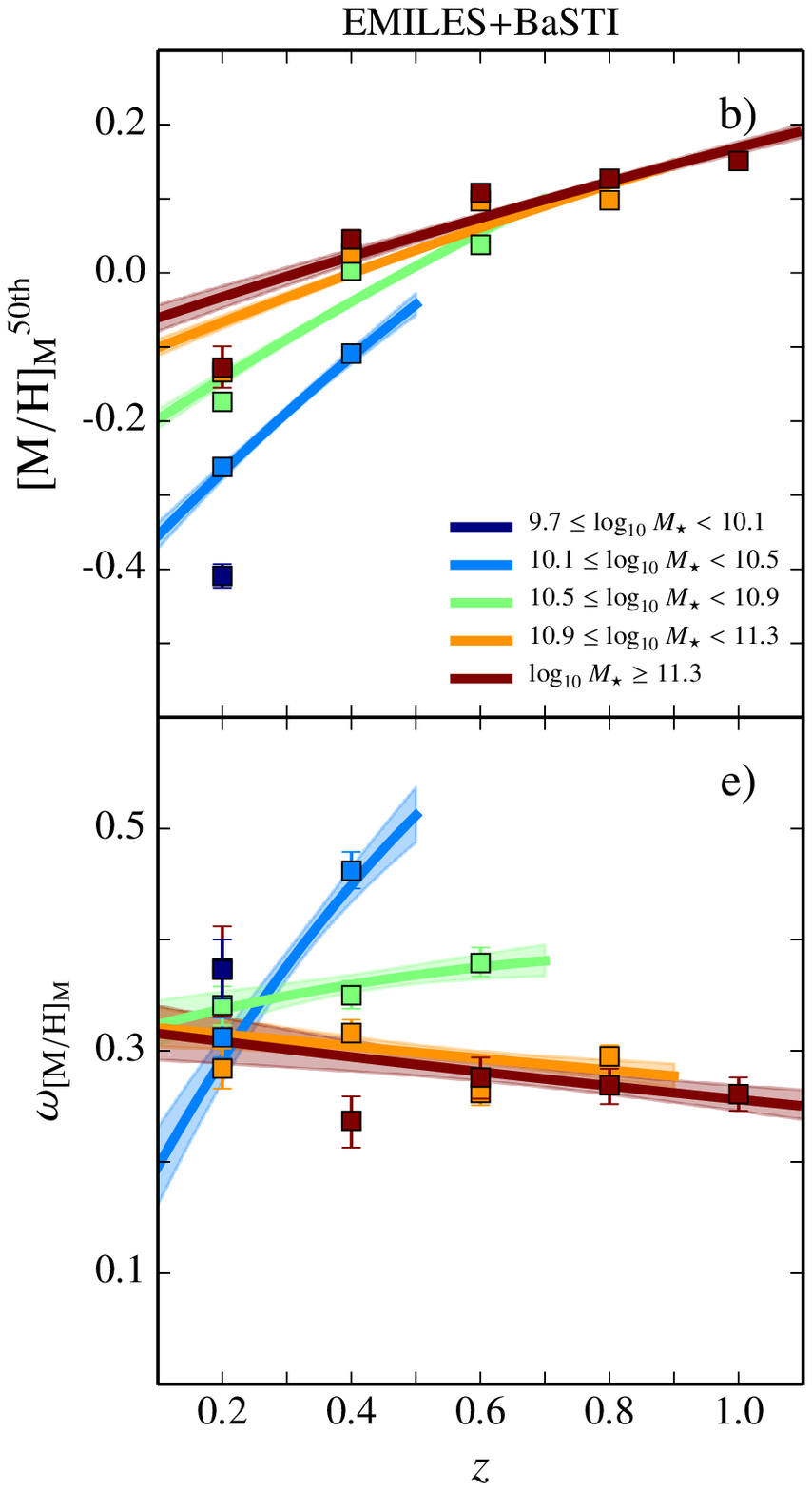}
\includegraphics[trim = 2.33cm 0 0.5cm 0, clip=True]{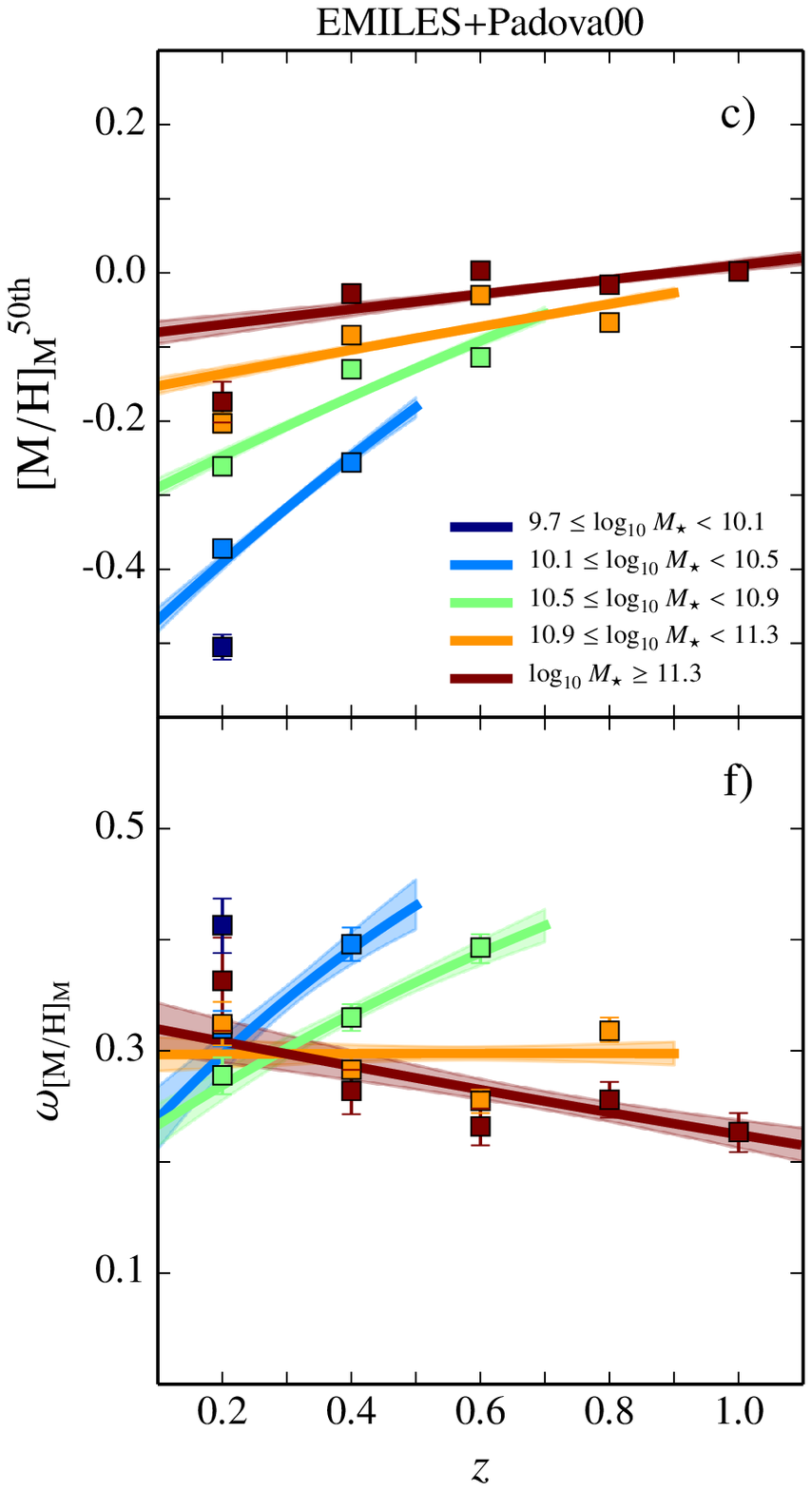}}
\caption{Evolution of the medians (\textit{panels a, b}, and \textit{c}) and widths (\textit{panels d, e}, and \textit{f}) of the mass-weighted metallicity PDFs  of the quiescent population along cosmic time for different stellar mass bins. \textit{From left to right}, results obtained  using the SSP models of BC03 (\textit{panels a and d}), EMILES with BaSTI isochrones (\textit{panels b} and \textit{e}), and EMILES with the Padova00 ones (\textit{panels c} and \textit{f}). The shaded regions enclose the $1~\sigma$ uncertainties of both parameters. Grey region limits the age of the Universe at any redshift. The square-shape markers illustrate the average median and width assuming for the MLE deconvolution $\mu_2 = \mu_1 = \sigma_2 = \sigma_1 = 0$.}
\label{fig:median_width_feh}
\end{figure*}

We turn now to the median and width of the mass-weighted metallicity PDFs (see panels a, b, and c in Fig.~\ref{fig:median_width_feh}). From the results shown in panels a--f of Fig.~\ref{fig:median_width_feh}, we infer a correlation between stellar mass and metallicity. At any redshift, the higher the galaxy mass, the larger the metal content. In general, quiescent galaxies present median metallicities around solar and super-solar values. Only the least massive galaxies at the lowest redshift in our sample exhibit sub-solar metallicities on average. The median of the mass-weighted metallicity PDF also exhibits a dependence with redshift (see panels a, b, and c in Fig.~\ref{fig:median_width_feh}). When using EMILES SSP models (panels b and c in Fig.~\ref{fig:median_width_feh}), there is evidence of a decrease in the median of the mass-weighted metallicity PDFs of quiescent galaxies since $z\sim 1$. This behaviour is intrinsic to the whole quiescent population and independent of stellar mass. In the most massive bin, $\log_{10}M_\star > 11.3$, this decrease amounts to $\Delta \mathrm{[M/H]_M^{50th}} \sim  0.2$~dex for EMILES and BaSTI isochrones, whereas for the Padova00 ones, the variation is $\Delta \mathrm{[M/H]_M^{50th}} \sim  0.1$~dex. At decreasing stellar mass, this decrease is steeper, suggesting that the slope of the MZR for quiescent galaxies varies with redshift. For BC03 SSP models (see panel a in Fig.~\ref{fig:median_width_feh}) the evolution of the median metallicity exhibits a maximum value at $z\sim 0.5$--$0.6$. A decrease of the median metallicity at earlier cosmic times is also present, but only since $z\sim 0.6$, noting that the uncertainty is $\Delta \mathrm{[M/H]_M^{50th}} \sim 0.2$~dex. Interestingly, this peak matches with the redshift range where  we report a lack of quiescent galaxies in ALHAMBRA owing to cosmic variance (see Sect.~\ref{sec:number} and Fig.~\ref{fig:density}), a result that appears more prominent with BC03 models. In the light of Fig.~\ref{fig:median_width_feh} (panels a--c), there are also quantitative discrepancies between metallicity predictions from the use of different SSP models, and these are more striking for metallicity than for the rest of stellar population properties. In brief, BC03 models predict quiescent galaxies that are typically more rich in metals than those in EMILES models at any redshift, where BaSTI isochrones provide larger metallicities than the Padova00 ones.

Regarding the width of the metallicity PDFs (panels d, e, and f in Fig.~\ref{fig:median_width_feh}), there is evidence that the metallicity distributions of quiescent galaxies are typically wider at lower mass, whereas the distribution is narrower at higher stellar mass. The less massive bins ($\log_{10} M_\star < 10.9$) present widths in the range $\omega_\mathrm{[M/H]_M} \sim 0.3$--$0.5$~dex up to $z\sim 0.5$; whereas for $\log_{10} M_\star \ge 10.9$, these are below $\omega_\mathrm{[M/H]_M} \lesssim 0.3$~dex. Nevertheless, for quiescent galaxies of $\log_{10} M_\star \ge 10.9$, there are discrepancies down to $0.2$~dex between the widths of the mass-weighted metallicity PDFs retrieved using BC03 and EMILES SSP models. While for BC03 these ones present a more relevant evolution with redshift, for EMILES they present a roughly constant width of $\omega_\mathrm{[M/H]_M} \sim 0.3$~dex, with a slight redshift dependence. In addition, BC03-derived metallicities show that the evolution of the width of the metallicity PDFs is independent of the stellar mass of quiescent galaxies, whereas the EMILES ones suggest a dependence with stellar mass.

\subsection{The extinction in the quiescent population}\label{sec:extinction}

\begin{figure*}
\centering
\resizebox{!}{10.8cm}{\includegraphics[trim = 0 0 0.5cm 0, clip=True]{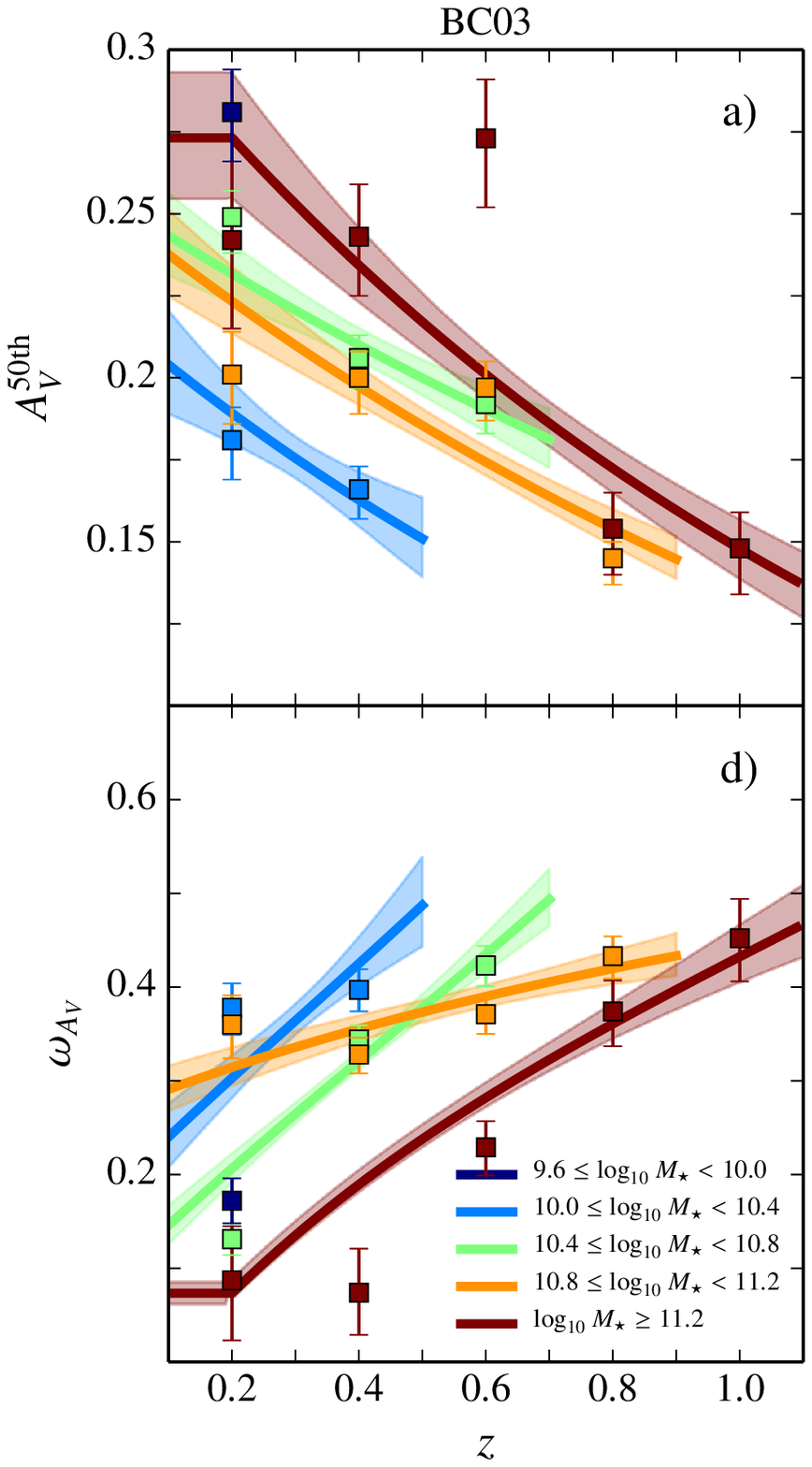}
\includegraphics[trim = 2.33cm 0 0.5cm 0, clip=True]{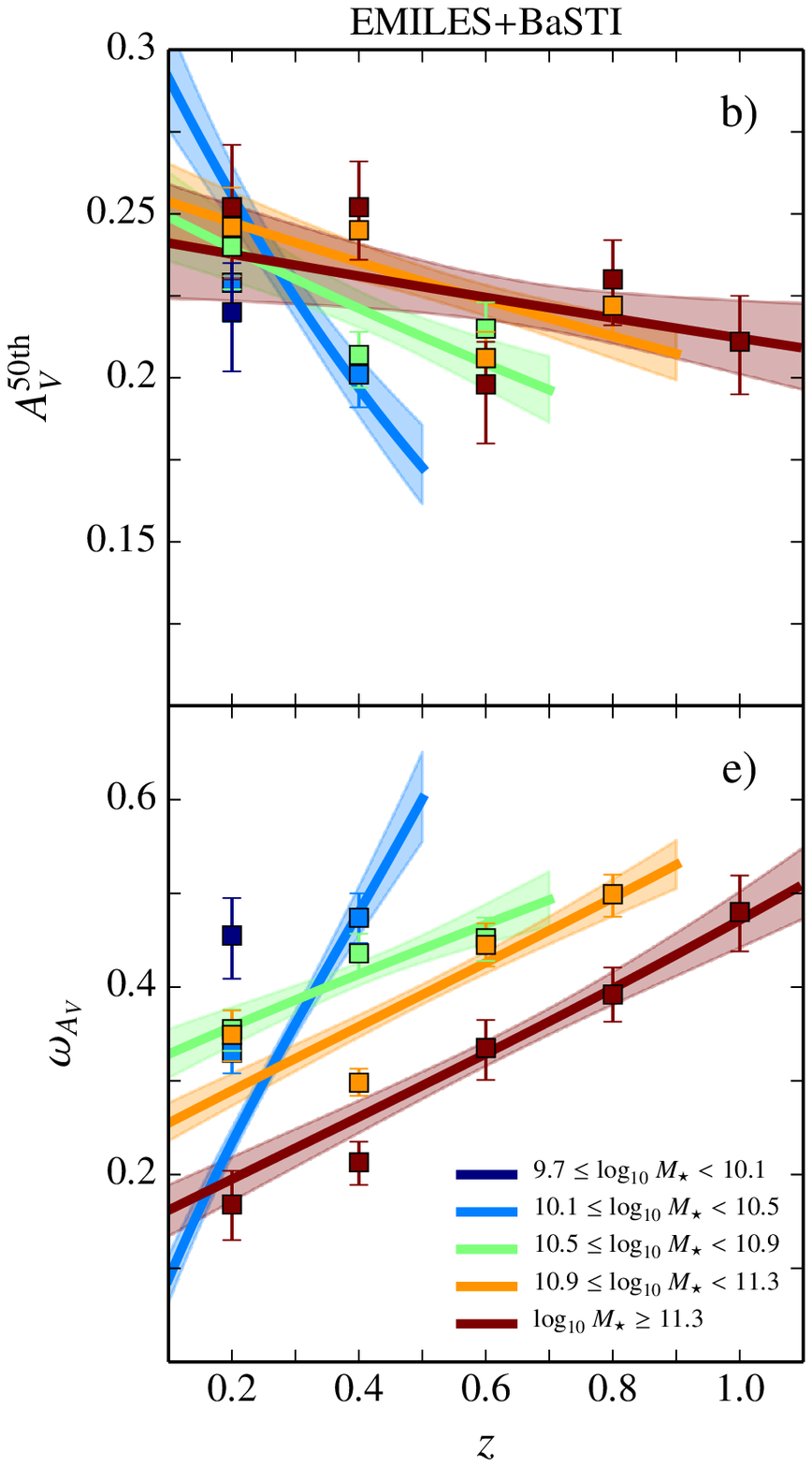}
\includegraphics[trim = 2.33cm 0 0.5cm 0, clip=True]{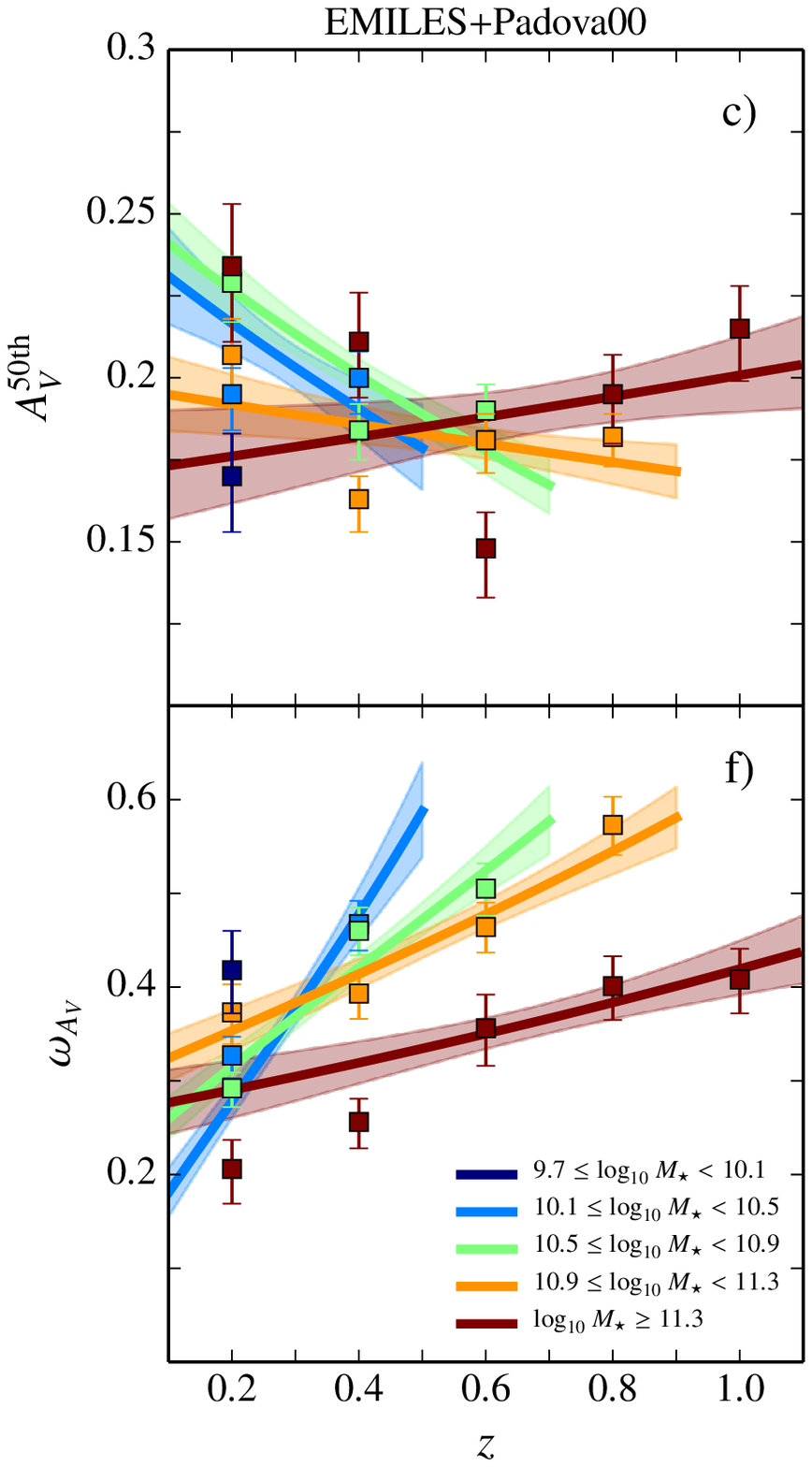}}
\caption{As Fig.~\ref{fig:median_width_age}, but for the extinction PDFs of quiescent galaxies.}
\label{fig:median_width_av}
\end{figure*}

Overall, the extinction PDFs of quiescent galaxies derived in this research, see panels a--f in Fig.~\ref{fig:median_width_av}, show predominant low values ($A_V \lesssim 0.6$) irrespectively of stellar mass or redshift. Indeed, the median of the extinction PDFs at all stellar masses and redshifts does not exceed $A_V^\mathrm{50th} = 0.3$.

From panels a, b, and c in Fig.~\ref{fig:median_width_av}, we infer a very subtle relation between the stellar mass and dust extinction of quiescent galaxies with differences $\Delta A_V^\mathrm{50th} < 0.1$ when using BC03 SSP models. For EMILES, we do not appreciate significant differences in the median values of the extinction PDFs amongst the stellar mass bins (see panels b and c in Fig.~\ref{fig:median_width_av}), where all the values retrieved are compatible at a $1$~$\sigma$ confidence level for both BaSTI and Padova00 results. Independently of the SSP model set used in this work, the median of extinction PDFs vary around  $A_V^\mathrm{50th} \sim 0.15$--$0.30$ up to $z= 1.1$. We also find  that there is a subtle increment of extinction in the lower-mass bins ($\log_{10}M_\star \le 10.8$) at decreasing redshift. On the other hand, quiescent galaxies with $\log_{10}M_\star \ge 10.8$ show that the median extinction remains constant $A_V^\mathrm{50th}\sim 0.2$ for EMILES and in the range $A_V^\mathrm{50th}\sim 0.15$--$0.30$ for BC03 models.

Both BC03 and EMILES estimates point out that there may be mild discrepancies between the widths of the extinction PDFs at different stellar masses (see panels d, e, and f in Fig.~\ref{fig:median_width_av}). In any case, these discrepancies amount to width differences below $0.1$. The lower the stellar mass, the larger the width of the PDF. At decreasing redshift the extinction PDFs get narrower across all  stellar masses, where less massive quiescent galaxies present broader probability distributions. Note that for the BC03-derived results at $0.1 \le z < 0.2$ and $\log_{10} M_\star \ge 11.2$, the assumption of linearity for $\sigma^\mathrm{int}(z)$ and $\mu(z)$ in the extinction case is too strict, and we imposed $\sigma^\mathrm{int}(0.1 \le z < 0.2)=\sigma^\mathrm{int}(z=0.2) = 0.14 \pm 0.03$ and $\mu(0.1 \le z < 0.2) = \mu(z=0.2) = -1.30 \pm 0.07$ (details in Appendix~\ref{appendix:pdf_bc03}).

\subsection{Constraints on the SFH}\label{sec:sfh_alt}

The different SFH assumptions adopted in the construction of composite models of stellar populations may have a potential impact on the results shown above. We studied these effects, repeating the full analysis with new sets of composite models with alternative SFH constraints. For this work, we explore the following assumptions on the SFH:

\begin{itemize}
\item[i)] Constant values of extinction for all  quiescent galaxies at any redshift and stellar mass.
\item[ii)] Fixed solar metallicity.
\item[iii)] Closed-box enrichment of metals. The young component in the mixture of the two SSP models of MUFFIT has to be more metal-rich than the old one.
\item[iv)] SSP mixtures with the same metallicity for the old and young components.
\item[v)] Infall of metal-poor cold gas from the cosmic web, that is, the young component in the mixture of SSPs is more metal poor than the old component.
\item[vi)] Quiescent galaxies exhibit a constant MZR with cosmic time, matching the one in the nearby Universe. This test is only performed for EMILES with BaSTI isochrones.
\end{itemize}

After studying the impact of these constraints on the stellar population parameters of quiescent galaxies and PDFs, we find that:
\begin{itemize}
\item[$\bullet$] None of these SFH assumptions alter our main conclusions, namely the evolution of quiescent galaxies is not compatible with a passive evolution and there is a continuous decrease in metallicity (trivially excluding the fixed solar metallicity assumption above) since $z\sim 1$.
\item[$\bullet$] SFH constraints introduce non-negligible systematics that quantitatively alter the age, metallicity, and extinction.
\item[$\bullet$] Constraints on the SFH are a source of (quantitative) uncertainties that can have a larger impact than the
  ``basic'' uncertainties obtained in the determination of the stellar population parameters.
\end{itemize}


\section{A global view on the evolution of quiescent galaxies since $z\sim 1$}\label{sec:discussion_evolution}

The PDFs of mass-weighted age, metallicity and dust extinction of quiescent galaxies since $z\sim 1$ constrain in an unprecedented way the evolution of these systems during the last $8$~Gyr of cosmic time. Thanks to the statistical deconvolution of uncertainty effects (MLE method, details in Appendix~\ref{sec:appendix_pdf}), and to the large and mass-complete set used here ($\sim8\,500$ galaxies), it is possible to explore the evolution of quiescent galaxies as a whole. In particular, the intrinsic dispersions of the PDFs constitute new observables to constrain the evolution of quiescent galaxies.

This work presents evidence suggesting the age evolution of quiescent galaxies departs from a passive scenario, showing on average milder ageing. This conclusion is obtained even when assuming different constraints on the SFHs during the MUFFIT analysis (Sect.~\ref{sec:sfh_alt}). Moreover, we find evidence for a slight decrease of the median of the metallicity PDF ($0.1$--$0.2$~dex) of quiescent galaxies since $z\sim 0.6$--$1.1$ (BC03 and EMILES respectively), consistently obtained under most of the adopted SFHs, and constitutes one of the most striking results of this work. Both a steeper MZR at larger cosmic time and the continuous decrease of the median metallicity with time additionally support that quiescent galaxies are continuously modifying their stellar content.

All these results are in conflict with strict passive evolution. Two alternative scenarios are discussed here to reconcile observations on the evolution of quiescent galaxies, as well as establishing the role of the ``progenitor'' bias:

\begin{itemize}
\item[i)] Mergers. The inclusion of new stars, formed ex-situ from less massive systems may be a potential mechanism to alter the stellar content of galaxies \citep[see e.~g.][]{Ferreras2005,Croton2006,Khochfar2006,Hopkins2008,Kaviraj2008,
Ferreras2009a,Ferreras2009c,Hopkins2009,Rogers2009,SanchezBlazquez2009a,vanderWel2009,DiazGarcia2013,LopezSanjuan2013,Skelton2012,
Ferreras2014}. In this case, the number of mergers involving quiescent galaxies is key to discern whether this mechanism can match the observed non-passive evolution.
\item[ii)] Remnants of star formation or ``frosting'' \citep[a term firstly introduced by][]{Trager2000}. Clouds of gas inside the galaxy or the infall of new gas from the cosmic web can originate new stellar populations with different properties than those already present in the galaxy \citep[][]{Ferreras2000,Trager2000,Schiavon2006,Kaviraj2007,
Rogers2007,Schiminovich2007,Serra2007,Somerville2008,Ferreras2009c,Rogers2009,Lonoce2014,Vazdekis2016}.
\item[iii)] The ``progenitor'' bias \citep[][]{vanDokkum2001}. Quiescent galaxies at high redshift provide a biased set with respect to their nearby counterparts, and therefore, not all the progenitors of the low redshift sample are included in the analysis \citep[see also][]{Valentinuzzi2010,Carollo2013,Belli2015}. Indeed, the generalized increase in the number of quiescent galaxies since $z\sim 1$ motivates the inclusion of the ``progenitor'' bias in the discussion.
\end{itemize}

Below, we discuss and detail the likely effects of mergers and ``frosting'' on the stellar content of quiescent galaxies, more precisely on age (Sect.~\ref{sec:discussion_sp_ages}), metallicity (Sect.~\ref{sec:discussion_sp_feh}), and extinction (Sect.~\ref{sec:discussion_sp_av}).

\subsection{On the ages of quiescent galaxies}\label{sec:discussion_sp_ages}

A galaxy undergoing a natural ageing would exhibit a constant formation epoch with cosmic time. Otherwise, we would conclude that the population of quiescent galaxies is being altered. Mergers and ``frosting'' may have a relevant role on the typical ages of quiescent galaxies because these processes introduce new stars in the stellar content of quiescent galaxies. In fact, these mechanisms can act in parallel, increasing their mutual effects. On the other hand, part of the progenitors of low-redshift quiescent galaxies are not included in our sample, and their stellar content can differ with respect to high-redshift quiescent galaxies. The consequences of each of these mechanisms and the ``progenitor'' bias mentioned above are discussed in the same order in which they were presented:

\begin{itemize}
\item[i)] Less massive systems are expected to contain younger stellar populations due to the age-mass relation \citep[see e.~g.][]{Gallazzi2005,Thomas2005,Peng2015}, in qualitative agreement with the ``downsizing'' scenario. Hence these systems are potential contributors to ``slow down'' the ageing of quiescent galaxies when they are accreted via mergers.
\item[ii)] The inclusion of new (in-situ) stars would easily explain why the evolution of these galaxies departs from passiveness. However, ``frosting'' is needed to affect the whole quiescent population to be a reliable mechanism.
\item[iii)] Part of the progenitors of the nearby quiescent galaxies from ALHAMBRA were star-forming galaxies in the high redshift bins explored in this work, that is, these progenitors were still forming new stars at that epoch. Consequently, galaxies quenching their star formation at more recent epochs will contain younger stellar populations. Hereby, the samples of high redshift quiescent galaxies are biased to the older parts of the age PDFs at lower redshift. This result would partly explain why the median ages of quiescent galaxies do not vary passively.
\end{itemize}

\subsection{On the metallicities of quiescent galaxies}\label{sec:discussion_sp_feh}

The evolution of the mass-weighted metallicity PDFs contributes an extra hint, pointing out that some of the mechanisms discussed above may be altering the typical stellar population parameters of quiescent galaxies. The consequences, listed separately for each of the scenarios considered,  would be the following:

\begin{itemize}
\item[i)] Mergers with less massive systems can contribute to the observed variation of the global metallicity, as less massive systems host metal poorer populations with respect to the more massive galaxy in the pair, owing to the stellar mass-metallicity relation \citep[see e.~g.][]{Peng2015,Jorgensen2017}. Again, the number of mergers would determine whether this mechanism is capable of altering the median of the mass-weighted metallicity PDF.
\item[ii)] A priori, ``frosting'' would not explain a decrease of the global metallicity in a monolithic collapse. However, in the chemo-evolutionary population synthesis model by \citet[][no interchange of any matter with the neighbourhood, or closed-box model]{Vazdekis1996}, the global metallicity of a galaxy may decrease at certain cosmic time. Owing to a very intense star formation, the available gas decreases very rapidly and the enrichment of metals quickly asymptotes to the net yield. Then, all the available gas mostly comes from the oldest and numerous low-metallicity stars, which is less processed, producing new stellar populations that can reduce the metallicity around $0.2$~dex at $\sim10$~Gyr after the initial star formation. For bottom heavy IMFs, this effect is even more prominent.
\item[iii)] If the progenitors of low redshift quiescent galaxies are biased at high redshift because they are still assembling their stellar content, the ``progenitor'' bias would only affect the results in the direction of more metal-poor metallicity. In fact, the evolution in number density of quiescent galaxies since $z\sim 1$ is more remarkable at decreasing stellar mass. Therefore, the progenitors of our sample of quiescent galaxies are likely more biased at lower masses and the metallicities would be more affected by the ``progenitor'' bias, i.~e.~presenting larger variations in metallicity, as observed in this work.
\end{itemize}

\subsection{Extinctions of quiescent galaxies}\label{sec:discussion_sp_av}

The extinction of quiescent galaxies presented in this paper are well constrained below $A_V < 0.6$ and they do not present large variations with cosmic time. Indeed, extinction is roughly constant ($A_V\sim 0.2$) for EMILES models. Consequently, the effects of mergers, ``frosting'' and the ``progenitor'' bias may be less remarkable here than for ages and metallicities (see Sects.~\ref{sec:discussion_sp_ages} and \ref{sec:discussion_sp_feh}), and they can even cancel each other. We would expect the following impact on the extinction PDFs:

\begin{itemize}
\item[i)] Mergers between quiescent and star-forming galaxies (the latter expected to feature larger extinction) will increase the overall extinction of the resulting galaxy. The merger orbit, as well as the distribution of dust in the progenitors would drive how the global extinction of the final galaxy is affected.
\item[ii)] Concerning ``frosting'', we would not expect that low levels of in-situ star formation can alter typical extinction values in quiescent galaxies, which are expected to have low reserves of available gas and where current star formation processes are not significant yet.
\item[iii)] Since star-forming galaxies are typically more reddened by dust than quiescent ones, our sample of quiescent galaxies at high redshift can be biased to lower extinctions. However, the mechanism responsible for shutting down the star formation is still unknown, as well as the typical extinction of galaxies quenching their star formation. In fact, the evolution of extinction in star forming galaxies is unclear, and the quenching mechanism can also be tightly related to it \citep[e.~g.~sudden removal of gas, ``strangulation'', heating of galaxy's gas by AGN, ][]{Silk1998,Balogh2000,Dekel2006,Hopkins2006,Nandra2007,Bundy2008,DiMatteo2008,Diamond2012,Peng2015}.
\end{itemize}

In general, mergers and ``frosting'' can potentially contribute to modify the width of the mass-weighted age, metallicity, and extinction PDFs of quiescent galaxies, which constitutes an additional hint towards the variation of their stellar content. In fact, a unique mechanism would not be able to produce all the changes revealed in this work, but rather a simultaneous combination of merger and ``frosting'' may be a more realistic scenario. Independently of the impact of each of the mechanisms mentioned above, any evolution in the widths of the stellar population PDFs, or intrinsic scatter, requires either a non-passive evolution or an external mechanism acting on the stellar content of quiescent galaxies. In addition, we have to take into account that part of the PDF evolution may be a consequence of a biased sample, as we only focus on the stellar content of quiescent galaxies over a wide redshift range (i.~e.~the ``progenitor'' bias).

\section{Comparison with previous studies}\label{sec:previous}

The evolution of the stellar populations of quiescent galaxies has been mainly tackled using spectroscopic data. In fact, this kind of studies usually require an analysis based on stacked spectra, which does not allow us to determine the intrinsic dispersion of the stellar population distributions. In this section and whenever possible, we compare the ages and metallicities retrieved in this research using the ALHAMBRA data with those from previous studies (see Table~\ref{tab:sp_pw} for further details). As the quiescent galaxies from ALHAMBRA comprise a wide range in cosmic time, we structured this section according to the redshift range explored (Sect.~\ref{sec:comp1}--\ref{sec:comp3}).

\begin{table*}
\caption{Brief description of previous studies trying to constrain the stellar population parameters of quiescent galaxies at intermediate redshift.}
\label{tab:sp_pw}
\centering
\begin{tabular}{lcccc}
\hline\hline
\multicolumn{1}{c}{\multirow{2}{*}{References}} & \multicolumn{1}{c}{\multirow{2}{*}{Redshift}} & \multicolumn{1}{c}{\multirow{2}{*}{Stellar mass}} & \multirow{2}{*}{Number} & \multirow{2}{*}{Parameters}\\
&&&& \\
\hline
&&&& \\
\citet{Schiavon2006}$^{\dag}$        & $0.7 < z < 1.0$     & $\sigma \gtrsim 170$          & $1\,160$ & Age, [M/H] \\
\citet{Ferreras2009c}                & $0.4 < z < 1.3$ &  $ 10^{9} < M_\star < 10^{12}$ & $228$ & Age, [M/H] \\
\citet{SanchezBlazquez2009}$^{\dag}$ & $0.4 < z < 0.8$     & $\sigma > 100$                & $215$    & Age, [M/H] \\
\citet{Whitaker2013}    & $1.4 < z < 2.2$     &     $M_\star > 10^{10.5}$            & $171$    & Age, [M/H] \\
\citet{Choi2014}         & $z < 0.7$     & $10^{\ \ 9.6} < M_\star < 10^{11.8}$                & $2\,400$  & Age, [M/H] \\
\citet{Gallazzi2014}     & $z \sim 0.7$  & $M_\star > 10^{10.5}$           & $33$      & Age, [M/H] \\
\citet{Belli2015}        & $1.0\ \ < z < 1.12$  & $M_\star > 10^{11.1}$    & $12$      & Age        \\
\citet{Peng2015}         & $0.05 < z < 0.09$    & $M_\star > 10^{\ \ 9.5}$ & $22\,168$ & Age, [M/H] \\
\citet{Fumagalli2016}    & $0.5 < z < 2.0$      & $M_\star > 10^{10.8}$    & $248$     & Age        \\
\citet{Jorgensen2017}$^{\dag}$    & $0.2 < z < 0.9$      & $M_\star > 10^{10.3} $    & $221$     & Age, [M/H]        \\
&&&& \\
\hline
\end{tabular}
\tablefoot{From left to right, reference of each study, redshift bin, stellar mass range of the sample, number of galaxies, and stellar population parameters explored. All the stellar masses (velocity dispersion, $\sigma$) are in solar units [M$_\odot$] ($\mathrm{km~s^{-1}}$). All the papers involve spectroscopic data. \\ ($\dag$) In the work by \citet{Schiavon2006,SanchezBlazquez2009,Jorgensen2017}, velocity dispersions are used as stellar mass proxy. We repeat the same colour selection than \citet{Schiavon2006} using the conversion factors from \citet{Willmer2006} to unveil that typical stellar masses in their sample ranges $10.8 < \log_{10}M_\star < 11$. For \citet{SanchezBlazquez2009}, we estimate stellar masses from velocity dispersions using eq.~($2$) in \citet{Thomas2005}.}
\end{table*}


\subsection{Ages and metallicities of quiescent galaxies in the nearby Universe}\label{sec:comp1}

Aimed at exploring the mechanisms responsible for quenching galaxies, \citet{Peng2015} studied both the metallicity- and age-stellar mass relations of quiescent galaxies at $0.05 \le z \le 0.085$. The quiescent sample was composed of $22\,618$ galaxy spectra from SDSS, where ages and metallicities were measured by a multiple fit to age and metallicity sensitive indices \citep[see also][]{Gallazzi2005}. After comparing the age estimations provided by MUFFIT and BC03 (EMILES+BaSTI) SSP models with the \citet{Peng2015} ones, the latter are $\sim1.5$~Gyr older (younger) with a similar correlation between age and stellar mass for $\log_{10}M_\star \ge 10$ (see panels a and b in Fig.~\ref{fig:comparison}, respectively). The EMILES+Padova00 age-stellar mass relation is steeper, but the range of ages for quiescent galaxies is qualitatively the same than the one provided by \citet[][see panel c in Fig.~\ref{fig:comparison}]{Peng2015}. Regarding metallicity, there is a shift of $\sim-0.2$~dex between our BC03 and EMILES+BaSTI values and those provided by \citet[][see panel d and e in Fig.~\ref{fig:comparison}]{Peng2015}. For the Padova00 models, at the lowest metallicity values, this shift is even larger ($\Delta\mathrm{[Fe/H]}\sim -0.3$~dex). Although if we account for the SDSS aperture effects \citep[$0.15$--$0.2$~dex, ][]{Gallazzi2005}, our metallicity predictions agree with those provided by \citet{Peng2015}.


\subsection{Quiescent galaxies at intermediate redshifts}\label{sec:comp2}

Our sample of quiescent galaxies is expected to be dominated by field galaxies. However, galaxies in dense environments may present systematic differences in their stellar contents with respect to field samples \citep[e.~g.][]{SanchezBlazquez2003,Eisenstein2003,Thomas2005,Trager2008,Thomas2010}. For this reason we distinguish between spectroscopic studies that include field quiescent galaxies (Sect.~\ref{sec:comp2a}) and those in clusters (Sect.~\ref{sec:comp2b}) to compare with our results.


\subsubsection{The stellar content of field quiescent galaxies}\label{sec:comp2a}

The spectroscopic study of \citet{Schiavon2006} also revealed that the ages of red sequence (RS) galaxies at $z\sim 0.9$ are not compatible with a passive evolution. More precisely, the authors found out that RS galaxies from the DEEP2 surveys \citep[][]{Davis2003} present luminosity-weighted SSP ages around $\sim1.5$~Gyr. These ages are significantly younger than the ones obtained by MUFFIT for quiescent galaxies (see panels a--c in Fig.~\ref{fig:comparison}, mass-weighted ages of $5$~Gyr). Nevertheless, this qualitative difference is in part explained by the use of both index-index diagrams (e.~g.~luminosity-weighted ages) and single stellar population models to estimate ages. Regarding metallicities, there is a reasonable agreement between the iron abundances of \citet{Schiavon2006} ($\mathrm{[Fe/H]}= 0.0$--$0.3$~dex) and the ones by MUFFIT ($0.12$, $0.13$, $-0.04$~dex for BC03, BaSTI, and Padova00; see panels d--f in Fig.~\ref{fig:comparison}). Finally, \citet{Schiavon2006} also concluded that either RS galaxies experienced a continuous low-level star formation, or there is an inclusion of new galaxies with younger stellar populations, in other words, ``frosting'' and the ``progenitor'' bias \citep[e.~g.][]{Trager2000,Bell2004,Faber2007}.

\citet{Ferreras2009c} used low-resolution slitless grism optical spectra from \textit{HST}/ACS to constrain the stellar populations of a sample of visually-selected ETGs at $0.4 < z < 1.3$. Overall, the authors found a strong correlation between stellar age and mass at intermediate redshift, where the age spread was roughly constant, at around $\sim1$~Gyr. \citet{Ferreras2009c} also divided their morphologically selected sample into red and blue galaxies. For a fair comparison, we selected all the red ETGs at $0.4 < z < 1.3$ and $10.6 < \log_{10}M_\star < 11.4$ to create a representative subsample to compare with our predictions \citep[a $\sim90$~\% fraction of ETGs at this mass range and redshift are quiescent galaxies, see][]{Moresco2013}. Red ETGs exhibit ages close to passive evolution, hence showing a better agreement with our predictions using EMILES+BaSTI SSP models and larger discrepancies with the BC03 models (see panels a--c in Fig.~\ref{fig:comparison}). Regarding metallicity, \citet{Ferreras2009c} present more metal-poor populations than our predictions, as expected from the use of a chemical enrichment model of stellar populations in comparison with other composite models \citep[e.~g.~$\tau$-models or two burst models; in any case compatible with a $\chi^2 \sim 1$ for all the composite models, see][]{Ferreras2009c}. However, we cannot state a clear decrease in metallicity at $0.4 < z < 1.3$ for the galaxies from \citet{Ferreras2009c}.

At a similar redshift range, \citet{Choi2014} analysed the stellar populations of quiescent galaxies up to $z<0.7$. They computed SSP equivalent ages and abundances of elements by stacked spectra. \citet{Choi2014} found that quiescent galaxies are older at lower redshift and compatible with both a ``downsizing'' scenario and a composed passive evolution. Regarding [Fe/H], unlike our predictions, the authors do not clearly retrieve either the MZR at lower redshift or a metallicity evolution with redshift. In addition, they state that the metallicity can be potentially affected by aperture effects (an excess of $\lesssim 0.1$~dex at the lowest redshifts). In fact, the metallicity decrease obtained in our work is $\lesssim0.15$~dex since $z=0.7$, that is, in the order of the aperture bias reported by \citet{Choi2014}.

Using a multiple fit to age- and metallicity-sensitive absorption features, \citet{Gallazzi2014} explored the ages and metallicities of $33$ quiescent galaxies at $z \sim 0.7$ with stellar mass $M_\star > 3\times 10^{10}$~$\mathrm{M_\odot}$. \citet{Gallazzi2014} found out that ages of quiescent galaxies at $z \sim 0.7$ are $\sim2$~Gyr younger than those obtained with a similar methodology at low-redshift \citep[][]{Gallazzi2005}, that is, less than the predicted by a passively evolving assumption. We find that the mass-weighted ages of MUFFIT+BC03 are in good agreement with those provided by \citet[][panel a in Fig.~\ref{fig:comparison}]{Gallazzi2014}. However, we obtain larger metallicities of $\Delta\mathrm{[Fe/H]}\sim0.1$~dex (see panel d in Fig.~\ref{fig:comparison}). For EMILES models, the ages provided by MUFFIT are $\sim1.5$~Gyr older (see panels b--c in Fig.~\ref{fig:comparison}). For BaSTI isochrones the agreement with metallicity is remarkable, mostly at the highest stellar mass bin.

Making use of spectroscopic data from the 3D-HST survey, \citet{Fumagalli2016} estimated the ages of $UVJ$ quiescent galaxies at $0.5 < z < 2.0$. Only the most massive galaxies ($\log_{10}M_\star > 10.8$) were selected in order to stack all the spectra in three redshift bins. As a result, the ages of quiescent galaxies present a large spread in age due to the use of different model sets, although typically below half of the age of the Universe. Our analysis at the same redshift ranges yields older mass-weighted ages. It is remarkable that the extrapolation at $z= 1.25$ of the mass-weighted ages (BC03 and EMILES) also match with the ages derived by \citet{Fumagalli2016} and BC03 models (see panel a in Fig.~\ref{fig:comparison}). It is worth mentioning that \citet{Fumagalli2016} only adopt solar metallicity models. This constraint may introduce substantial systematics in the retrieved ages that can lead to passive evolution of the stellar content.


\subsubsection{Quiescent galaxies in clusters up to intermediate redshifts}\label{sec:comp2b}

\begin{figure*}
\centering
\resizebox{\hsize}{!}{
\includegraphics[trim= 0 1.35cm 1.62cm 0cm,clip=True]{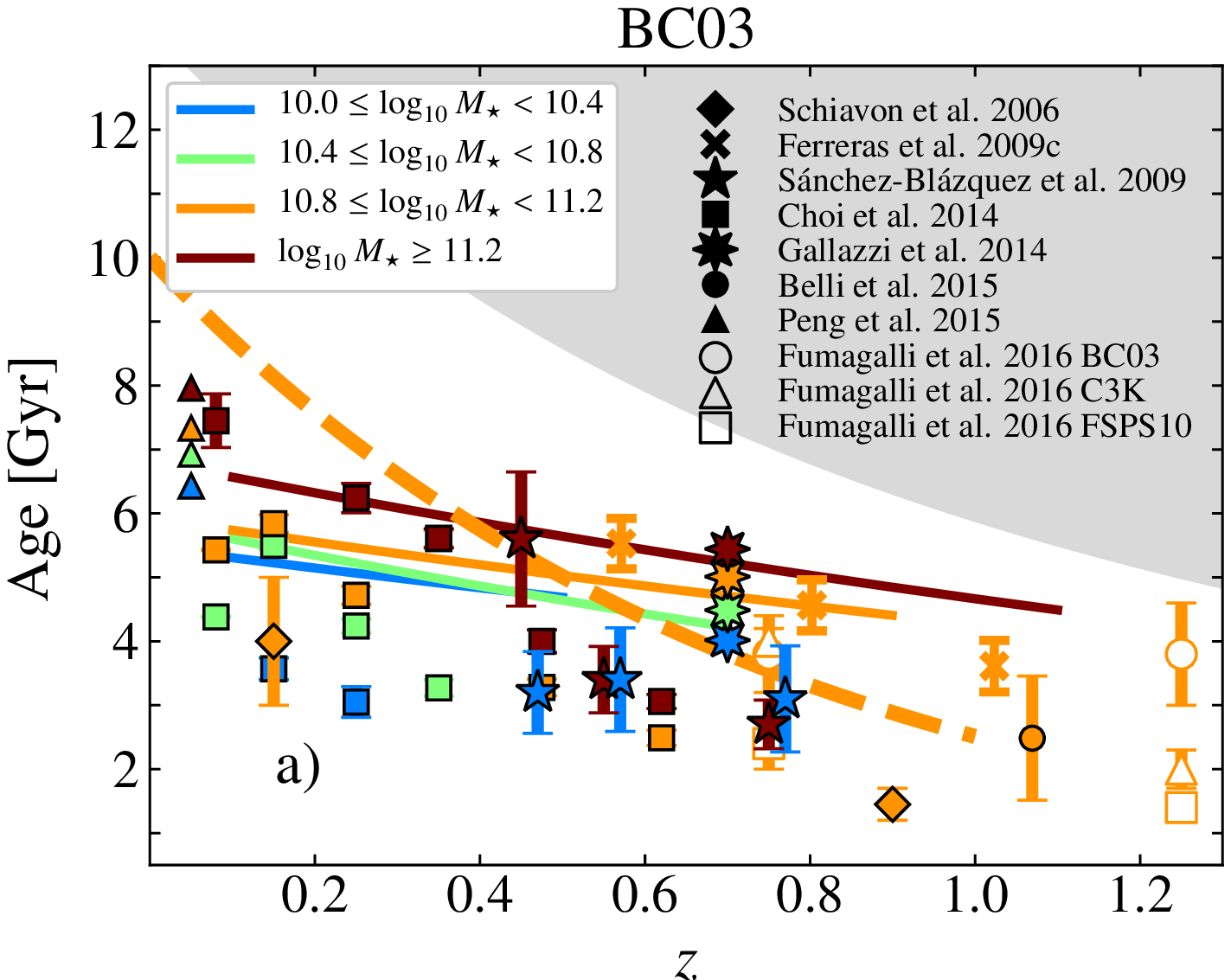}
\includegraphics[trim= 2.03cm 1.35cm 1.62cm 0,clip=True]{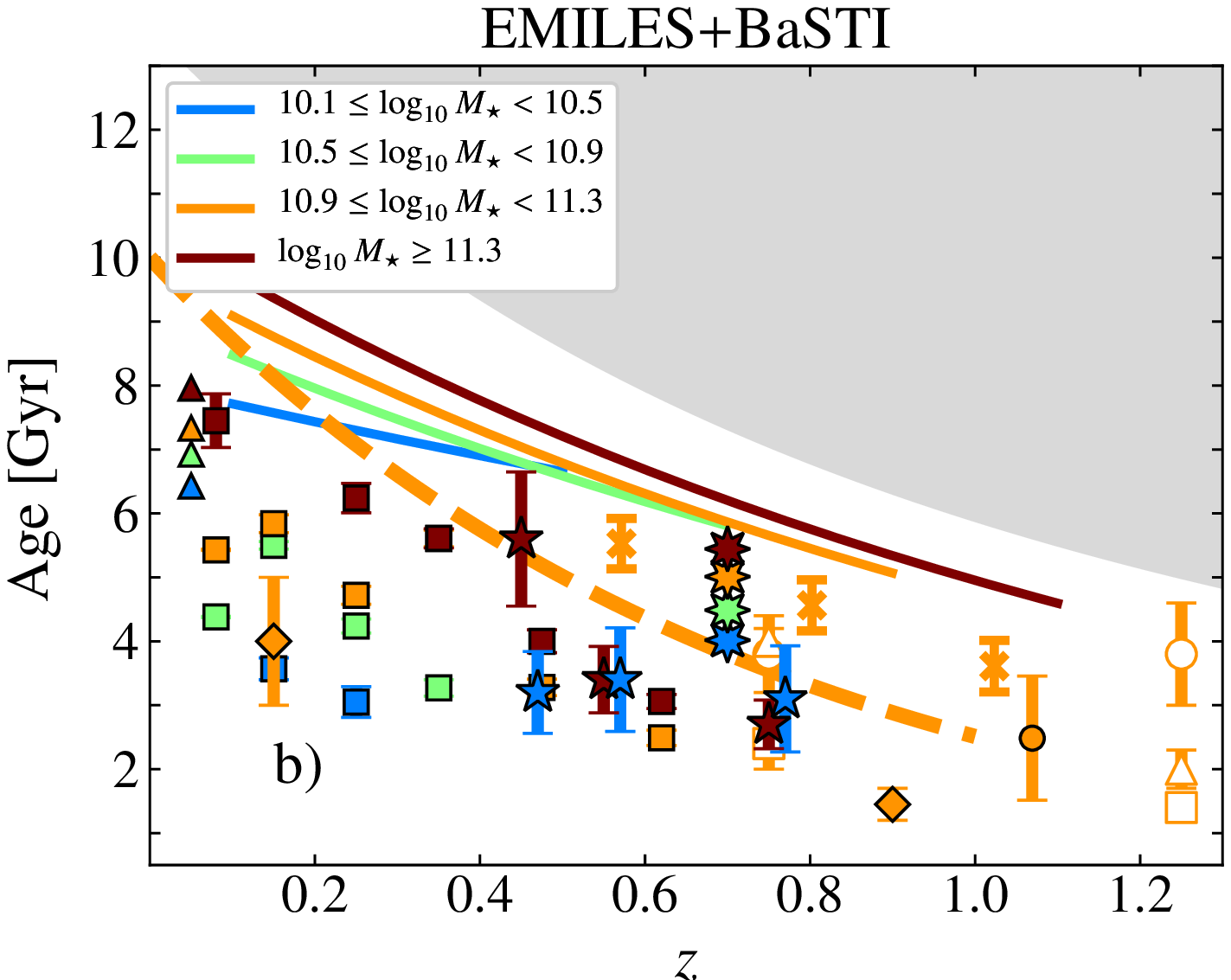}
\includegraphics[trim= 2.03cm 1.35cm 1.62cm 0,clip=True]{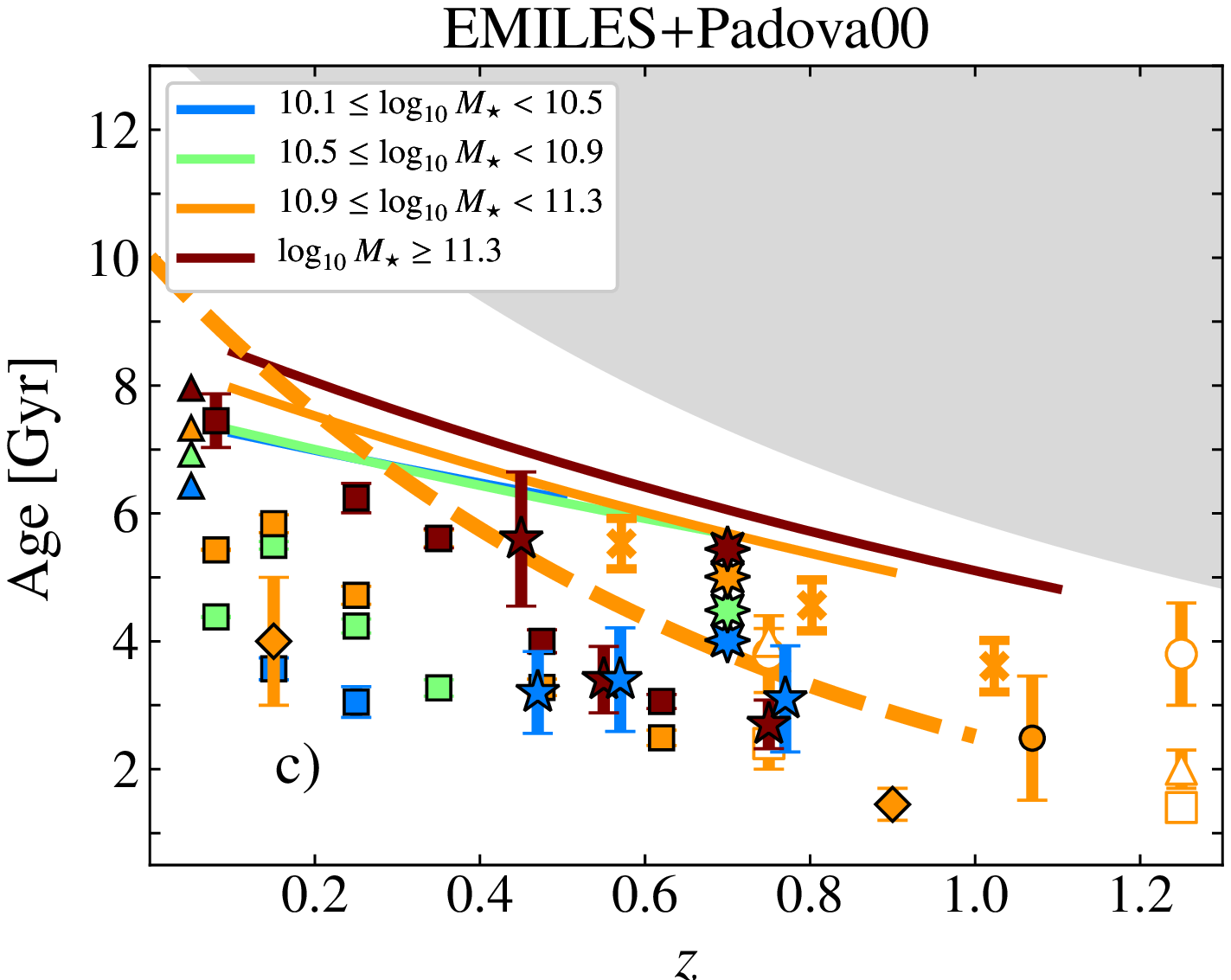}}\newline
\resizebox{\hsize}{!}{
\includegraphics[trim= 0 0 1.62cm 0.57cm,clip=True]{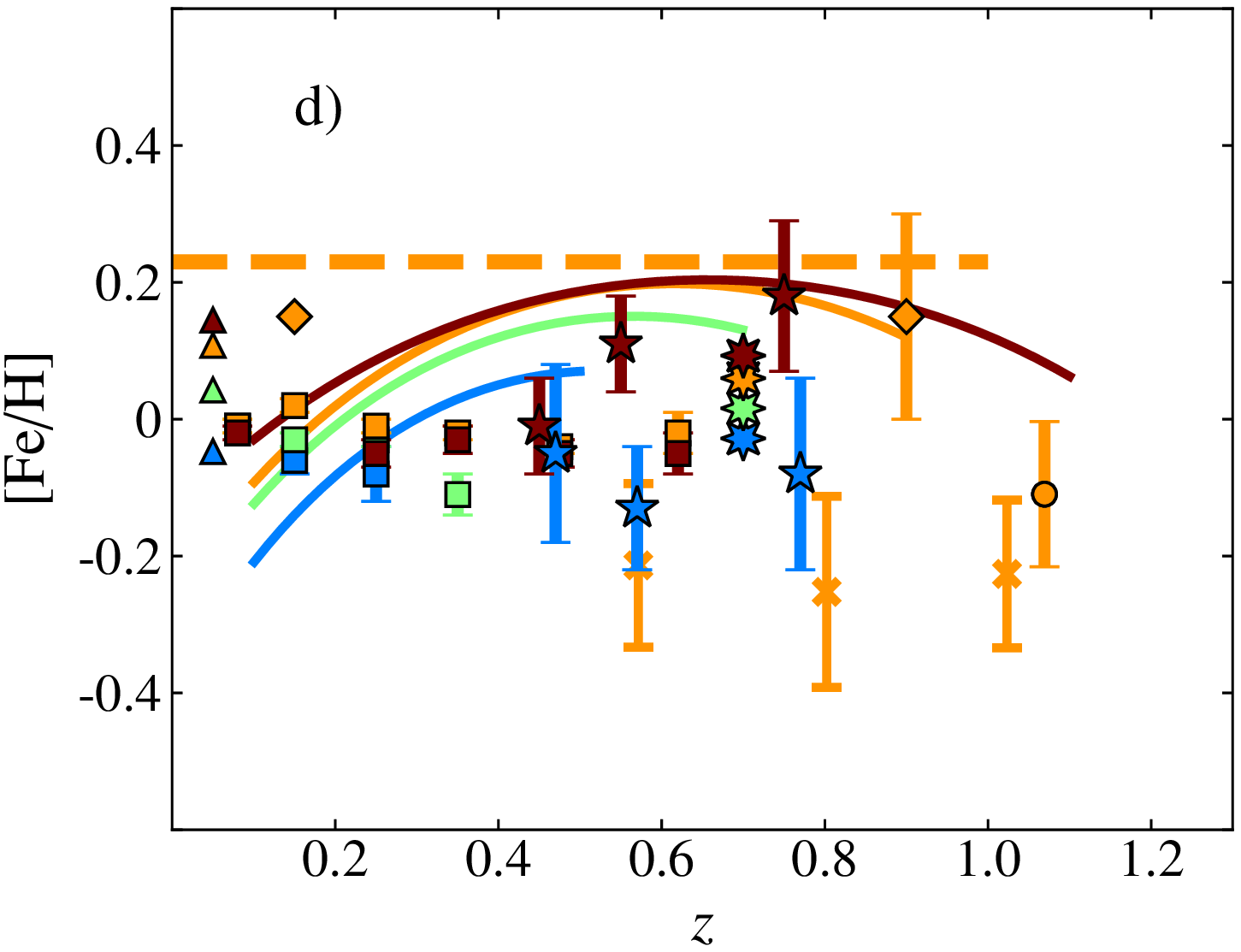}
\includegraphics[trim= 2.03cm 0 1.62cm 0.57cm,clip=True]{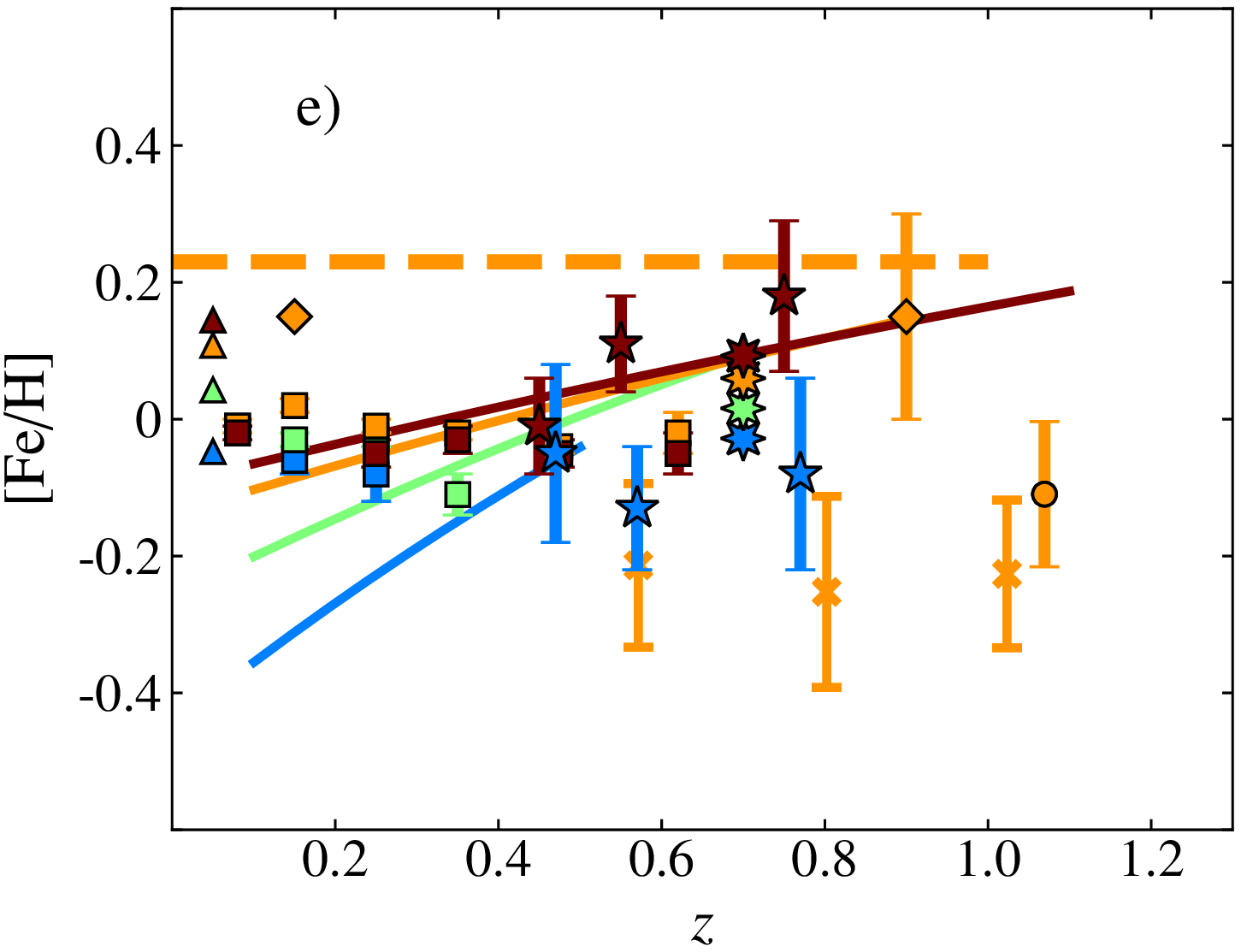}
\includegraphics[trim= 2.03cm 0 1.62cm 0.57cm,clip=True]{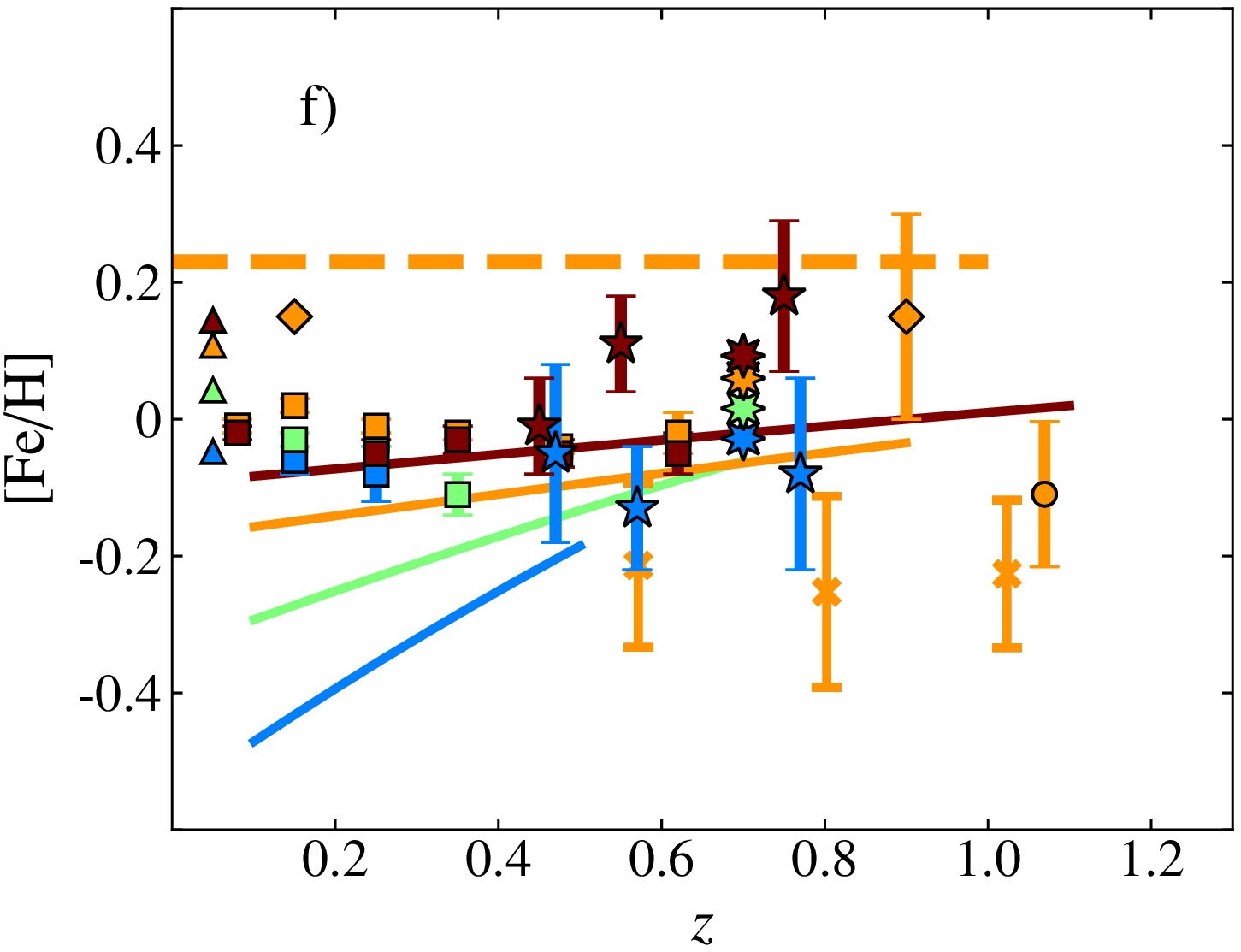}}\newline
\caption{Comparison of ages (\textit{panels a}, \textit{b}, and \textit{c}) and metallicities (\textit{panels d}, \textit{e}, and \textit{f}) of quiescent galaxies from several work (details in the text, see legend for references) and our results from ALHAMBRA at the redshift range $0 < z < 1.2$. \textit{From left to right}, the median mass-weighted ages and metallicities retrieved using MUFFIT (solid lines) and BC03 (\textit{panels a} and \textit{d}), EMILES with BaSTI (\textit{panels b} and \textit{e}) and Padova00 isochrones (\textit{panels c} and \textit{f}), respectively. The ages and metallicities retrieved from other work are colour coded in concordance with their proximity to the stellar mass bins of our quiescent sample (see inset). Dashed orange lines illustrate the average age and metallicity in galaxy clusters obtained by \citet{Jorgensen2017}. Grey region limits the Universe age.}
\label{fig:comparison}
\end{figure*}

The stellar population properties of RS galaxies in clusters and groups at $0.4 < z < 0.8$ were studied by \citet{SanchezBlazquez2009}. The authors found that quiescent galaxies have older ages at lower redshift (see panels a--c in Fig.~\ref{fig:comparison}), which were compatible with a passive evolution at $1$~$\sigma$ uncertainty level (formation redshift $z_\mathrm{f} > 1.4$). Interestingly, the most massive galaxies in \citet{SanchezBlazquez2009} exhibit hints for a decrease in metallicity at lower redshift, which would be in qualitative agreement with our results (see panels d--f in Fig.~\ref{fig:comparison}). Regarding the less massive systems, they showed neither age nor metallicity evolution within the errors. Overall, they concluded that massive red galaxies are compatible with passive evolution, whereas the low mass systems either need a continuous level of star formation to maintain a constant age or the RS is in a continuous build-up adding new stars. Nevertheless, as commented above, the stellar population properties of galaxies in clusters may slightly differ with respect to those in the field \citep[e.~g.][]{SanchezBlazquez2003,Eisenstein2003,Thomas2005,Trager2008,Thomas2010}.

\citet{Jorgensen2017} analysed the stellar populations of passive galaxies in seven massive clusters at $0.2 < z < 0.9$ \citep[see also][]{Jorgensen2005,Chiboucas2009,Jorgensen2013}. This study comprises multiple absorption-line strengths to interpret ages, metallicities and abundance ratios [$\alpha$/Fe] of composite spectra using the models of \citet{Thomas2011}. It is worth noting that the age of their passive galaxies does not present a correlation with velocity dispersion. However, there is a steep relation between velocity dispersion and metallicity, which is constant at any redshift. This fact reflects that the MZR is also observed up to $z\sim1 $, but differs from our results in that the MZR does not depend on redshift (see panels d--f in Fig.~\ref{fig:comparison}). \citet{Jorgensen2017} discuss that their local reference sample includes galaxies that are too young to be the descendant of the passive galaxies at intermediate redshift if only passive evolution is at work. They also reveal intrinsic scatter on the age and metallicity of $0.08$--$0.15$~dex. The scatter in age agrees with our results, but it is lower than ours for metallicity. However, the authors used velocity dispersions instead of stellar mass to build their relations, which would explain these discrepancies \citep{Trager2000,Gallazzi2006,Graves2010,Cappellari2013}.

\subsection{Spectroscopic quiescent galaxies at $z\gtrsim 1$}\label{sec:comp3}

By use of $Keck$ LRIS spectra and photometric data, \citet{Belli2015} retrieved the ages, metallicities, and extinctions of $51$ quiescent galaxies at $1<z<1.6$. In this sample, there are $12$ galaxies at $1.0 < z < 1.12$ with median stellar mass equal to $\log_{10}M_\star \sim 11.1$). The median age, metallicity, and extinction of this subsample is $2.3$~Gyr, $-0.1$~dex, and $0.36$, respectively. At the redshift upper-limit, MUFFIT suggests that massive quiescent galaxies ($10.8\le \log_{10}M_\star < 11.2$) exhibit a median age of $\sim4.2$~Gyr, $\mathrm{[Fe/H]}=-0.05$, $0.0$, and $0.18$~dex (BC03, EMILES+Padova00, and EMILES+BaSTI, respectively; see panels d--f in Fig.~\ref{fig:comparison}), and extinctions of $A_V\sim0.2$. \citet{Belli2015} assumed metallicities around solar values, which would partly explain the discrepancies between ages and extinctions owing to colour degeneracies.

Making use of $171$ quiescent galaxies at $1.4 < z < 2.2$ from the 3D-HST grism survey \citep{Brammer2012}, \citet{Whitaker2013} explore the stellar content of quiescent galaxies via stacked spectra. The authors found that there are prominent absorption features in the G-band and some metal lines, indicative of evolved stellar populations with an average age ranging from $0.9$ to $1.6$~Gyr. Although they present features of evolved populations, the authors also found H$\beta$ in emission as well as [\ion{O}{III}] in the residuals when subtracting the best-fit model. This would indicate a presence of residual star formation, or even the presence of a AGN.

In summary, there is a general consensus regarding the evolution scenario since $z\sim 1$, in good agreement with our results, whereby quiescent galaxies get older with cosmic time independently of their stellar mass. In addition, the ``downsizing'' scenario is well reproduced by several of the studies in this section, as well as the results obtained by MUFFIT using the ALHAMBRA dataset and different sets of SSP models (BC03 and EMILES). Despite the good agreement amongst the studies shown above, there is a large spread of the stellar ages that are strongly related to the use of different techniques, stellar population models, and SFH assumptions. Concerning metallicity, there are fewer measurements, but most results point out that quiescent galaxies have around solar and super-solar metallicities, with hints that the MZR is already in place at earlier epochs. The difficulty to disentangle the evolution with redshift of the metal content in quiescent galaxies is, primarily, caused by the fact that in these galaxies this evolution is expected to be mild ($\sim0.1$~dex), as the bulk of the star formation must have happened at higher redshift.


\section{Summary and conclusions}\label{sec:conclusions}

By means of the catalogue of galaxies published by \citet[][]{DiazGarcia2017a}, we select all the galaxies classified as quiescent via the dust-corrected stellar mass-colour diagram (MCDE). This catalogue is complete in stellar mass and in magnitude down to $I=23$ and it contains $\sim8\,500$ quiescent galaxies with stellar population properties obtained at redshift $0.1 \le z \le 1.1$. This catalogue provides stellar population parameters such as mass- and luminosity-weighted ages and metallicities, stellar masses, extinctions, photo-$z$, rest-frame luminosities, colours corrected for extinction, and uncertainties of the parameters. These parameters were computed through the multi-filter fitting tool for stellar population diagnostics MUFFIT \citep[][]{DiazGarcia2015}, only using the photometric data from the ALHAMBRA survey, and the SSP models of \citet{Bruzual2003} and EMILES (including the two sets of isochrones of BaSTI and Padova00).

We explore the comoving number density of quiescent galaxies at $0.1\le z \le 1.1$ and we find an increasing number of quiescent galaxies from high redshift to the present time. The evolution of the number densities is well reproduced by a power-law function \citep[as proposed by][]{Moustakas2013}. We find that the increase in number density for less massive quiescent galaxies is higher than for the massive case ($\sim52$~\% and $12$~\% fraction between $z=0.4$ and $z=0.2$ for $10 \le \log_{10}M_\star < 10.4$ and $\log_{10}M_\star \ge 11.2$, respectively). The above numbers agree within a ``downsizing'' picture in which less massive galaxies were formed in more recent epochs than the massive ones.

Finally, we studied the evolution of the stellar population parameters of quiescent galaxies since $z=1.1$ based on the results provided by MUFFIT and ALHAMBRA photometry. We construct, for the first time, the probability distribution functions (PDF) of mass-weighted age/formation epoch, metallicity, and extinction during the last $8$~Gyr, using a maximum likelihood estimator in order to deconvolve the uncertainty effects from these distributions and to parametrize them as a function of redshift and stellar mass. This allows us to determine the evolution of the typical parameters (age, metallicity, and extinction), as well as to explore the intrinsic dispersions of the distributions of parameters. As expected, we find that the PDFs of galaxy properties are strongly dependent on the SSP models used during the diagnostic SED-fitting process (BC03 and EMILES), although we still retrieve very interesting and striking results in common for both model sets, namely:

\begin{itemize}

\item[$\bullet$]
Quiescent galaxies are older at larger cosmic times, but these values are not compatible with a simple passive evolution, that is, there must be an incoming of new galaxies from the blue cloud and/or a mechanism including new stars of young and intermediate ages in their stellar populations (e.~g.~mergers, frosting, infall of cold gas, etc.) that slows down their ageing. In addition, the higher the galaxy mass the older the stellar population at any redshift, supporting the ``downsizing'' scenario up to $z\sim 1$. Regarding the widths of the mass-weighted age PDFs (intrinsic dispersions), these present values of $\omega_\mathrm{Age_M}\sim 1$--$2$~Gyr and a slight correlation with the stellar mass and redshift.
\item[$\bullet$]
For BC03 and EMILES SSP models, quiescent galaxies show predominantly solar and super-solar metallicities, except for the low redshift and less massive quiescent galaxies, which reveal sub-solar metallicities on average. Furthermore, the galaxy mass-metallicity relation seems to be present since earlier times, with hints of being steeper at lower redshift. 
\item[$\bullet$]
We find evidence of a decrease of the median of the metallicity PDF of quiescent galaxies since $z\sim 0.6$--$1.1$, depending on the SSP models (BC03 and EMILES, respectively). This decrement amounts to $0.1$--$0.2$ dex and it is consistently recovered irrespective of the SSP models and isochrones employed during the analysis. At decreasing stellar masses the range of mass-weighted metallicities increases, and therefore, there is a dependence of the width of this PDF with stellar mass. The width of the mass-weighted metallicity PDF also depends on redshift, being broader at higher redshift.
\item[$\bullet$]
All the quiescent galaxies present dust extinction values below $A_V < 0.6$, with median values in the range $A_V=0.15$--$0.3$. For BC03 SSP models, there is a dependence with stellar mass and redshift, in the sense that more massive galaxies also present larger extinctions by dust and the extinction increases at lower redshift. However for EMILES SSP models, all the quiescent galaxies exhibit extinction values of $A_V\sim 0.2$ independently of the stellar mass or redshift bins explored. As in the metallicity case, the width of the extinction PDF becomes larger at higher redshift.

\end{itemize}

The consistency of these results is studied imposing different constraints on the SFH of the models as well, such as constant extinction, constant solar metallicity, a closed-box enrichment of metals, infall of metal-poor cold gas, and more. Furthermore, we determine the systematic effects and the modifications of the stellar population predictions as a result of these constraints. In some cases, these ones can alter well studied relations as the MZR or the ``downsizing'' scenario. Moreover, the use of different SSP model sets and/or SFH assumptions introduce systematics that can be larger than the uncertainties involved in the determination of the stellar population properties, but we find these potential systematics do not alter the conclusions of this work. Our results are compared with those obtained from previous studies, including spectroscopic data and different analysis techniques (e.~g.~Lick indices), providing a good agreement with many of the conclusions and aspects treated in this work. Consequently, SED-fitting techniques involving multi-filter photometric surveys offer a great opportunity to explore the stellar population properties of galaxies. This is particularly interesting for metallicity, whose predictions can be used to constrain or complement spectroscopic measurements.

In the light of our results, we find essential and observationally justified the inclusion of scenarios such as mergers and/or ``frosting'' (remnants of star formation) to reconcile the observed trends, especially at larger masses where the number density of quiescent galaxies does not change remarkably. We propose that the quiescent population must undergo an evolutive pathway including such scenarios, where each one can play a role that depends on stellar mass and redshift. Indeed, we believe that it is hardly possible that a unique scenario is able to alter all the observables studied in this work (number density, median and width of the PDFs), but rather a simultaneous contribution of them.


\begin{acknowledgements}

The authors are grateful to the anonymous referee for his/her fruitful comments, which contributed to improve the present research. This work has been partly supported by the “Programa Nacional de Astronom\'ia y Astrof\'isica” of the Spanish Ministry of Economy and Competitiveness (MINECO, grants AYA2012-30789 and AYA2015-66211-C2-1-P), by the Ministry of Science and Technology of Taiwan (grant MOST 106-2628-M-001-003-MY3), Academia Sinica (grant AS-IA-107-M01) and the Government of Arag\'on (Research Group E103). L.~A.~D.~G.~also thanks the support of I.~F. for offering the opportunity to develop part of this research at the Mullard Space Science Laboratory (MSSL). We also acknowledge support from the Spanish Ministry for Economy and Competitiveness and FEDER funds through grants AYA2010-15081, AYA2010-15169, AYA2010-22111-C03-01, AYA2010-22111-C03-02, AYA2011-29517-C03-01, AYA2012-39620, AYA2013-40611-P, AYA2013-42227-P, AYA2013-43188-P, AYA2013-48623-C2-1, AYA2013-48623-C2-2, ESP2013-48274, AYA2014-57490-P, AYA2014-58861-C3-1, AYA2016-76682-C3-1-P, AYA2016-77846-P, AYA2016-81065-C2-1,AYA2016-81065-C2-2, Generalitat Valenciana projects Prometeo 2009/064 and PROMETEOII/2014/060, Junta de Andaluc\'{\i}a grants TIC114, JA2828, P10-FQM-6444, and Generalitat de Catalunya project SGR-1398. Authors acknowledges Y.~Peng and A.~Citro for sharing their stellar population numerical results. Throughout this research, we made use of the \texttt{Matplotlib} package \citep{Hunter2007}, a 2D graphics package used for \texttt{Python} which is designed for interactive scripting and quality image generation. This paper is dedicated to M.~A.~L.~C., for being there when I needed her most and for her patience and continuous encouragement until finishing my Ph.~D.

\end{acknowledgements}


\bibliographystyle{aa}
\bibliography{alh_sp_editor} 

%
%
\begin{appendix}
%
%

\section{Probability distribution functions of age, metallicity, and extinction: the MLE method}\label{sec:appendix_pdf}

Maximum likelihood estimator methods (MLE) have been successfully used for different purposes in Astronomy \citep[e.~g.][]{Naylor2006,Makarov2006,Arzner2007,LopezSanjuan2008,LopezSanjuan2015}. In particular, we adapt the MLE methodology developed by \citet[][]{LopezSanjuan2014} to deconvolve uncertainty effects on the observed distributions of age, metallicity, and extinction.

For the present MLE, we find that the observed distributions of stellar-population parameters of quiescent galaxies (age, metallicity, and extinction) are properly described by Gaussian-like probability distributions in the log-space, meaning log-normal distributions in the real space. However, these distributions include observational errors, owing to Gaussian uncertainties in the stellar-population properties of each individual galaxy. The intrinsic distributions or probability distribution functions (PDF) are therefore described by two parameters: the mean ($\mu$) and the intrinsic standard deviation ($\sigma^\mathrm{int}$). As our main goal is to constrain how quiescent galaxies evolve since $z \sim 1$, we adopt that $\mu$ and $\sigma^\mathrm{int}$ are redshift dependent: $\mu (z) = \mu_2 \times z^2 + \mu_1 \times z + \mu_0$ and $\sigma^\mathrm{int} (z) = \sigma_2 \times z^2 + \sigma_1 \times z + \sigma_0$. Thereby, we search the set of $\mu_2$, $\mu_1$, $\mu_0$, $\sigma_2$, $\sigma_1$, and $\sigma_0$ values that maximize the likelihood for this work:
\begin{equation}
\mathcal{L} = - \frac{1}{2} \sum\limits_{j} \left[ \ln \left( {{p_{\mathrm{e},j}}^2+\sigma^\mathrm{int}_p(z_j)}^2 \right) + \frac{{\left( \mu_p(z_j) - p_j \right)}^2}{{{p_{\mathrm{e},j}}^2 + \sigma^\mathrm{int}_p(z_j)}^2} \right]\ ,
\label{eq:likelihood_MLE}
\end{equation}
where $p_j$ is the stellar-population property (age, metallicity, or extinction) of the $j$th galaxy in the sample, $p_{\mathrm{e},j}$ its uncertainty, $z_j$ its photo~\textit{z}. We note that the dependence on redshift of $\mu (z)$ and $\sigma^\mathrm{int}_p(z)$ favours the determination of these parameters at $0.1\le z < 0.3$ and $\log_{10} M_\star \ge 11.2$~dex, where our sample is more limited in number. In a general case, this assumption may introduce correlations between the stellar population parameters and photo $z$. In the light of the results of \citet{DiazGarcia2015}, the photo-$z$ uncertainties of ALHAMBRA lead negligible effects on the stellar population parameters obtained via SED-fitting. i.~e.~no correlations with redshift. Furthermore, the redshift range for this work ($0.1\le z \le 1.1$) is much larger than the photo-$z$ uncertainties, i.~e.~$\sigma_z << z_\mathrm{max}-z_\mathrm{min}$.

The process of maximization of Eq.~(\ref{eq:likelihood_MLE}) was carried out by \textsc{emcee}\footnote{http://dan.iel.fm/emcee} \citep[][]{ForemanMackey2013}, namely a \texttt{Python} implementation of an affine invariant sampling algorithm for a Markov chain Monte Carlo method (MCMC), which also provides uncertainties and correlations for the parameters that maximize such equation. Consequently, the MLE for this research is equivalent to deconvolve uncertainty effects on the observed stellar-population distributions. 

\begin{figure}
\centering
\resizebox{\hsize}{!}{\includegraphics[trim=0 15 0 3mm,clip=True]{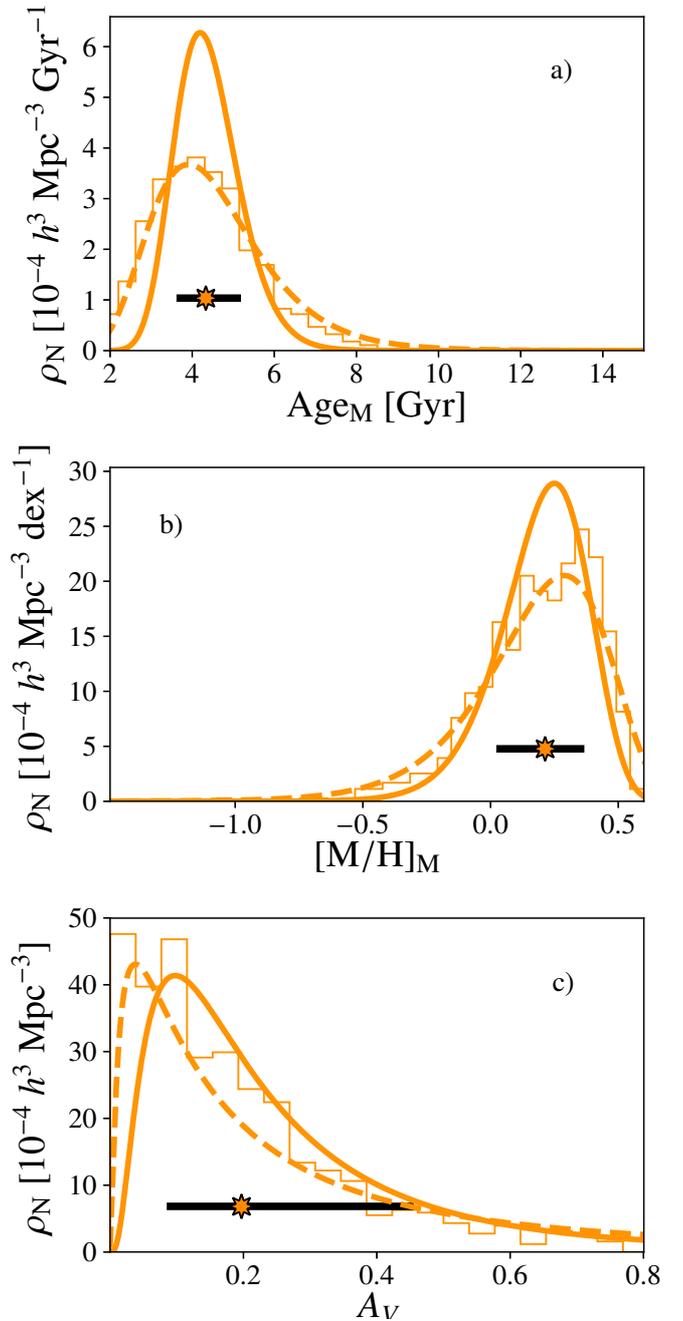}}
\caption{Histograms of stellar-population parameters (thin line, derived using BC03 SSP models) of the $1480$ quiescent galaxies with stellar mass $10.8 \le \log_{10}M_\star < 11.2$~dex and at redshift $0.5 \le z < 0.7$. \textit{From top to bottom}, mass-weighted age, mass-weighted metallicity, and extinction (\textit{panels a}, \textit{b}, and \textit{c}, respectively). The dashed line is the distribution fit, whereas the orange solid line is the parameter distributions after applying the MLE method to deconvolve uncertainty effects (further details in the text). The star-shape marker and the solid black line illustrate the median and the $1$~$\sigma$ width of the distributions, respectively. All the curves were normalised to the fitted number density (see Sect.~\ref{sec:number} and Table~\ref{tab:density_par}).}
\label{fig:MLE}
\end{figure}

Finally, we normalise the log-normal distributions (in the real space) of all stellar-population parameters to the fitted number densities obtained in Sect.~\ref{sec:number} (see Table~\ref{tab:density_par}). This allows us to provide PDFs of age, metallicity, and extinction for quiescent galaxies up to $z \sim 1.1$. Due to the nature of the parameters, the analytical form of the PDF of age, metallicity, and extinction is formally expressed as
\begin{align}
\mathrm{PDF}(\mathrm{Age}&,z,M_\star) = \frac{\rho_\mathrm{N}(z,M_\star)}{\sqrt{2 \pi}\ \mathrm{Age}\ \sigma^\mathrm{int}_\mathrm{Age}(z,M_\star)} \times  \nonumber \\
& \times \exp \left[-\frac{{\left(\ln\mathrm{Age}-\mu_\mathrm{Age}(z,M_\star)\right)}^2}{2{\sigma^\mathrm{int}_\mathrm{Age}(z,M_\star)}^2} \right] \ ,
\label{eq:pdf_age}
\end{align}
\begin{align}
\mathrm{PDF}([\mathrm{M/H}&],z,M_\star) =  \frac{\rho_\mathrm{N}(z,M_\star)}{\sqrt{2 \pi}\ (1-\mathrm{[M/H]})\ \sigma^\mathrm{int}_\mathrm{[M/H]}(z,M_\star)} \times \nonumber \\
& \times \exp \left[-\frac{{\left(\ln\{1-\mathrm{[M/H]}\}-\mu_\mathrm{[M/H]}(z,M_\star)\right)}^2}{2{\sigma^\mathrm{int}_\mathrm{[M/H]}(z,M_\star)}^2} \right] \ ,
\label{eq:pdf_feh}
\end{align}
\begin{align}
\mathrm{PDF}(A_V&,z,M_\star) = \frac{\rho_\mathrm{N}(z,M_\star)}{\sqrt{2 \pi}\ A_V\ \sigma^\mathrm{int}_{A_V}(z,M_\star)} \times  \nonumber \\
& \times \exp \left[-\frac{{\left(\ln A_V-\mu_{A_V}(z,M_\star)\right)}^2}{2{\sigma^\mathrm{int}_{A_V}(z,M_\star)}^2} \right] \ .
\label{eq:pdf_av}
\end{align}
From these definitions, the medians and widths ($\omega$, defined as the difference between the $84^\mathrm{th}$ and $16^\mathrm{th}$ percentiles) of age, metallicity, and extinction are expressed as:
\begin{equation}
\mathrm{Age^{50th}}(z,M_\star) = \exp \left\{ \mu_\mathrm{Age}(z,M_\star) \right\}\ ,
\label{eq:median_age}
\end{equation}
\begin{equation}
A_V^\mathrm{50th}(z,M_\star) = \exp \left\{ \mu_{A_V}(z,M_\star) \right\} \ ,
\label{eq:median_av}
\end{equation}
\begin{equation}
\mathrm{[M/H]^{50th}}(z,M_\star) = 1 - \exp \left\{ \mu_\mathrm{[M/H]}(z,M_\star) \right\} \ ,
\label{eq:median_feh}
\end{equation}
\begin{align}
\omega_p(z,M_\star) =& \exp \left\{ \mu_p(z,M_\star) + \sigma^\mathrm{int}_p(z,M_\star)\right\} - \nonumber \\
& -\ \exp \left\{ \mu_p(z,M_\star) - \sigma^\mathrm{int}_p(z,M_\star)\right\} \ ,
\label{eq:width}
\end{align}
where the superscript 50th denotes median and $p$ refers to age, metallicity, and extinction.

In Fig.~\ref{fig:MLE}, we present an illustrative case of our MLE methodology to deconvolve uncertainty effects and build the PDFs of age, metallicty, and extinction. We performed our MLE methodology over all the quiescent galaxies in our sample at $0.7 \le z < 0.9$ and $10.8 \le \log_{10}M_\star < 11.2$~dex (BC03 SSP models), amounting to a total of $1480$ galaxies. Only for this illustrative case, we assume that $\mu(z)$ and $\sigma^\mathrm{int}(z)$ are redshift independent (i.~e.~$\mu_2 = \mu_1 = \sigma_2 = \sigma_1 = 0$). The histograms of the observed stellar-population parameters (see Fig.~\ref{fig:MLE}) are fitted to a log-normal distribution correctly (i.~e.~before deconvolving observational errors, see dashed line in Fig.~\ref{fig:MLE}), supporting our initial assumption. After deconvolving uncertainty effects, we find that the PDFs of age, metallicity, and extinction (solid lines in panels a, b, and c of Fig.~\ref{fig:MLE}, respectively) are narrower than the observed distributions, whereas the median is poorly affected.

As we explore potential systematics on stellar population properties due to the use of different population synthesis models, we perform the MLE deconvolution of uncertainty effects on the distributions of age, metallicity, and extinction for BC03 and EMILES (including both BaSTI and Padova00 isochrones) SSP models. In addition, we provide uncertainties for $\mu(z)$ and $\sigma^\mathrm{int}(z)$, making use of the \textsc{emcee} Markov chains. The formation epoch, age, metallicity and extinction PDFs for the ALHAMBRA quiescent galaxies are presented in Appendix~\ref{appendix:pdf_bc03}, \ref{appendix:pdf_basti}, and \ref{appendix:pdf_padova} for BC03, EMILES+BaSTI isochrones, and EMILES+Padova00 isochrones SSP models, respectively. For the analysis of these distributions, we refer readers to Sect.~\ref{sec:results}.


\subsection{Stellar-population PDFs of the ALHAMBRA quiescent galaxies: BC03 SSP models}\label{appendix:pdf_bc03}

For BC03 SSP models, we adopted a linear dependence of $\mu(z)$ and $\sigma^\mathrm{int}(z)$ for ages and extinctions (i.~e.~$\mu_2 = \sigma_2 = 0$) and a quadratic one for metallicity. We note that we provide the PDFs of stellar population properties at different stellar mass ranges, because these properties may depend on stellar mass in a general case. Only for the mass bin $10.0 \le \log_{10}M_\star < 10.4$ and metallicity, we adopt $\sigma_2 = 0$ because $\sigma^\mathrm{int}(z)$ is properly reproduced by a linear function and this reduces degeneracy effects during the maximization of Eq.~(\ref{eq:likelihood_MLE}). In addition, we compare the $\mu(z)$ and $\sigma^\mathrm{int}(z)$ curves (obtained after carrying out the MLE method) with the values obtained from the same analysis in redshift bins of $\Delta z = 0.2$ and setting $\mu_2 = \mu_1 = \sigma_2 = \sigma_1 = 0$ (i.~e.~the average value of $\mu(z)$ and $\sigma^\mathrm{int}(z)$ at that redshift bin). As a result, we obtain that the linear and quadratic assumptions are fairly suitable in each case. All the parameters and uncertainties obtained after the maximization of Eq.~(\ref{eq:likelihood_MLE}) that are necessary to build the stellar population PDFs are provided in Tables~\ref{tab:par_age}--\ref{tab:par_av} (see also Eqs.~(\ref{eq:pdf_age})--(\ref{eq:pdf_av})). Notice that the parameters to compute the age and formation epoch PDFs in the log-space are also provided in Tables~\ref{tab:par_logage} and \ref{tab:par_logagetl}. Comoving number densities for the PDF normalization can be found in Table~\ref{tab:density_par} (details in Sect.~\ref{sec:number}). It is worth mentioning that at $z\gtrsim1.1$, the reliability of the parameters may be questionable. Indeed, we cannot confirm whether $\mu(z)$ and $\sigma^\mathrm{int}(z)$ feature linear or quadratic functional forms with redshift.

Owing to the low number of galaxies at $9.6 \le \log_{10}M_\star < 10.0$ and $0.1 \le z < 0.3$, the MLE deconvolution of uncertainty effects is highly uncertain and degenerated. Only at this mass range, we assume that the median and intrinsic dispersion of the PDFs is constant across cosmic time ($\mu_2 = \mu_1 = \sigma_2 = \sigma_1 = 0$; see Figs.~\ref{fig:median_width_age}--\ref{fig:median_width_av}). Regarding extinction at $0.1 \le z < 0.2$ and $\log_{10}M_\star \ge 11.2$, we find that the assumption of linearity for $\sigma^\mathrm{int}(z)$ is not physical ($\sigma^\mathrm{int} \sim 0.0$ at $z\sim0.1$), but only mathematically motivated. Consequently, only for quiescent galaxies at $0.1 \le z < 0.2$ and $\log_{10}M_\star \ge 11.2$, we adopt a constant value of $\sigma^\mathrm{int}(0.1 \le z < 0.2)=\sigma^\mathrm{int}(z=0.2) = 0.14 \pm 0.03$ and $\mu(0.1 \le z < 0.2) = \mu(z=0.2) = -1.30 \pm 0.07$. We note that for the lowest extinction values in Eq.~(\ref{eq:pdf_av}) ($A_V \sim 0$), the probability might be underestimated because the log-normal function falls to zero at this regime of values.


\subsection[Stellar-population PDFs of the ALHAMBRA quiescent galaxies: BaSTI isochrones]{Stellar-population PDFs of the ALHAMBRA quiescent galaxies: EMILES and BaSTI isochrones}\label{appendix:pdf_basti}
For EMILES SSP models and BaSTI isochrones, we find qualitatively similar trends to the obtained in Appendix~\ref{appendix:pdf_bc03}. However as expected, there is a quantitative systematic caused by the differing model prescriptions between both model sets. For age, formation epoch and extinction distributions, a linear dependence with redshift is adopted for $\mu(z)$ and $\sigma^\mathrm{int}(z)$ instead, i.~e.~$\mu_2=\sigma_2=0$. As in Appendix~\ref{appendix:pdf_bc03}, the assumption of linearity for $\mu(z)$ and $\sigma^\mathrm{int}(z)$ is supported after comparing with the MLE solutions when values of $\mu_2=\mu_1=\sigma_2=\sigma_1=0$ are adopted at redshift bins of $\Delta z = 0.2$ (see Fig.~\ref{fig:median_width_age}). Nevertheless, the assumption of a quadratic form for $\mu(z)$ and $\sigma^\mathrm{int}(z)$ is not needed for metallicity and we assume a linear dependence with redshift for these parameters. The parameters and uncertainties obtained after deconvolving the stellar population distributions provided by MUFFIT using EMILES+BaSTI isochrones to build the stellar population PDFs (Eqs.~(\ref{eq:pdf_age})--(\ref{eq:pdf_av})) of quiescent galaxies are provided in Tables~\ref{tab:par_age}--\ref{tab:par_av}.

As in the Sect.~\ref{appendix:pdf_bc03}, for quiescent galaxies at $9.7 \le \log_{10}M_\star < 10.1$ we perform the MLE deconvolution of uncertainty effects adopting a constant value for $\mu(z)$ and $\sigma^\mathrm{int}(z)$ at $0.1 \le z \le 0.3$.


\subsection[Stellar-population PDFs of the ALHAMBRA quiescent galaxies: Padova00 isochrones]{Stellar-population PDFs of the ALHAMBRA quiescent galaxies: EMILES and Padova00 isochrones}\label{appendix:pdf_padova}
For Padova00 isochrones, we only appreciate mild quantitative discrepancies with respect the BaSTI ones. As in Appendix~\ref{appendix:pdf_basti}, we assumed a linear dependence for $\mu(z)$ and $\sigma^\mathrm{int}(z)$ for all the stellar-population parameters during the MLE process. The values to compute $\mu(z)$ and $\sigma^\mathrm{int}(z)$ along with their uncertainties, meaning the PDFs of age, metallicity, and extinction of quiescent galaxies, are shown in Tables~\ref{tab:par_age}--\ref{tab:par_av}.

For those quiescent galaxies with stellar mass at $9.7 \le \log_{10}M_\star < 10.1$ (restricted by completeness reasons at $0.1 \le z \le 0.3$), we provide reference values for its stellar population PDFs assuming $\mu_2=\mu_1=\sigma_2=\sigma_1=0$ (for further details, see Sect.~\ref{sec:results} and Figs.~\ref{fig:median_width_age}--\ref{fig:median_width_av}).


\newpage


\begin{table*}
\caption{Parameters $\mu(z,M_\star)$ and $\sigma^\mathrm{int}(z,M_\star)$ of the probability distribution functions of luminosity and mass-weighted ages in the real space (Gyr units, see Eq.~(\ref{eq:pdf_age})), which were derived using BC03, EMILES+BaSTI, and EMILES+Padova00 SSP models, respectively.}
\label{tab:par_age}
\centering
\renewcommand{\arraystretch}{1.42}
\begin{tabular}{crccccc}
\hline\hline
\multicolumn{2}{c}{\multirow{2}{*}{$\mathrm{Age_L}\ \mathrm{[Gyr]}$}} & \multirow{2}{*}{$N_\mathrm{gal}$} & \multicolumn{2}{c}{$\mu_{\mathrm{Age_L}}$} & \multicolumn{2}{c}{$\sigma^\mathrm{int}_{\mathrm{Age_L}}$} \\ 
\cline{4-5} 
\cline{6-7} 
 & & & $\mu_1$ & $\mu_0$ & $\sigma_1$ & $\sigma_0$ \\ 
\hline
\parbox[t]{2mm}{\multirow{6}{*}{\rotatebox[origin=c]{90}{BC03}}} &&&&&& \\
&$10.0 \le \log_{10} M_\star < 10.4$ & $1429$ & $-0.64^{+0.09}_{-0.09}$ & $1.57^{+0.03}_{-0.03}$ & $0.33^{+0.10}_{-0.09}$ & $0.08^{+0.03}_{-0.03}$\\ 
&$10.4 \le \log_{10} M_\star < 10.8$ & $2526$ & $-0.64^{+0.04}_{-0.04}$ & $1.59^{+0.02}_{-0.02}$ & $0.32^{+0.05}_{-0.04}$ & $0.05^{+0.02}_{-0.02}$\\ 
&$10.8 \le \log_{10} M_\star < 11.2$ & $3181$ & $-0.37^{+0.03}_{-0.03}$ & $1.54^{+0.02}_{-0.02}$ & $0.17^{+0.03}_{-0.03}$ & $0.11^{+0.02}_{-0.02}$\\ 
&$\log_{10} M_\star \ge 11.2$ & $1122$ & $-0.34^{+0.04}_{-0.05}$ & $1.62^{+0.04}_{-0.04}$ & $0.02^{+0.04}_{-0.04}$ & $0.20^{+0.03}_{-0.03}$\\ 
&&&&&& \\
\hline
\parbox[t]{2mm}{\multirow{6}{*}{\rotatebox[origin=c]{90}{EMILES+BaSTI}}} &&&&&& \\
&$10.1 \le \log_{10} M_\star < 10.5$ & $1284$ & $-0.72^{+0.09}_{-0.08}$ & $1.96^{+0.03}_{-0.03}$ & $-0.04^{+0.08}_{-0.08}$ & $0.19^{+0.03}_{-0.03}$\\ 
&$10.5 \le \log_{10} M_\star < 10.9$ & $2370$ & $-0.82^{+0.04}_{-0.04}$ & $2.04^{+0.02}_{-0.02}$ & $-0.02^{+0.03}_{-0.03}$ & $0.16^{+0.02}_{-0.02}$\\ 
&$10.9 \le \log_{10} M_\star < 11.3$ & $3089$ & $-0.87^{+0.02}_{-0.02}$ & $2.10^{+0.02}_{-0.02}$ & $-0.01^{+0.02}_{-0.02}$ & $0.15^{+0.01}_{-0.01}$\\ 
&$\log_{10} M_\star \ge 11.3$ & $1159$ & $-0.86^{+0.03}_{-0.03}$ & $2.19^{+0.03}_{-0.03}$ & $-0.01^{+0.02}_{-0.02}$ & $0.15^{+0.02}_{-0.02}$\\
&&&&&& \\
\hline
\parbox[t]{2mm}{\multirow{6}{*}{\rotatebox[origin=c]{90}{EMILES+Padova00}}} &&&&&& \\
&$10.1 \le \log_{10} M_\star < 10.5$ & $1329$ & $-0.71^{+0.10}_{-0.10}$ & $1.86^{+0.03}_{-0.04}$ & $-0.10^{+0.08}_{-0.08}$ & $0.28^{+0.03}_{-0.03}$\\ 
&$10.5 \le \log_{10} M_\star < 10.9$ & $2335$ & $-0.61^{+0.04}_{-0.04}$ & $1.84^{+0.02}_{-0.02}$ & $-0.01^{+0.03}_{-0.03}$ & $0.21^{+0.02}_{-0.02}$\\ 
&$10.9 \le \log_{10} M_\star < 11.3$ & $2945$ & $-0.72^{+0.03}_{-0.02}$ & $1.95^{+0.02}_{-0.02}$ & $-0.00^{+0.02}_{-0.02}$ & $0.17^{+0.01}_{-0.01}$\\ 
&$\log_{10} M_\star \ge 11.3$ & $1141$ & $-0.74^{+0.03}_{-0.03}$ & $2.07^{+0.02}_{-0.02}$ & $0.01^{+0.02}_{-0.02}$ & $0.15^{+0.02}_{-0.02}$\\
&&&&&& \\
\hline
\end{tabular}
\begin{tabular}{crccccc}
\multicolumn{2}{c}{\multirow{2}{*}{$\mathrm{Age_M}\ \mathrm{[Gyr]}$}} & \multirow{2}{*}{$N_\mathrm{gal}$} & \multicolumn{2}{c}{$\mu_{\mathrm{Age_M}}$} & \multicolumn{2}{c}{$\sigma^\mathrm{int}_{\mathrm{Age_M}}$} \\ 
\cline{4-5} 
\cline{6-7} 
 & & & $\mu_1$ & $\mu_0$ & $\sigma_1$ & $\sigma_0$ \\ 
\hline
\parbox[t]{2mm}{\multirow{6}{*}{\rotatebox[origin=c]{90}{BC03}}} &&&&&& \\
&$10.0 \le \log_{10} M_\star < 10.4$ & $1429$ & $-0.31^{+0.10}_{-0.09}$ & $1.70^{+0.03}_{-0.03}$ & $0.27^{+0.11}_{-0.10}$ & $0.06^{+0.04}_{-0.04}$\\ 
&$10.4 \le \log_{10} M_\star < 10.8$ & $2526$ & $-0.47^{+0.04}_{-0.05}$ & $1.77^{+0.02}_{-0.02}$ & $0.17^{+0.04}_{-0.04}$ & $0.10^{+0.02}_{-0.02}$\\ 
&$10.8 \le \log_{10} M_\star < 11.2$ & $3181$ & $-0.33^{+0.03}_{-0.03}$ & $1.78^{+0.02}_{-0.02}$ & $0.09^{+0.03}_{-0.03}$ & $0.13^{+0.02}_{-0.02}$\\ 
&$\log_{10} M_\star \ge 11.2$ & $1122$ & $-0.38^{+0.04}_{-0.04}$ & $1.92^{+0.04}_{-0.04}$ & $-0.04^{+0.04}_{-0.03}$ & $0.21^{+0.03}_{-0.03}$\\ 
&&&&&& \\
\hline
\parbox[t]{2mm}{\multirow{6}{*}{\rotatebox[origin=c]{90}{EMILES+BaSTI}}} &&&&&& \\
&$10.1 \le \log_{10} M_\star < 10.5$ & $1284$ & $-0.37^{+0.08}_{-0.08}$ & $2.08^{+0.03}_{-0.03}$ & $-0.00^{+0.06}_{-0.07}$ & $0.12^{+0.02}_{-0.02}$\\ 
&$10.5 \le \log_{10} M_\star < 10.9$ & $2370$ & $-0.63^{+0.04}_{-0.04}$ & $2.20^{+0.02}_{-0.02}$ & $-0.03^{+0.03}_{-0.03}$ & $0.11^{+0.02}_{-0.02}$\\ 
&$10.9 \le \log_{10} M_\star < 11.3$ & $3089$ & $-0.73^{+0.02}_{-0.02}$ & $2.28^{+0.01}_{-0.01}$ & $0.01^{+0.02}_{-0.02}$ & $0.08^{+0.01}_{-0.01}$\\ 
&$\log_{10} M_\star \ge 11.3$ & $1159$ & $-0.75^{+0.03}_{-0.03}$ & $2.35^{+0.02}_{-0.02}$ & $0.01^{+0.02}_{-0.02}$ & $0.08^{+0.02}_{-0.02}$\\
&&&&&& \\
\hline
\parbox[t]{2mm}{\multirow{6}{*}{\rotatebox[origin=c]{90}{EMILES+Padova00}}} &&&&&& \\
&$10.1 \le \log_{10} M_\star < 10.5$ & $1329$ & $-0.38^{+0.08}_{-0.08}$ & $2.02^{+0.03}_{-0.03}$ & $-0.18^{+0.07}_{-0.07}$ & $0.21^{+0.03}_{-0.02}$\\ 
&$10.5 \le \log_{10} M_\star < 10.9$ & $2335$ & $-0.42^{+0.04}_{-0.04}$ & $2.03^{+0.02}_{-0.02}$ & $-0.03^{+0.03}_{-0.03}$ & $0.13^{+0.01}_{-0.01}$\\ 
&$10.9 \le \log_{10} M_\star < 11.3$ & $2945$ & $-0.56^{+0.02}_{-0.02}$ & $2.13^{+0.01}_{-0.02}$ & $-0.04^{+0.02}_{-0.02}$ & $0.13^{+0.01}_{-0.01}$\\ 
&$\log_{10} M_\star \ge 11.3$ & $1141$ & $-0.57^{+0.03}_{-0.03}$ & $2.20^{+0.02}_{-0.02}$ & $-0.03^{+0.02}_{-0.02}$ & $0.12^{+0.01}_{-0.01}$\\
&&&&&& \\
\hline
\end{tabular}
\tablefoot{The top and bottom numbers stablish the $1~\sigma$ confidence level of the parameters.}
\end{table*}


\begin{table*}
\caption{As Table~\ref{tab:par_age}, but for the probability distribution functions of luminosity and mass-weighted in the log-space (age in yr units).}
\label{tab:par_logage}
\centering
\renewcommand{\arraystretch}{1.42}
\begin{tabular}{crccccc}
\hline\hline
\multicolumn{2}{c}{\multirow{2}{*}{$\mathrm{\log_{10}Age_L}\ \mathrm{[yr]}$}} & \multirow{2}{*}{$N_\mathrm{gal}$} & \multicolumn{2}{c}{$\mu_{\mathrm{\log_{10}Age_L}}$} & \multicolumn{2}{c}{$\sigma^\mathrm{int}_{\mathrm{\log_{10}Age_L}}$} \\ 
\cline{4-5} 
\cline{6-7} 
 & & & $\mu_1~[10^{-2}]$ & $\mu_0~[10^{-2}]$ & $\sigma_1~[10^{-2}]$ & $\sigma_0~[10^{-2}]$ \\ 
\hline
\parbox[t]{2mm}{\multirow{6}{*}{\rotatebox[origin=c]{90}{BC03}}} &&&&&& \\
&$10.0 \le \log_{10} M_\star < 10.4$ & $1429$ & $-2.87^{+0.41}_{-0.42}$ & $227.03^{+0.14}_{-0.14}$ & $1.57^{+0.45}_{-0.43}$ & $0.31^{+0.15}_{-0.16}$\\ 
&$10.4 \le \log_{10} M_\star < 10.8$ & $2526$ & $-2.89^{+0.20}_{-0.20}$ & $227.16^{+0.09}_{-0.09}$ & $1.47^{+0.24}_{-0.22}$ & $0.23^{+0.11}_{-0.12}$\\ 
&$10.8 \le \log_{10} M_\star < 11.2$ & $3181$ & $-1.68^{+0.14}_{-0.15}$ & $226.92^{+0.10}_{-0.10}$ & $0.81^{+0.14}_{-0.14}$ & $0.48^{+0.09}_{-0.09}$\\ 
&$\log_{10} M_\star \ge 11.2$ & $1122$ & $-1.55^{+0.21}_{-0.21}$ & $227.29^{+0.17}_{-0.17}$ & $0.09^{+0.18}_{-0.19}$ & $0.90^{+0.16}_{-0.15}$\\ 
&&&&&& \\
\hline
\parbox[t]{2mm}{\multirow{6}{*}{\rotatebox[origin=c]{90}{EMILES+BaSTI}}} &&&&&& \\
&$10.1 \le \log_{10} M_\star < 10.5$ & $1284$ & $-3.13^{+0.39}_{-0.40}$ & $228.80^{+0.14}_{-0.14}$ & $-0.10^{+0.34}_{-0.34}$ & $0.83^{+0.12}_{-0.12}$\\ 
&$10.5 \le \log_{10} M_\star < 10.9$ & $2370$ & $-3.67^{+0.18}_{-0.18}$ & $229.15^{+0.09}_{-0.09}$ & $-0.07^{+0.15}_{-0.15}$ & $0.69^{+0.08}_{-0.07}$\\ 
&$10.9 \le \log_{10} M_\star < 11.3$ & $3089$ & $-3.92^{+0.11}_{-0.11}$ & $229.45^{+0.07}_{-0.07}$ & $-0.01^{+0.09}_{-0.09}$ & $0.63^{+0.06}_{-0.06}$\\ 
&$\log_{10} M_\star \ge 11.3$ & $1159$ & $-3.86^{+0.14}_{-0.15}$ & $229.83^{+0.12}_{-0.11}$ & $-0.03^{+0.11}_{-0.10}$ & $0.63^{+0.08}_{-0.09}$\\ 
&&&&&& \\
\hline
\parbox[t]{2mm}{\multirow{6}{*}{\rotatebox[origin=c]{90}{EMILES+Padova00}}} &&&&&& \\
&$10.1 \le \log_{10} M_\star < 10.5$ & $1329$ & $-3.15^{+0.42}_{-0.43}$ & $228.34^{+0.15}_{-0.15}$ & $-0.44^{+0.34}_{-0.35}$ & $1.26^{+0.12}_{-0.12}$\\ 
&$10.5 \le \log_{10} M_\star < 10.9$ & $2335$ & $-2.71^{+0.20}_{-0.19}$ & $228.26^{+0.09}_{-0.10}$ & $-0.01^{+0.15}_{-0.15}$ & $0.91^{+0.08}_{-0.08}$\\ 
&$10.9 \le \log_{10} M_\star < 11.3$ & $2945$ & $-3.21^{+0.11}_{-0.11}$ & $228.76^{+0.07}_{-0.07}$ & $0.01^{+0.08}_{-0.09}$ & $0.73^{+0.06}_{-0.06}$\\ 
&$\log_{10} M_\star \ge 11.3$ & $1141$ & $-3.30^{+0.13}_{-0.14}$ & $229.30^{+0.10}_{-0.11}$ & $0.05^{+0.09}_{-0.09}$ & $0.64^{+0.07}_{-0.07}$\\ 
&&&&&& \\
\hline
\end{tabular}
\begin{tabular}{crccccc}
\multicolumn{2}{c}{\multirow{2}{*}{$\log_{10}\mathrm{Age_M}\ \mathrm{[yr]}$}} & \multirow{2}{*}{$N_\mathrm{gal}$} & \multicolumn{2}{c}{$\mu_{\log_{10}\mathrm{Age_M}}$} & \multicolumn{2}{c}{$\sigma^\mathrm{int}_{\log_{10}\mathrm{Age_M}}$} \\ 
\cline{4-5} 
\cline{6-7} 
 & & & $\mu_1~[10^{-2}]$ & $\mu_0~[10^{-2}]$ & $\sigma_1~[10^{-2}]$ & $\sigma_0~[10^{-2}]$ \\ 
\hline
\parbox[t]{2mm}{\multirow{6}{*}{\rotatebox[origin=c]{90}{BC03}}} &&&&&& \\
&$10.0 \le \log_{10} M_\star < 10.4$ & $1429$ & $-1.26^{+0.43}_{-0.45}$ & $227.60^{+0.15}_{-0.15}$ & $1.21^{+0.47}_{-0.45}$ & $0.29^{+0.16}_{-0.16}$\\ 
&$10.4 \le \log_{10} M_\star < 10.8$ & $2526$ & $-2.02^{+0.21}_{-0.21}$ & $227.92^{+0.10}_{-0.10}$ & $0.73^{+0.21}_{-0.20}$ & $0.46^{+0.10}_{-0.10}$\\ 
&$10.8 \le \log_{10} M_\star < 11.2$ & $3181$ & $-1.46^{+0.15}_{-0.15}$ & $228.00^{+0.10}_{-0.10}$ & $0.41^{+0.14}_{-0.13}$ & $0.56^{+0.09}_{-0.09}$\\ 
&$\log_{10} M_\star \ge 11.2$ & $1122$ & $-1.73^{+0.20}_{-0.20}$ & $228.64^{+0.16}_{-0.17}$ & $-0.19^{+0.15}_{-0.16}$ & $0.92^{+0.14}_{-0.12}$\\ 
&&&&&& \\
\hline
\parbox[t]{2mm}{\multirow{6}{*}{\rotatebox[origin=c]{90}{EMILES+BaSTI}}} &&&&&& \\
&$10.1 \le \log_{10} M_\star < 10.5$ & $1284$ & $-1.53^{+0.33}_{-0.34}$ & $229.28^{+0.12}_{-0.11}$ & $-0.06^{+0.29}_{-0.29}$ & $0.51^{+0.10}_{-0.10}$\\ 
&$10.5 \le \log_{10} M_\star < 10.9$ & $2370$ & $-2.77^{+0.15}_{-0.15}$ & $229.84^{+0.07}_{-0.07}$ & $-0.11^{+0.15}_{-0.15}$ & $0.45^{+0.07}_{-0.07}$\\ 
&$10.9 \le \log_{10} M_\star < 11.3$ & $3089$ & $-3.27^{+0.10}_{-0.09}$ & $230.21^{+0.06}_{-0.06}$ & $0.08^{+0.08}_{-0.08}$ & $0.35^{+0.05}_{-0.05}$\\ 
&$\log_{10} M_\star \ge 11.3$ & $1159$ & $-3.36^{+0.12}_{-0.12}$ & $230.51^{+0.09}_{-0.10}$ & $0.06^{+0.09}_{-0.10}$ & $0.33^{+0.08}_{-0.07}$\\ 
&&&&&& \\
\hline
\parbox[t]{2mm}{\multirow{6}{*}{\rotatebox[origin=c]{90}{EMILES+Padova00}}} &&&&&& \\
&$10.1 \le \log_{10} M_\star < 10.5$ & $1329$ & $-1.65^{+0.37}_{-0.36}$ & $229.04^{+0.13}_{-0.13}$ & $-0.79^{+0.32}_{-0.31}$ & $0.92^{+0.11}_{-0.11}$\\ 
&$10.5 \le \log_{10} M_\star < 10.9$ & $2335$ & $-1.84^{+0.16}_{-0.16}$ & $229.11^{+0.08}_{-0.08}$ & $-0.14^{+0.12}_{-0.13}$ & $0.57^{+0.06}_{-0.06}$\\ 
&$10.9 \le \log_{10} M_\star < 11.3$ & $2945$ & $-2.49^{+0.10}_{-0.10}$ & $229.51^{+0.07}_{-0.07}$ & $-0.17^{+0.08}_{-0.08}$ & $0.58^{+0.05}_{-0.05}$\\ 
&$\log_{10} M_\star \ge 11.3$ & $1141$ & $-2.52^{+0.12}_{-0.12}$ & $229.84^{+0.10}_{-0.10}$ & $-0.14^{+0.08}_{-0.08}$ & $0.52^{+0.06}_{-0.06}$\\ 
&&&&&& \\
\hline
\end{tabular}
\end{table*}


\begin{table*}
\caption{As Table~\ref{tab:par_age}, but for the probability distribution functions of luminosity and mass-weighted formation epochs in the real-space (Gyr units, see Eq.~(\ref{eq:pdf_age})).}
\label{tab:par_agetl}
\centering
\renewcommand{\arraystretch}{1.42}
\begin{tabular}{crccccc}
\hline\hline
\multicolumn{2}{c}{\multirow{2}{*}{$\mathrm{Age_L}+t_\mathrm{LB}\ \mathrm{[Gyr]}$}} & \multirow{2}{*}{$N_\mathrm{gal}$} & \multicolumn{2}{c}{$\mu_{\mathrm{Age_L}+t_\mathrm{LB}}$} & \multicolumn{2}{c}{$\sigma^\mathrm{int}_{\mathrm{Age_L}+t_\mathrm{LB}}$} \\ 
\cline{4-5} 
\cline{6-7} 
 & & & $\mu_1$ & $\mu_0$ & $\sigma_1$ & $\sigma_0$ \\ 
\hline
\parbox[t]{2mm}{\multirow{6}{*}{\rotatebox[origin=c]{90}{BC03}}} &&&&&& \\
&$10.0 \le \log_{10} M_\star < 10.4$ & $1429$ & $0.88^{+0.05}_{-0.05}$ & $1.70^{+0.02}_{-0.02}$ & $-0.01^{+0.05}_{-0.05}$ & $0.10^{+0.02}_{-0.02}$\\ 
&$10.4 \le \log_{10} M_\star < 10.8$ & $2526$ & $0.68^{+0.02}_{-0.02}$ & $1.78^{+0.01}_{-0.01}$ & $0.04^{+0.02}_{-0.02}$ & $0.07^{+0.01}_{-0.01}$\\ 
&$10.8 \le \log_{10} M_\star < 11.2$ & $3181$ & $0.62^{+0.01}_{-0.01}$ & $1.83^{+0.01}_{-0.01}$ & $-0.01^{+0.01}_{-0.01}$ & $0.09^{+0.01}_{-0.01}$\\ 
&$\log_{10} M_\star \ge 11.2$ & $1122$ & $0.45^{+0.02}_{-0.02}$ & $1.98^{+0.02}_{-0.02}$ & $-0.05^{+0.02}_{-0.02}$ & $0.13^{+0.02}_{-0.01}$\\ 
&&&&&& \\
\hline
\parbox[t]{2mm}{\multirow{6}{*}{\rotatebox[origin=c]{90}{EMILES+BaSTI}}} &&&&&& \\
&$10.1 \le \log_{10} M_\star < 10.5$ & $1284$ & $0.55^{+0.06}_{-0.06}$ & $2.02^{+0.02}_{-0.02}$ & $-0.20^{+0.05}_{-0.05}$ & $0.19^{+0.02}_{-0.02}$\\ 
&$10.5 \le \log_{10} M_\star < 10.9$ & $2370$ & $0.33^{+0.02}_{-0.02}$ & $2.12^{+0.01}_{-0.01}$ & $-0.10^{+0.02}_{-0.02}$ & $0.14^{+0.01}_{-0.01}$\\ 
&$10.9 \le \log_{10} M_\star < 11.3$ & $3089$ & $0.23^{+0.01}_{-0.01}$ & $2.20^{+0.01}_{-0.01}$ & $-0.09^{+0.01}_{-0.01}$ & $0.12^{+0.01}_{-0.01}$\\ 
&$\log_{10} M_\star \ge 11.3$ & $1159$ & $0.16^{+0.02}_{-0.01}$ & $2.29^{+0.01}_{-0.01}$ & $-0.07^{+0.01}_{-0.01}$ & $0.11^{+0.01}_{-0.01}$\\
&&&&&& \\
\hline
\parbox[t]{2mm}{\multirow{6}{*}{\rotatebox[origin=c]{90}{EMILES+Padova00}}} &&&&&& \\
&$10.1 \le \log_{10} M_\star < 10.5$ & $1329$ & $0.60^{+0.06}_{-0.06}$ & $1.96^{+0.02}_{-0.02}$ & $-0.24^{+0.04}_{-0.05}$ & $0.24^{+0.02}_{-0.02}$\\ 
&$10.5 \le \log_{10} M_\star < 10.9$ & $2335$ & $0.51^{+0.02}_{-0.02}$ & $2.00^{+0.01}_{-0.01}$ & $-0.11^{+0.02}_{-0.02}$ & $0.16^{+0.01}_{-0.01}$\\ 
&$10.9 \le \log_{10} M_\star < 11.3$ & $2945$ & $0.32^{+0.01}_{-0.01}$ & $2.12^{+0.01}_{-0.01}$ & $-0.08^{+0.01}_{-0.01}$ & $0.13^{+0.01}_{-0.01}$\\ 
&$\log_{10} M_\star \ge 11.3$ & $1141$ & $0.22^{+0.01}_{-0.01}$ & $2.23^{+0.01}_{-0.01}$ & $-0.06^{+0.01}_{-0.01}$ & $0.11^{+0.01}_{-0.01}$\\
&&&&&& \\
\hline
\end{tabular}
\begin{tabular}{crccccc}
\multicolumn{2}{c}{\multirow{2}{*}{$\mathrm{Age_M}+t_\mathrm{LB}\ \mathrm{[Gyr]}$}} & \multirow{2}{*}{$N_\mathrm{gal}$} & \multicolumn{2}{c}{$\mu_{\mathrm{Age_M}+t_\mathrm{LB}}$} & \multicolumn{2}{c}{$\sigma^\mathrm{int}_{\mathrm{Age_M}+t_\mathrm{LB}}$} \\ 
\cline{4-5} 
\cline{6-7} 
 & & & $\mu_1$ & $\mu_0$ & $\sigma_1$ & $\sigma_0$ \\ 
\hline
\parbox[t]{2mm}{\multirow{6}{*}{\rotatebox[origin=c]{90}{BC03}}} &&&&&& \\
&$10.0 \le \log_{10} M_\star < 10.4$ & $1429$ & $0.87^{+0.06}_{-0.06}$ & $1.83^{+0.02}_{-0.02}$ & $0.04^{+0.06}_{-0.06}$ & $0.08^{+0.02}_{-0.02}$\\ 
&$10.4 \le \log_{10} M_\star < 10.8$ & $2526$ & $0.58^{+0.03}_{-0.03}$ & $1.94^{+0.01}_{-0.01}$ & $-0.01^{+0.02}_{-0.02}$ & $0.10^{+0.01}_{-0.01}$\\ 
&$10.8 \le \log_{10} M_\star < 11.2$ & $3181$ & $0.54^{+0.02}_{-0.02}$ & $2.00^{+0.01}_{-0.01}$ & $-0.04^{+0.01}_{-0.01}$ & $0.11^{+0.01}_{-0.01}$\\ 
&$\log_{10} M_\star \ge 11.2$ & $1122$ & $0.34^{+0.02}_{-0.02}$ & $2.17^{+0.02}_{-0.02}$ & $-0.08^{+0.02}_{-0.02}$ & $0.15^{+0.02}_{-0.02}$\\ 
&&&&&& \\
\hline
\parbox[t]{2mm}{\multirow{6}{*}{\rotatebox[origin=c]{90}{EMILES+BaSTI}}} &&&&&& \\
&$10.1 \le \log_{10} M_\star < 10.5$ & $1284$ & $0.60^{+0.05}_{-0.05}$ & $2.15^{+0.02}_{-0.02}$ & $-0.09^{+0.04}_{-0.05}$ & $0.11^{+0.02}_{-0.02}$\\ 
&$10.5 \le \log_{10} M_\star < 10.9$ & $2370$ & $0.29^{+0.02}_{-0.02}$ & $2.28^{+0.01}_{-0.01}$ & $-0.07^{+0.02}_{-0.02}$ & $0.10^{+0.01}_{-0.01}$\\ 
&$10.9 \le \log_{10} M_\star < 11.3$ & $3089$ & $0.16^{+0.01}_{-0.01}$ & $2.37^{+0.01}_{-0.01}$ & $-0.05^{+0.01}_{-0.01}$ & $0.08^{+0.01}_{-0.01}$\\ 
&$\log_{10} M_\star \ge 11.3$ & $1159$ & $0.11^{+0.01}_{-0.01}$ & $2.43^{+0.01}_{-0.01}$ & $-0.04^{+0.01}_{-0.01}$ & $0.07^{+0.01}_{-0.01}$\\
&&&&&& \\
\hline
\parbox[t]{2mm}{\multirow{6}{*}{\rotatebox[origin=c]{90}{EMILES+Padova00}}} &&&&&& \\
&$10.1 \le \log_{10} M_\star < 10.5$ & $1329$ & $0.62^{+0.05}_{-0.05}$ & $2.10^{+0.02}_{-0.02}$ & $-0.22^{+0.05}_{-0.05}$ & $0.18^{+0.02}_{-0.02}$\\ 
&$10.5 \le \log_{10} M_\star < 10.9$ & $2335$ & $0.45^{+0.02}_{-0.02}$ & $2.17^{+0.01}_{-0.01}$ & $-0.06^{+0.02}_{-0.02}$ & $0.10^{+0.01}_{-0.01}$\\ 
&$10.9 \le \log_{10} M_\star < 11.3$ & $2945$ & $0.27^{+0.01}_{-0.01}$ & $2.28^{+0.01}_{-0.01}$ & $-0.07^{+0.01}_{-0.01}$ & $0.10^{+0.01}_{-0.01}$\\ 
&$\log_{10} M_\star \ge 11.3$ & $1141$ & $0.19^{+0.01}_{-0.01}$ & $2.36^{+0.01}_{-0.01}$ & $-0.05^{+0.01}_{-0.01}$ & $0.08^{+0.01}_{-0.01}$\\
&&&&&& \\
\hline
\end{tabular}
\end{table*}


\begin{table*}
\caption{As Table~\ref{tab:par_age}, but for the probability distribution functions of luminosity and mass-weighted formation epochs in the log-space (age in yr units).}
\label{tab:par_logagetl}
\centering
\renewcommand{\arraystretch}{1.42}
\begin{tabular}{crccccc}
\hline\hline
\multicolumn{2}{c}{\multirow{2}{*}{$\log_{10}\mathrm{Age_L}+t_\mathrm{LB}\ \mathrm{[yr]}$}} & \multirow{2}{*}{$N_\mathrm{gal}$} & \multicolumn{2}{c}{$\mu_{\log_{10}\mathrm{Age_L}+t_\mathrm{LB}}$} & \multicolumn{2}{c}{$\sigma^\mathrm{int}_{\log_{10}\mathrm{Age_L}+t_\mathrm{LB}}$} \\ 
\cline{4-5} 
\cline{6-7} 
 & & & $\mu_1~[10^{-2}]$ & $\mu_0~[10^{-2}]$ & $\sigma_1~[10^{-2}]$ & $\sigma_0~[10^{-2}]$ \\ 
\hline
\parbox[t]{2mm}{\multirow{6}{*}{\rotatebox[origin=c]{90}{BC03}}} &&&&&& \\
&$10.0 \le \log_{10} M_\star < 10.4$ & $1429$ & $3.88^{+0.22}_{-0.21}$ & $227.61^{+0.08}_{-0.08}$ & $-0.04^{+0.23}_{-0.23}$ & $0.45^{+0.09}_{-0.09}$\\ 
&$10.4 \le \log_{10} M_\star < 10.8$ & $2526$ & $2.95^{+0.09}_{-0.09}$ & $228.00^{+0.05}_{-0.05}$ & $0.18^{+0.10}_{-0.09}$ & $0.30^{+0.05}_{-0.05}$\\ 
&$10.8 \le \log_{10} M_\star < 11.2$ & $3181$ & $2.72^{+0.06}_{-0.06}$ & $228.20^{+0.04}_{-0.04}$ & $-0.04^{+0.05}_{-0.05}$ & $0.39^{+0.04}_{-0.04}$\\ 
&$\log_{10} M_\star \ge 11.2$ & $1122$ & $1.98^{+0.09}_{-0.09}$ & $228.85^{+0.08}_{-0.08}$ & $-0.24^{+0.07}_{-0.07}$ & $0.56^{+0.06}_{-0.06}$\\ 
&&&&&& \\
\hline
\parbox[t]{2mm}{\multirow{6}{*}{\rotatebox[origin=c]{90}{EMILES+BaSTI}}} &&&&&& \\
&$10.1 \le \log_{10} M_\star < 10.5$ & $1284$ & $2.42^{+0.23}_{-0.24}$ & $229.05^{+0.09}_{-0.09}$ & $-0.87^{+0.21}_{-0.20}$ & $0.82^{+0.08}_{-0.08}$\\ 
&$10.5 \le \log_{10} M_\star < 10.9$ & $2370$ & $1.44^{+0.10}_{-0.10}$ & $229.50^{+0.05}_{-0.05}$ & $-0.45^{+0.08}_{-0.08}$ & $0.59^{+0.04}_{-0.04}$\\ 
&$10.9 \le \log_{10} M_\star < 11.3$ & $3089$ & $0.98^{+0.05}_{-0.05}$ & $229.83^{+0.04}_{-0.04}$ & $-0.37^{+0.04}_{-0.04}$ & $0.53^{+0.03}_{-0.03}$\\ 
&$\log_{10} M_\star \ge 11.3$ & $1159$ & $0.69^{+0.07}_{-0.07}$ & $230.20^{+0.06}_{-0.06}$ & $-0.29^{+0.05}_{-0.05}$ & $0.48^{+0.04}_{-0.04}$\\ 
&&&&&& \\
\hline
\parbox[t]{2mm}{\multirow{6}{*}{\rotatebox[origin=c]{90}{EMILES+Padova00}}} &&&&&& \\
&$10.1 \le \log_{10} M_\star < 10.5$ & $1329$ & $2.63^{+0.25}_{-0.24}$ & $228.75^{+0.09}_{-0.09}$ & $-1.08^{+0.19}_{-0.19}$ & $1.03^{+0.07}_{-0.07}$\\ 
&$10.5 \le \log_{10} M_\star < 10.9$ & $2335$ & $2.22^{+0.10}_{-0.10}$ & $228.93^{+0.05}_{-0.05}$ & $-0.48^{+0.08}_{-0.08}$ & $0.70^{+0.04}_{-0.04}$\\ 
&$10.9 \le \log_{10} M_\star < 11.3$ & $2945$ & $1.40^{+0.05}_{-0.05}$ & $229.46^{+0.04}_{-0.04}$ & $-0.34^{+0.04}_{-0.04}$ & $0.55^{+0.03}_{-0.03}$\\ 
&$\log_{10} M_\star \ge 11.3$ & $1141$ & $0.95^{+0.06}_{-0.06}$ & $229.95^{+0.05}_{-0.05}$ & $-0.26^{+0.05}_{-0.04}$ & $0.49^{+0.04}_{-0.04}$\\ 
&&&&&& \\
\hline
\end{tabular}
\begin{tabular}{crccccc}
\multicolumn{2}{c}{\multirow{2}{*}{$\log_{10}\mathrm{Age_M}+t_\mathrm{LB}\ \mathrm{[yr]}$}} & \multirow{2}{*}{$N_\mathrm{gal}$} & \multicolumn{2}{c}{$\mu_{\log_{10}\mathrm{Age_M}+t_\mathrm{LB}}$} & \multicolumn{2}{c}{$\sigma^\mathrm{int}_{\log_{10}\mathrm{Age_M}+t_\mathrm{LB}}$} \\ 
\cline{4-5} 
\cline{6-7} 
 & & & $\mu_1~[10^{-2}]$ & $\mu_0~[10^{-2}]$ & $\sigma_1~[10^{-2}]$ & $\sigma_0~[10^{-2}]$ \\ 
\hline
\parbox[t]{2mm}{\multirow{6}{*}{\rotatebox[origin=c]{90}{BC03}}} &&&&&& \\
&$10.0 \le \log_{10} M_\star < 10.4$ & $1429$ & $3.78^{+0.26}_{-0.25}$ & $228.22^{+0.09}_{-0.10}$ & $0.18^{+0.25}_{-0.25}$ & $0.35^{+0.09}_{-0.09}$\\ 
&$10.4 \le \log_{10} M_\star < 10.8$ & $2526$ & $2.54^{+0.11}_{-0.11}$ & $228.70^{+0.06}_{-0.06}$ & $-0.05^{+0.10}_{-0.10}$ & $0.43^{+0.06}_{-0.06}$\\ 
&$10.8 \le \log_{10} M_\star < 11.2$ & $3181$ & $2.33^{+0.07}_{-0.07}$ & $228.93^{+0.05}_{-0.05}$ & $-0.18^{+0.06}_{-0.06}$ & $0.49^{+0.05}_{-0.04}$\\ 
&$\log_{10} M_\star \ge 11.2$ & $1122$ & $1.46^{+0.10}_{-0.10}$ & $229.70^{+0.09}_{-0.09}$ & $-0.35^{+0.08}_{-0.08}$ & $0.64^{+0.07}_{-0.07}$\\ 
&&&&&& \\
\hline
\parbox[t]{2mm}{\multirow{6}{*}{\rotatebox[origin=c]{90}{EMILES+BaSTI}}} &&&&&& \\
&$10.1 \le \log_{10} M_\star < 10.5$ & $1284$ & $2.61^{+0.23}_{-0.22}$ & $229.62^{+0.08}_{-0.08}$ & $-0.39^{+0.19}_{-0.20}$ & $0.49^{+0.08}_{-0.07}$\\ 
&$10.5 \le \log_{10} M_\star < 10.9$ & $2370$ & $1.24^{+0.09}_{-0.09}$ & $230.20^{+0.05}_{-0.05}$ & $-0.30^{+0.09}_{-0.09}$ & $0.41^{+0.05}_{-0.05}$\\ 
&$10.9 \le \log_{10} M_\star < 11.3$ & $3089$ & $0.69^{+0.05}_{-0.05}$ & $230.55^{+0.04}_{-0.04}$ & $-0.19^{+0.04}_{-0.04}$ & $0.35^{+0.03}_{-0.03}$\\ 
&$\log_{10} M_\star \ge 11.3$ & $1159$ & $0.46^{+0.06}_{-0.06}$ & $230.83^{+0.05}_{-0.05}$ & $-0.16^{+0.05}_{-0.05}$ & $0.30^{+0.04}_{-0.04}$\\ 
&&&&&& \\
\hline
\parbox[t]{2mm}{\multirow{6}{*}{\rotatebox[origin=c]{90}{EMILES+Padova00}}} &&&&&& \\
&$10.1 \le \log_{10} M_\star < 10.5$ & $1329$ & $2.70^{+0.22}_{-0.23}$ & $229.41^{+0.08}_{-0.08}$ & $-0.96^{+0.19}_{-0.20}$ & $0.79^{+0.07}_{-0.07}$\\ 
&$10.5 \le \log_{10} M_\star < 10.9$ & $2335$ & $1.95^{+0.09}_{-0.10}$ & $229.68^{+0.05}_{-0.05}$ & $-0.25^{+0.08}_{-0.08}$ & $0.45^{+0.04}_{-0.04}$\\ 
&$10.9 \le \log_{10} M_\star < 11.3$ & $2945$ & $1.15^{+0.05}_{-0.05}$ & $230.17^{+0.04}_{-0.04}$ & $-0.29^{+0.04}_{-0.04}$ & $0.41^{+0.03}_{-0.03}$\\ 
&$\log_{10} M_\star \ge 11.3$ & $1141$ & $0.83^{+0.06}_{-0.06}$ & $230.52^{+0.05}_{-0.05}$ & $-0.21^{+0.04}_{-0.04}$ & $0.36^{+0.03}_{-0.03}$\\ 
&&&&&& \\
\hline
\end{tabular}
\end{table*}


\begin{table*}
\caption{As Table~\ref{tab:par_age}, but for the probability distribution functions of luminosity and mass-weighted metallicities (see Eq.~(\ref{eq:pdf_feh})).}
\label{tab:par_feh}
\centering
\renewcommand{\arraystretch}{1.42}
\begin{tabular}{crccccccc}
\hline\hline
\multicolumn{2}{c}{\multirow{2}{*}{$\mathrm{[M/H]_L}$}} & \multirow{2}{*}{$N_\mathrm{gal}$} & \multicolumn{3}{c}{$\mu_{\mathrm{[M/H]_L}}$} & \multicolumn{3}{c}{$\sigma^\mathrm{int}_{\mathrm{[M/H]_L}}$} \\ 
\cline{4-6} 
\cline{7-9} 
& & & $\mu_2$ & $\mu_1$ & $\mu_0$ & $\sigma_2$ & $\sigma_1$ & $\sigma_0$ \\ 
\hline
\parbox[t]{2mm}{\multirow{6}{*}{\rotatebox[origin=c]{90}{BC03}}} &&&&&&&& \\
&$10.0 \le \log_{10} M_\star < 10.4$ & $1429$ & $1.73^{+0.61}_{-0.58}$ & $-1.38^{+0.35}_{-0.36}$ & $0.29^{+0.05}_{-0.05}$ & \tablefootmark{*}$0.00^{+0.00}_{-0.00}$ & $0.59^{+0.07}_{-0.06}$ & $0.01^{+0.02}_{-0.02}$\\ 
&$10.4 \le \log_{10} M_\star < 10.8$ & $2526$ & $1.38^{+0.21}_{-0.20}$ & $-1.22^{+0.17}_{-0.18}$ & $0.22^{+0.03}_{-0.03}$ & $-0.49^{+0.20}_{-0.19}$ & $0.71^{+0.16}_{-0.17}$ & $0.02^{+0.03}_{-0.03}$\\ 
&$10.8 \le \log_{10} M_\star < 11.2$ & $3181$ & $1.49^{+0.11}_{-0.10}$ & $-1.34^{+0.12}_{-0.12}$ & $0.21^{+0.03}_{-0.03}$ & $-0.83^{+0.10}_{-0.10}$ & $0.98^{+0.11}_{-0.11}$ & $-0.04^{+0.03}_{-0.03}$\\ 
&$\log_{10} M_\star \ge 11.2$ & $1122$ & $0.77^{+0.12}_{-0.12}$ & $-0.60^{+0.16}_{-0.15}$ & $0.00^{+0.04}_{-0.05}$ & $-0.49^{+0.09}_{-0.08}$ & $0.73^{+0.10}_{-0.12}$ & $-0.05^{+0.04}_{-0.02}$\\ 
&&&&&&&&\\
\hline
\parbox[t]{2mm}{\multirow{6}{*}{\rotatebox[origin=c]{90}{EMILES+BaSTI}}} &&&&&&&& \\
&$10.1 \le \log_{10} M_\star < 10.5$ & $1284$ & -- & $-0.45^{+0.05}_{-0.05}$ & $0.39^{+0.01}_{-0.01}$ & -- & $0.42^{+0.05}_{-0.05}$ & $0.00^{+0.01}_{-0.01}$\\ 
&$10.5 \le \log_{10} M_\star < 10.9$ & $2370$ & -- & $-0.26^{+0.03}_{-0.03}$ & $0.29^{+0.01}_{-0.01}$ & -- & $0.21^{+0.02}_{-0.02}$ & $0.07^{+0.01}_{-0.01}$\\ 
&$10.9 \le \log_{10} M_\star < 11.3$ & $3089$ & -- & $-0.00^{+0.02}_{-0.02}$ & $0.18^{+0.01}_{-0.01}$ & -- & $0.08^{+0.02}_{-0.02}$ & $0.10^{+0.01}_{-0.01}$\\ 
&$\log_{10} M_\star \ge 11.3$ & $1159$ & -- & $0.07^{+0.02}_{-0.02}$ & $0.10^{+0.02}_{-0.02}$ & -- & $0.03^{+0.02}_{-0.02}$ & $0.13^{+0.02}_{-0.01}$\\
&&&&&&&&\\
\hline
\parbox[t]{2mm}{\multirow{6}{*}{\rotatebox[origin=c]{90}{EMILES+Padova00}}} &&&&&&&& \\
&$10.1 \le \log_{10} M_\star < 10.5$ & $1329$ & -- & $-0.29^{+0.04}_{-0.04}$ & $0.37^{+0.01}_{-0.01}$ & -- & $0.27^{+0.04}_{-0.04}$ & $0.04^{+0.01}_{-0.01}$\\ 
&$10.5 \le \log_{10} M_\star < 10.9$ & $2335$ & -- & $-0.05^{+0.02}_{-0.02}$ & $0.25^{+0.01}_{-0.01}$ & -- & $0.19^{+0.02}_{-0.02}$ & $0.05^{+0.01}_{-0.01}$\\ 
&$10.9 \le \log_{10} M_\star < 11.3$ & $2945$ & -- & $0.09^{+0.02}_{-0.02}$ & $0.18^{+0.01}_{-0.01}$ & -- & $0.04^{+0.01}_{-0.01}$ & $0.10^{+0.01}_{-0.01}$\\ 
&$\log_{10} M_\star \ge 11.3$ & $1141$ & -- & $0.13^{+0.02}_{-0.02}$ & $0.12^{+0.01}_{-0.02}$ & -- & $-0.02^{+0.02}_{-0.02}$ & $0.14^{+0.01}_{-0.01}$\\
&&&&&&&&\\
\hline
\end{tabular}
\begin{tabular}{crccccccc}
\multicolumn{2}{c}{\multirow{2}{*}{$\mathrm{[M/H]_M}$}} & \multirow{2}{*}{$N_\mathrm{gal}$} & \multicolumn{3}{c}{$\mu_{\mathrm{[M/H]_M}}$} & \multicolumn{3}{c}{$\sigma^\mathrm{int}_{\mathrm{[M/H]_M}}$} \\ 
\cline{4-6} 
\cline{7-9} 
& & & $\mu_2$ & $\mu_1$ & $\mu_0$ & $\sigma_2$ & $\sigma_1$ & $\sigma_0$ \\ 
\hline
\parbox[t]{2mm}{\multirow{6}{*}{\rotatebox[origin=c]{90}{BC03}}} &&&&&&&& \\
&$10.0 \le \log_{10} M_\star < 10.4$ & $1429$ & $1.49^{+0.66}_{-0.64}$ & $-1.55^{+0.38}_{-0.41}$ & $0.33^{+0.06}_{-0.06}$ & \tablefootmark{*}$0.00^{+0.00}_{-0.00}$ & $0.40^{+0.07}_{-0.07}$ & $0.07^{+0.02}_{-0.02}$\\ 
&$10.4 \le \log_{10} M_\star < 10.8$ & $2526$ & $1.29^{+0.22}_{-0.22}$ & $-1.46^{+0.19}_{-0.19}$ & $0.25^{+0.04}_{-0.04}$ & $-0.26^{+0.20}_{-0.20}$ & $0.45^{+0.18}_{-0.18}$ & $0.07^{+0.03}_{-0.04}$\\ 
&$10.8 \le \log_{10} M_\star < 11.2$ & $3181$ & $1.16^{+0.11}_{-0.12}$ & $-1.43^{+0.14}_{-0.13}$ & $0.22^{+0.03}_{-0.03}$ & $-0.29^{+0.11}_{-0.11}$ & $0.49^{+0.12}_{-0.12}$ & $0.03^{+0.03}_{-0.03}$\\ 
&$\log_{10} M_\star \ge 11.2$ & $1122$ & $0.83^{+0.12}_{-0.12}$ & $-1.09^{+0.16}_{-0.15}$ & $0.13^{+0.04}_{-0.04}$ & $-0.32^{+0.09}_{-0.08}$ & $0.57^{+0.10}_{-0.12}$ & $-0.04^{+0.04}_{-0.02}$\\ 
&&&&&&&&\\
\hline
\parbox[t]{2mm}{\multirow{6}{*}{\rotatebox[origin=c]{90}{EMILES+BaSTI}}} &&&&&&&& \\
&$10.1 \le \log_{10} M_\star < 10.5$ & $1284$ & -- & $-0.66^{+0.06}_{-0.06}$ & $0.37^{+0.02}_{-0.02}$ & -- & $0.43^{+0.05}_{-0.05}$ & $0.03^{+0.02}_{-0.02}$\\ 
&$10.5 \le \log_{10} M_\star < 10.9$ & $2370$ & -- & $-0.47^{+0.03}_{-0.03}$ & $0.23^{+0.01}_{-0.01}$ & -- & $0.12^{+0.02}_{-0.02}$ & $0.12^{+0.01}_{-0.01}$\\ 
&$10.9 \le \log_{10} M_\star < 11.3$ & $3089$ & -- & $-0.32^{+0.02}_{-0.02}$ & $0.13^{+0.01}_{-0.01}$ & -- & $0.02^{+0.02}_{-0.02}$ & $0.14^{+0.01}_{-0.01}$\\ 
&$\log_{10} M_\star \ge 11.3$ & $1159$ & -- & $-0.27^{+0.03}_{-0.03}$ & $0.09^{+0.02}_{-0.02}$ & -- & $0.01^{+0.02}_{-0.02}$ & $0.15^{+0.01}_{-0.01}$\\
&&&&&&&&\\
\hline
\parbox[t]{2mm}{\multirow{6}{*}{\rotatebox[origin=c]{90}{EMILES+Padova00}}} &&&&&&&& \\
&$10.1 \le \log_{10} M_\star < 10.5$ & $1329$ & -- & $-0.54^{+0.05}_{-0.05}$ & $0.44^{+0.02}_{-0.02}$ & -- & $0.26^{+0.04}_{-0.04}$ & $0.05^{+0.01}_{-0.01}$\\ 
&$10.5 \le \log_{10} M_\star < 10.9$ & $2335$ & -- & $-0.33^{+0.03}_{-0.03}$ & $0.29^{+0.01}_{-0.01}$ & -- & $0.17^{+0.02}_{-0.02}$ & $0.07^{+0.01}_{-0.01}$\\ 
&$10.9 \le \log_{10} M_\star < 11.3$ & $2945$ & -- & $-0.14^{+0.02}_{-0.02}$ & $0.16^{+0.01}_{-0.01}$ & -- & $0.02^{+0.01}_{-0.01}$ & $0.13^{+0.01}_{-0.01}$\\ 
&$\log_{10} M_\star \ge 11.3$ & $1141$ & -- & $-0.10^{+0.02}_{-0.02}$ & $0.09^{+0.02}_{-0.02}$ & -- & $-0.04^{+0.02}_{-0.02}$ & $0.15^{+0.01}_{-0.01}$\\
&&&&&&&&\\
\hline
\end{tabular}
\tablefoot{\tablefoottext{*}{For $10.0 \le \log_{10} M_\star < 10.4$ and BC03 SSP models, a linear redshift-dependence of $\sigma^\mathrm{int}_\mathrm{[M/H]}$ was assumed.}}
\end{table*}


\begin{table*}
\caption{As Table~\ref{tab:par_age}, but for the probability distribution functions of extinctions (see Eq.~(\ref{eq:pdf_av})).}
\label{tab:par_av}
\centering
\renewcommand{\arraystretch}{1.42}
\begin{tabular}{crccccc}
\hline\hline
\multicolumn{2}{c}{\multirow{2}{*}{$A_V$}} & \multirow{2}{*}{$N_\mathrm{gal}$} & \multicolumn{2}{c}{$\mu_{A_V}$} & \multicolumn{2}{c}{$\sigma^\mathrm{int}_{A_V}$}\\
\cline{4-5} 
\cline{6-7} 
& & & $\mu_1$ & $\mu_0$ & $\sigma_1$ & $\sigma_0$ \\ 
\hline
\parbox[t]{2mm}{\multirow{6}{*}{\rotatebox[origin=c]{90}{BC03}}} &&&&&& \\
&$10.0 \le \log_{10} M_\star < 10.4$ & $1429$ & $-0.78^{+0.38}_{-0.35}$ & $-1.50^{+0.11}_{-0.12}$ & $1.71^{+0.30}_{-0.29}$ & $0.39^{+0.10}_{-0.10}$\\ 
&$10.4 \le \log_{10} M_\star < 10.8$ & $2526$ & $-0.50^{+0.15}_{-0.14}$ & $-1.36^{+0.06}_{-0.06}$ & $1.36^{+0.13}_{-0.12}$ & $0.16^{+0.06}_{-0.06}$\\ 
&$10.8 \le \log_{10} M_\star < 11.2$ & $3181$ & $-0.61^{+0.11}_{-0.11}$ & $-1.37^{+0.07}_{-0.07}$ & $0.76^{+0.07}_{-0.07}$ & $0.50^{+0.04}_{-0.04}$\\ 
&$^{*}\log_{10} M_\star \ge 11.2$ & $1122$ & $-0.76^{+0.14}_{-0.13}$ & $-1.14^{+0.09}_{-0.09}$ & $1.29^{+0.05}_{-0.05}$ & $-0.12^{+0.03}_{-0.02}$\\ 
&&&&&&\\
\hline
\parbox[t]{2mm}{\multirow{6}{*}{\rotatebox[origin=c]{90}{EMILES+BaSTI}}} &&&&&& \\
&$10.1 \le \log_{10} M_\star < 10.5$ & $1284$ & $-1.31^{+0.27}_{-0.27}$ & $-1.10^{+0.08}_{-0.08}$ & $2.91^{+0.22}_{-0.22}$ & $-0.14^{+0.07}_{-0.06}$\\ 
&$10.5 \le \log_{10} M_\star < 10.9$ & $2370$ & $-0.39^{+0.15}_{-0.16}$ & $-1.35^{+0.07}_{-0.07}$ & $0.72^{+0.11}_{-0.11}$ & $0.54^{+0.05}_{-0.05}$\\ 
&$10.9 \le \log_{10} M_\star < 11.3$ & $3089$ & $-0.25^{+0.09}_{-0.09}$ & $-1.34^{+0.06}_{-0.05}$ & $0.72^{+0.07}_{-0.07}$ & $0.41^{+0.04}_{-0.04}$\\ 
&$\log_{10} M_\star \ge 11.3$ & $1159$ & $-0.14^{+0.12}_{-0.12}$ & $-1.40^{+0.08}_{-0.09}$ & $0.68^{+0.09}_{-0.09}$ & $0.27^{+0.07}_{-0.06}$\\
&&&&&&\\
\hline
\parbox[t]{2mm}{\multirow{6}{*}{\rotatebox[origin=c]{90}{EMILES+Padova00}}} &&&&&& \\
&$10.1 \le \log_{10} M_\star < 10.5$ & $1329$ & $-0.68^{+0.30}_{-0.30}$ & $-1.39^{+0.09}_{-0.09}$ & $2.22^{+0.25}_{-0.26}$ & $0.16^{+0.08}_{-0.08}$\\ 
&$10.5 \le \log_{10} M_\star < 10.9$ & $2335$ & $-0.61^{+0.16}_{-0.16}$ & $-1.36^{+0.07}_{-0.07}$ & $1.33^{+0.12}_{-0.11}$ & $0.38^{+0.05}_{-0.05}$\\ 
&$10.9 \le \log_{10} M_\star < 11.3$ & $2945$ & $-0.16^{+0.12}_{-0.12}$ & $-1.61^{+0.07}_{-0.07}$ & $0.67^{+0.08}_{-0.08}$ & $0.69^{+0.05}_{-0.05}$\\ 
&$\log_{10} M_\star \ge 11.3$ & $1141$ & $0.17^{+0.15}_{-0.15}$ & $-1.77^{+0.11}_{-0.11}$ & $0.21^{+0.11}_{-0.12}$ & $0.70^{+0.09}_{-0.08}$\\
&&&&&&\\
\hline
\end{tabular}
\tablefoot{\tablefoottext{*}{For quiescent galaxies and BC03 SSP models at $0.1 \le z < 0.2$ and $\log_{10} M_\star \ge 11.2$, the assumption of linear dependence for $\sigma^\mathrm{int}(z)$ and $\mu(z)$ is too strict, and we impose $\sigma^\mathrm{int}(0.1 \le z < 0.2)=\sigma^\mathrm{int}(z=0.2) = 0.14 \pm 0.03$ and $\mu(0.1 \le z < 0.2) = \mu(z=0.2) = -1.29 \pm 0.09$ (details in text).}}
\end{table*}

\end{appendix}
\end{document}